\documentclass[11pt,preprint]{aastex}
\usepackage{float}
\usepackage{graphics,graphicx}
\usepackage[small,sc]{caption2}
\usepackage{multicol}
\usepackage{paralist}
\usepackage{url}
\usepackage[usenames,dvipsnames]{color}
\usepackage{mdframed}
\usepackage{environ}
\usepackage{calc}
\usepackage{lipsum}
\usepackage[normalem]{ulem}   
\usepackage{bold-extra}
\usepackage{lscape}
\usepackage{enumitem}
\usepackage[margin=1in, top=0.9in, headsep=.2in]{geometry}
\usepackage{courier}
\usepackage{booktabs}
\usepackage{longtable}
\usepackage{array}
\usepackage{indentfirst}
\usepackage{mathtools}
\usepackage{tensor}
\definecolor{linkcolor}{rgb}{0.2,0.2,0.6}
\usepackage{dcolumn}    
\usepackage{bm}         
\usepackage{tabularx}
\usepackage{fullpage}
\usepackage[breaklinks,colorlinks=true, citecolor=blue, linkcolor=blue, urlcolor=blue]{hyperref}
\newcommand{\ra}[1]{\renewcommand{\arraystretch}{#1}}
\usepackage{natbib}
\usepackage{etoolbox}
\usepackage{array}
\newcolumntype{L}[1]{>{\raggedright\let\newline\\\arraybackslash\hspace{0pt}}m{#1}}
\newcolumntype{C}[1]{>{\centering\let\newline\\\arraybackslash\hspace{0pt}}m{#1}}
\newcolumntype{R}[1]{>{\raggedleft\let\newline\\\arraybackslash\hspace{0pt}}m{#1}}

\makeatletter
\renewcommand{\section}{\@startsection%
{section}{1}{0mm}{-\baselineskip}%
{0.1\baselineskip}{\normalfont\Large\bfseries}}%
\makeatother

\newcommand{\ltsima}{$\; \buildrel < \over \sim \;$}
\newcommand{\simlt}{\lower.5ex\hbox{\ltsima}}

\newcommand{\adsurl}[1]{\href{#1}{ADS}} 
\providecommand{\url}[1]{\href{#1}{#1}}

\newcommand{\co}{\mbox{$\rm CO$}}
\newcommand{\thirteenco}{\mbox{$\rm ^{13}CO$}}
\newcommand{\water}{\mbox{$\rm H_{2}O$}}
\newcommand{\hcoplus}{\mbox{$\rm HCO^{+}$}}
\newcommand{\chplus}{\mbox{$\rm CH^{+}$}}
\newcommand{\owater}{\mbox{o-$\rm H_{2}O$}}
\newcommand{\pwater}{\mbox{p-$\rm H_{2}O$}}
\newcommand{\coul}[2]{\mbox{${\rm CO}~J=#1\rightarrow#2$}}
\newcommand{\tcoul}[2]{\mbox{${\rm ^{13}CO}~J=#1\rightarrow#2$}}
\newcommand{\hcoul}[2]{\mbox{${\rm HCO^{+}}~J=#1\rightarrow#2$}}
\newcommand{\chul}[2]{\mbox{${\rm CH^{+}}~J=#1\rightarrow#2$}}
\newcommand{\ohul}[3]{\mbox{$\rm ~^{2}\Pi_{#1,~#2}^{#3}$}}
\newcommand{\lsmm}{\mbox{$L_{\rm smm}$}} 
\newcommand{\lbol}{\mbox{$L_{\rm bol}$}} 
\newcommand{\tbol}{\mbox{$T_{\rm bol}$}} 
\newcommand{\jj}[2]{\mbox{$J = #1\rightarrow#2$}}

\newcommand{\trot}{\mbox{$T_{\rm rot}$}}
\newcommand{\OI}{\mbox{[\ion{O}{1}]}}
\newcommand{\CI}{\mbox{[\ion{C}{1}]}}
\newcommand{\CII}{\mbox{[\ion{C}{2}]}}
\newcommand{\NII}{\mbox{[\ion{N}{2}]}}

\shorttitle{CDF Archive}
\shortauthors{Green et al.}

\begin{document}

\title{The CDF Archive: Herschel PACS and SPIRE Spectroscopic Data Pipeline and Products for Protostars and Young Stellar Objects}

\author{
Joel D. Green \altaffilmark{1,2},
Yao-Lun Yang\altaffilmark{1},
Neal J. Evans II\altaffilmark{1},
Agata Karska\altaffilmark{3},
Gregory Herczeg\altaffilmark{4},
Ewine F. van Dishoeck\altaffilmark{5,6},
Jeong-Eun Lee\altaffilmark{7},
Rebecca L. Larson\altaffilmark{1},
\& Jeroen Bouwman\altaffilmark{8}
}

\affil{
1.  The University of Texas at Austin, Department of
Astronomy, 2515 Speedway, Stop C1400,
Austin, TX 78712-1205, USA \\
2. Space Telescope Science Institute, Baltimore, MD, USA \\
3. Astronomical Observatory Institute, Faculty of Physics, A. Mickiewicz University, Sloneczna 36, 60-286 Poznan, Poland \\
4. Kavli Institute for Astronomy and Astrophysics, Peking University, Yi He Yuan Lu 5, Haidian Qu, 100871 Beijing, China \\
5. Leiden Observatory, Leiden University, Netherlands \\
6. Max Planck Institute for Extraterrestrial Physics, Garching, Germany \\
7. Department of Astronomy \& Space Science, Kyung Hee University, Gyeonggi 446-701, Korea  \\
School of Space Research, Kyung Hee University, Yongin-shi, Kyungki-do 449-701, Korea \\
8.  Max Planck Institute for Astronomy, Heidelberg, Germany \\
 }

\begin{abstract}

We present the COPS-DIGIT-FOOSH (CDF) {\it Herschel} spectroscopy data product archive, and related ancillary data products, along with data fidelity assessments, and a user-created archive in collaboration with the Herschel-PACS and SPIRE ICC groups.  Our products include datacubes, contour maps, automated line fitting results, and best 1-D spectra products for all protostellar and disk sources observed with PACS in RangeScan mode for two observing programs: the DIGIT Open Time Key Program (KPOT\_nevans\_1 and SDP\_nevans\_1; PI: N.~Evans), and the FOOSH Open Time Program (OT1\_jgreen02\_2; PI: J.~Green).  In addition, we provide our best SPIRE-FTS spectroscopic products for the COPS Open Time Program (OT2\_jgreen02\_6; PI: J.~Green) and FOOSH sources.  We include details of data processing, descriptions of output products, and tests of their reliability for user applications.  We  identify the parts of the dataset to be used with caution.  The resulting absolute flux calibration has improved in almost all cases.   Compared to previous reductions, the resulting rotational temperatures and numbers of
CO molecules have changed substantially in some sources. On average, however, the rotational temperatures have not changed substantially ($< 2$\%), but the number of
warm ($\trot \sim 300$ K) CO molecules has increased by about $18$\%.

\end{abstract}
\keywords{Astronomical databases —- catalogs —- surveys —- stars: pre-main
sequence}

\section{Introduction}

It has been long established that protostars and young stars form from the collapse of a dense molecular core, developing through different stages as the relative density, temperature, and composition of the envelope, disk, and protostar shift over time.  However, many uncertainties remain in the details of the collapse. Specific conditions during infall and outflow throughout the protostellar stage may drive the conditions in the resulting circumstellar disks and protoplanetary systems, influencing the compositional properties of the planet-forming material.  Particularly important is the role played by accretion-driven heating events; whether the accretion process is characterized by gradually diminishing accretion \citep{offner11} or episodic accretion \citep[e.g.,][]{dunham10, kimhj12b,dunham14a} can have a large impact.  In either case, the final disk mass and chemistry may be set by the time spent in relatively high accretion phases; in the case of episodic accretion, the disk mass at
the end of the protostellar phase may be determined by the phasing
of the last burst of accretion onto the star and the end of infall.
Observational constraints on the physical processes in these systems are gained from 
a multi-wavelength understanding of molecular, atomic, and ionic tracers  through optical,
infrared, and millimeter-wave telescopes.

Many protostars in relatively nearby ($d \leq 300$ pc) clouds
have been studied extensively in the infrared. The Infrared Space Observatory's Long Wavelength Spectrograph detected gas phase H$_2$O, high-$J$ CO rotational transitions, and fine
structure emission lines toward protostars and related sources
\citep[e.g.,][]{lorenzetti99,giannini99,ceccarelli99,lorenzetti00,giannini01,nisini02}.  These lines trace the innermost regions of the protostellar envelope, 
exposed to heating by the central object, and the outflow cavity region, 
where winds and jets may interact with the envelope and the surrounding cloud.  

The Spitzer c2d (``Cores to Disks'')  and Gould Belt Legacy surveys 
(\citealt{evans09}, Dunham et al. submitted), 
along with a survey of the Taurus cloud 
\citep{2010ApJS..186..259R},
have produced a rather complete list of young stellar objects within 300 pc.
These source lists in turn informed key program surveys in the far-infrared and submillimeter, with the Herschel Space Observatory, an ESA space-based  telescope with a 3.5-meter primary mirror optimized for far-infrared and 
submillimeter observations.  {\it Herschel}-SPIRE \citep[Spectral and Photometric Imaging REceiver, 194-670 $\mu$m;][]{griffin10} allowed low resolution spectroscopy over the entire submillimeter domain; the PACS (Photodetector Array Camera and Spectrometer; \citealt{poglitsch10}) 
instrument provided low-resolution spectroscopy over the far-infrared range.  {\it Herschel} was sensitive to dust continuum, and had access to the full suite of mid-$J$ emission lines from CO, HCO$^+$, $^{13}$CO, and several low-lying energy states of H$_2$O, which trace the shocked gas in the outflow and the surrounding envelope. 

In a previous paper \citep{green13b} we used data from the PACS spectrograph to characterize a sample of well-studied protostars, selected primarily from the c2d sample, including both Class 0 and Class I objects.
Class 0 and Class I sources are characterized observationally by rising spectral 
energy distributions (SEDs)
between near-infrared and mid-infrared wavelengths.
In addition to the continuum emission, 
the far-infrared/submillimeter bands contain
numerous pure rotational transitions of the CO ladder,
as well as low-lying lines of H$_2$O, OH, HCO$^+$, atomic lines (\CI, \OI), and ionic lines (\CII, \NII), all potential tracers of gas
content and properties.   The
transitions and collisional rates of these simple molecules are
well-understood
\citep[see, e.g.,][for a recent update on CO]{yang10,neufeld12}.  Thus these lines make
excellent diagnostics of opacity, density, temperature, and shock velocities
\citep[e.g.,][]{kaufman96,flower10} of the
gas surrounding these systems. 
In a second previous paper \citep{green13c},  we combined PACS, SPIRE, and ground-based spectroscopy of stars undergoing episodic accretion events (FU Orionis objects, hereafter FUors) in order to determine their gas and envelope properties post-outburst.  Both previous papers used an earlier data reduction pipeline; we will compare results from those papers to those obtained with the 
new reduction presented here for a few characteristic quantities in \S 5.3.

This paper describes the CDF (COPS-DIGIT-FOOSH) archive, with {\it Herschel}-PACS and SPIRE spectroscopic observations of 70 objects (protostars, young stellar objects, and FUors) from the ``Dust, Ice, and Gas in Time'' (DIGIT Key Project), ``FU Orionis Objects Surveyed with {\it Herschel}'' (FOOSH OT1), and ``CO in Protostars'' (COPS OT2) {\it Herschel} programs. These have been delivered
to the {\it Herschel} archive and are available. Here we describe the reduction
methods and the data products.

We use data products from the HIPE 13 / CalTree 65 pipeline, provided by the Herschel Science Center, the most current version at the time of reprocessing during summer 2014. The most notable new feature in this reduction of PACS spectroscopy is a correction for pointing and jitter offsets during observations.  The SPIRE spectra are also enhanced by
the use of a correction for semi-extended sources
\citep{2013A&A...556A.116W}. 
The spectra are reduced and analyzed using automated routines optimized for {\it Herschel} to detect line emission and continuum properties.  We provide spectroscopic data cubes, line fluxes, continuum analysis, and error analysis across all spatial positions observed in these 70 sources.  In addition to this paper, a full description of the pipeline and details of this archive can be found in the web release of the data to the Herschel Science Archive (HSA) User Provided Data Products\footnote{\href{http://www.cosmos.esa.int/web/herschel/user-provided-data-products}{http://www.cosmos.esa.int/web/herschel/user-provided-data-products}}, upcoming analysis papers (\textcolor{blue}{Green et al.}, \textcolor{blue}{in prep.}), and a paper on 3-D radiative transfer modeling of a single source (the protostar BHR71; \textcolor{blue}{Yang~et~al.}, \textcolor{blue}{in prep.}). The corresponding {\it Spitzer} data can mostly be found in IRS\_Disks and c2d spectroscopic 
programs
\citep{lahuis06}.  
Finally we note that this analysis could be expanded to include all objects observed in similar modes, and the automated routines adapted to other (non-Herschel) datasets.

\section{Observations }\label{obs}

\subsection{Dust, Ice, and Gas In Time (DIGIT)}

\begin{figure}
	\centering
	\includegraphics[height=2.5in]{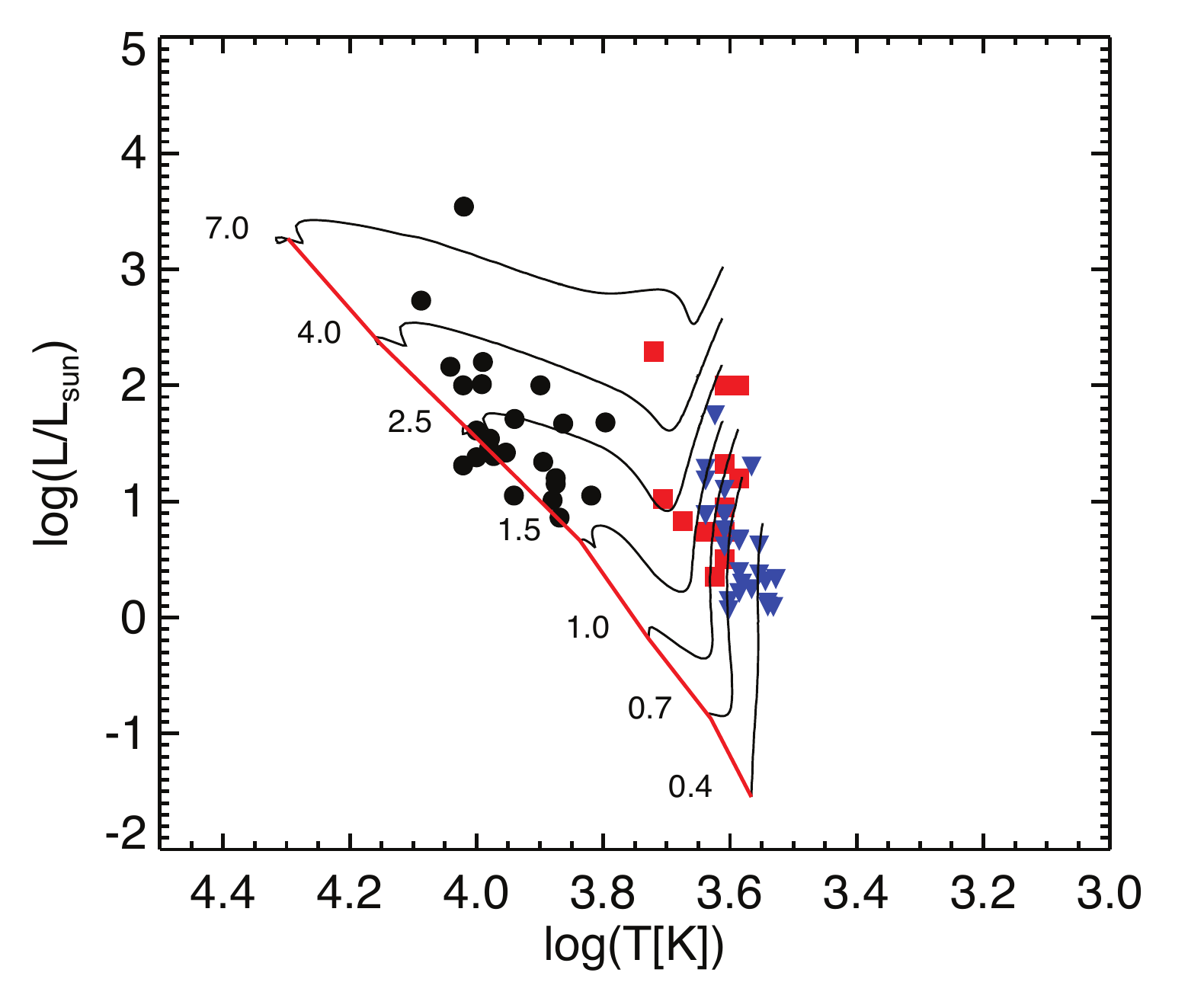}
	\includegraphics[height=2.5in]{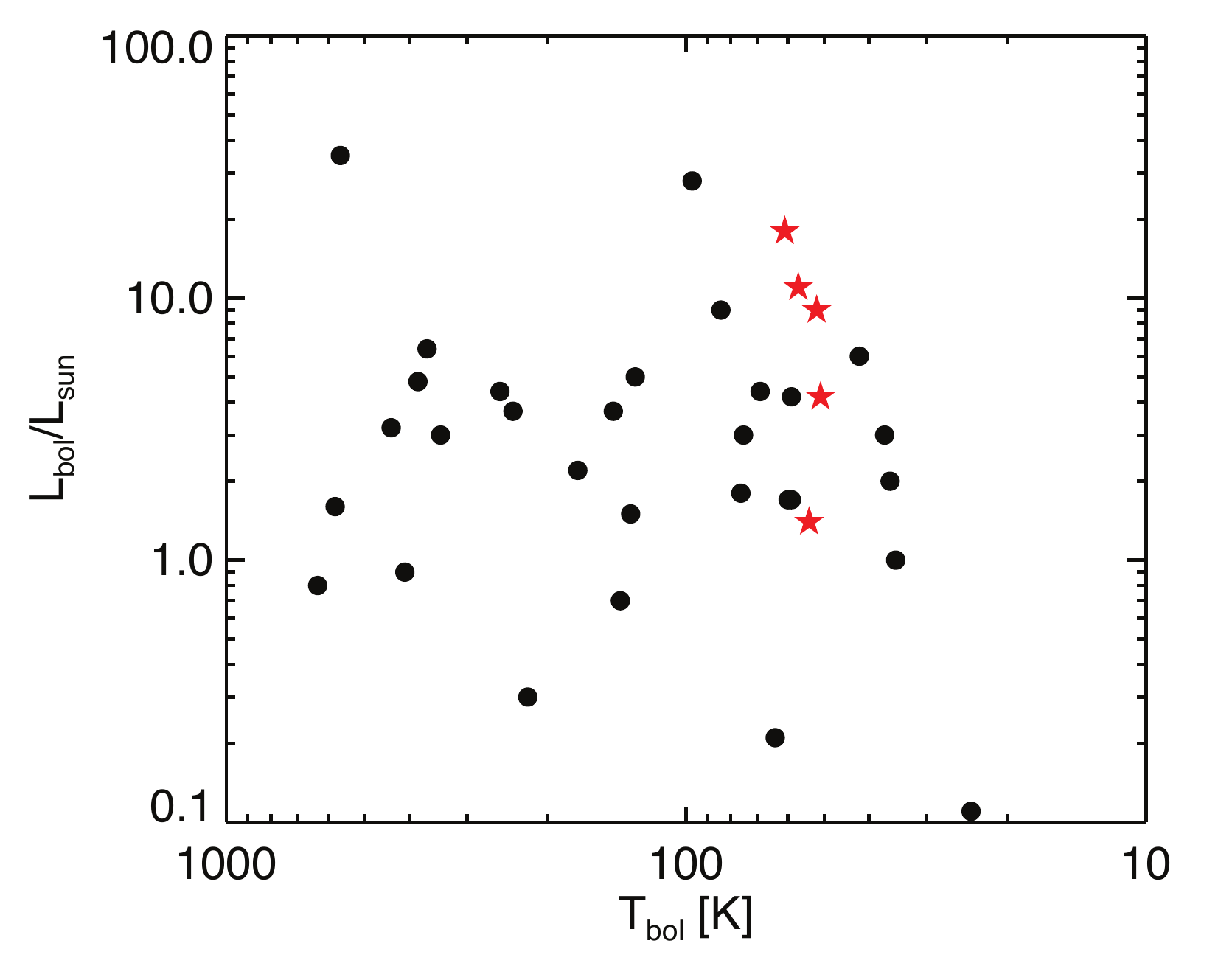}
	\caption{{\bf Left:} Range of stellar temperature and \lbol\ for the disk sample, superimposed on the HR diagram with evolutionary tracks by \citet{1993ApJ...418..414P}. The filled black circles indicate our Herbig Ae sample, the red squares indicate the HR diagram positions of our classical T Tauri star (cTTS)  sample, and the blue triangles denote the positions of our wTTS sample. {\bf Right:} Range of \tbol\ and \lbol\ for the embedded sample. Red stars indicate sources observed by the ``WISH'' key program with PACS full spectral scans.}
	\label{fig:bol}
\end{figure}

The full DIGIT spectroscopic sample consists of 63 sources: 24 Herbig Ae/Be stars (intermediate mass sources with circumstellar disks),
9 T Tauri stars (low mass young stars with circumstellar disks), and 30 protostars (young stars with significant envelope emission) observed with PACS spectroscopy.  The distribution in luminosity and characteristic temperature are shown in Figure \ref{fig:bol}. DIGIT also included an additional wTTS (weak-line T Tauri star) sample that was observed photometrically and delivered separately. The wTTS sample is fully described by \citet{2013ApJ...762..100C}.  


The full DIGIT embedded protostellar sample consisted of 30 Class 0/I targets, drawn from previous studies, focusing on protostars with high-quality {\it Spitzer}-IRS 5-40 $\mu$m spectroscopy (summarized by \citealt{lahuis06}), and UV, optical, infrared, and submillimeter complementary data.  Two sources (IRS44 and IRS46) were observed in a single pointing centered on IRS46.  These objects are selected from some of the nearest and best-studied molecular clouds: Taurus (140 pc; 6 targets), Ophiuchus (125 pc; 7 targets), Perseus (230-250 pc; 7 targets), R Corona Australis (130 pc; 3 targets), Serpens (429 pc; 2 targets), Chamaeleon (178 pc, 1 target), and 4 additional isolated cores.  The sources span two orders of magnitude in bolometric luminosity, from $\lbol = 0.11$ to 27.8 L$_{\odot}$.  The photometric fluxes of embedded protostellar sample at 100 $\mu$m span two orders of magnitude as well, from 1.0 to 240.1 Jy. In terms of the usual tracers of evolutionary development, the  bolometric temperatures range from $\tbol = 27$ to 592 K, spanning the Class 0/I divide at $\tbol = 70$ K. The ratio of \lbol\ to the luminosity at $\lambda \geq 350$ \micron\ ranges from $\lbol/\lsmm = 5$ to $> 10000$, with most ratios (18 of 22 with well-constrained submillimeter data)  falling between 10 and 1000.  The disk sources were selected as bright targets for PACS spectral scans.

PACS is a
5$\times$5 array of 9.4\arcsec$\times$9.4\arcsec\ spatial pixels
(hereinafter referred
to as ``spaxels'') covering the spectral range from 50-210 $\mu$m with
$\lambda$/$\Delta\lambda$ $\sim$ 1000-3000, divided into four segments,
covering
$\lambda \sim$ 50-75, 70-105, 100-145, and 140-210 $\mu$m.
The PACS spatial resolution ranges from $\sim$ 9$\arcsec$ at the shortest
wavelengths (50 $\mu$m) to $\sim$ 18$\arcsec$ at the longest (210 $\mu$m),
corresponding to 1000 to 4500 AU at the distances of most sources.
The nominal pointing RMS of the telescope is 2$\arcsec$.  

For the DIGIT embedded protostars sample we utilized the full range of PACS 
(50-210 $\mu$m) in
two linked, pointed, chop/nod rangescans: a blue scan covering 50-75 and
100-150 $\mu$m (SED B2A + short R1); and a red scan covering 70-105 and
140-210 $\mu$m (SED B2B + long R1). We used 6 and 4 range repetitions
respectively, for integration times
of 6853 and 9088 seconds (a total of $\sim$ 16000 seconds per target for
the entire 50-210 $\mu$m scan).  Excluding overhead, 50\% of the integration 
time is spent on source and 50\% on sky.  Thus the effective on-source integration 
times are 3088 and 4180 seconds, for the blue and red scans, respectively. 
The total on-source integration time to achieve the entire 50-210 $\mu$m 
scan is then 7268 seconds.

The telescope and sky background emission was subtracted using two nod
positions 6\arcmin\ from the source in opposite directions.
The telescope chopped between the source and nod positions, cycling every
1/8 of a second in a
pre-determined pattern of on and off positions \citep{poglitsch10} during
the integration.

Most (21 of 33) disk sources were observed with the same procedure as the embedded objects.  The other 12 sources have only partial spectral coverage:
  8 Herbig Ae/Be sources (HD 35187, HD 203024, HD 245906, HD 142666, HD 144432, HD 141569, HD98922, and HD 150193) and 4 T~Tauri sources (HT~Lup, RU~Lup, RY~Lup, and RNO90) were observed using {\it only} the blue scans (i.e. achieving a wavelength coverage only from SED B2A + short R1, 100-150 $\mu$m).  9 of these 12 sources (all except HD 35187, HD 203024, and HD 245906)  were observed in a further limited wavelength range (60-72 + 120 - 134 $\mu$m; referred to as ``forsterite only'' scans for their focus on the 69 $\mu$m forsterite dust feature).  This procedure allowed shorter integration times to achieve the same SNR in the covered regions.

\subsection{FU Orionis Objects Surveyed with Herschel (FOOSH)}

FUors are  low-mass pre-main
sequence objects named after the archetype FU Orionis (hereafter, FU Ori), which
produced a 6 magnitude outburst at $B$-band in 1936 and has remained close to peak
brightness ever since.  The FOOSH program consisted of 21 hrs of Herschel observing time: V1057~Cyg, V1331~Cyg, V1515~Cyg, V1735~Cyg, and FU~Ori were observed as part of FOOSH and analyzed from preliminary data reduction \citep{green13c}.

For the FOOSH sample we again utilized the full range of PACS 
(50-210 $\mu$m) in
two linked, pointed, chop/nod rangescans: a blue scan covering 50-75 and
100-150 $\mu$m (SED B2A + short R1); and a red scan covering 70-105 and
140-210 $\mu$m (SED B2B + long R1). We used 6 and 4 range repetitions
respectively, for integration times
of 3530 and 4620 seconds (a total of $\sim$ 8000 seconds per target and off-positions combined, for
the entire 50-210 $\mu$m scan; the on-source integration time  is $\sim$ 3000 seconds).  The telescope sky background was subtracted using two nod positions 6\arcmin\ from the source. 

The SPIRE-Fourier Transform Spectrometer (FTS) data were taken in a single pointing with sparse image sampling, high spectral resolution mode, over 1 hr of integration time.  
The spectrum is divided into two orders covering the spectral ranges 194 -- 325 $\mu$m (``SSW''; Spectrograph Short Wavelengths) and 320 -- 690 $\mu$m (``SLW''; Spectrograph Long Wavelengths), with a resolution, $\Delta\nu$ of  1.44 GHz and
resolving power, $\lambda$/$\Delta\lambda$ $\sim$ 300--800, increasing at shorter wavelengths \citep{griffin10}.  The FOOSH data were observed in a single pointing with sparse image sampling, high spectral resolution, in 1 hr of integration time per source. 
  
\subsection{CO in ProtoStars (COPS)}

The sample of 31 ``COPS'' protostars observed with SPIRE-FTS includes 25 sources from the DIGIT and 6 from the WISH
(Water in Star-forming regions with {\it Herschel}, PI: E.~van~Dischoek; \citealt{vandishoeck11}; see also \citealt{nisini10,kristensen12,karska13,wampfler13}) key programs.  A nearly identical sample was observed in CO \jj{16}{15} with HIFI (PI: L.~Kristensen) and is presented in \textcolor{blue}{Kristensen~et~al.} (\textcolor{blue}{in prep.}).  This dataset (COPS: SPIRE-FTS) is analyzed in a forthcoming paper (\textcolor{blue}{Green~et~al.}, \textcolor{blue}{in prep.}).  The SPIRE beamsize ranges from 17-40$\arcsec$, equivalent to physical sizes of $\sim$ 2000-10000 AU at the distances of the COPS sources, comparable to the size of a typical core \citep{ward07} but smaller than many collimated outflows.

The COPS SPIRE-FTS data were observed identically to the FOOSH SPIRE data, in a single pointing with sparse image sampling, high spectral resolution, in 1 hr of integration time per source, with one exception: the IRS 44/46 data were observed in medium image sampling (e.g. complete spatial coverage within the inner 2 rings of spaxels), in 1.5 hr, in order to better distinguish IRS 44 (the comparatively brighter IR source; \citealt{green13b}, \textcolor{blue}{Green~et~al.}, \textcolor{blue}{in prep.}) from IRS 46. 
 
\subsection{Overview of the Products}

The basic data products we produce for each source observed with PACS and SPIRE spectroscopy are the following:

\begin{itemize}

\item Datacubes (wavelength vs. flux density vs. spatial position), with advanced corrections from HIPE 13.

\item Best-calibration 1-D spectra for point sources and marginally extended sources within the data cube, including flat spectra (continuum-subtracted), continuum (line-subtracted), and residual spectra.

\item Line identification and fitted parameters obtained with an automated fitting routine (for both 2-D and 1-D spectra), organized by spatial position for each source, including a complete, sortable linelist.

\item Contour plots for extended line and continuum emission.

\item EPS files for the last three products above, allowing users to verify all automated fits.

\end{itemize}

In this document, we review the archive data processing and products in detail.  First we describe the custom data reduction pipeline we used in \S \ref{sec:pipe}.  Second, we describe the automatic line fitting routine and the line fitting results in detail in \S \ref{sec:line_fitting}.  Third, we describe the data products and output file format, including post-processing data products in \S \ref{sec:format}. Lastly, we summarize the results of this archive in \S \ref{sec:conclusion}.  Our line source libraries and a full inventory of sources included in this archive are shown at the end of this paper, in the Appendix (\S \ref{append}).

\section{Pipeline settings}  
\label{sec:pipe}
\subsection{PACS Pipeline for Pointlike and Small Sources}
\label{sec:pacs_pipe}
We begin with the {\tt hipe13.0.1006.ChopNodPointingCorrection.py} pipeline script (the ``Point Source Background Normalization'' script).  This script was made available in HIPE 13; any version of HIPE 13 should work for this purpose.  We then made the following modifications:

\begin{itemize}

\item The script is run as a loop over numerous OBSIDs for blue and red cameras.  

\item  We make an adjustment in the pipeline to account for the effects of pointing jitter (changes in pointing offset during the course of the observation). The chi-squared difference between the flux of the ``point'' source and the flux of the centered beam, via a map/grid, is determined using the beams of calibration products.  We define the ``oversample'' (the samples per spatial element in each dimension, improving precision of positional determination) and the ``smoothFactor'' (smoothing in the spectral domain, in terms of grating positions rather than wavelength).  For the pipeline, we set these to oversample $=$ 8 and smoothFactor $=$ 7; using larger values yielded no apparent benefit.  

\item  In the specFlatFieldRange command, we decrease ``polyOrder'' to 4 (from the default of 5).  This slightly improves memory usage, although the effect is minor.  There is no impact on the success of the flatfielding.

\item  We select excludeLeaks $=$ True so that we do not use the light-leakage affected regions for flatfielding.  These regions are not reported in the final spectral product, which removes the 95-105 $\mu$m range (and a few spectral lines including CO \jj{25}{24} in that region).  The advantage to removing these data is that we avoid the typically erratic continuum in that region when determining the flatfielding correction.

\item  We change the ``gaussianFilterWidth'' to 35 (instead of 50).  This parameter is the number of wavelength points over which we assess the scaling of the central spaxel to the level of the 3x3 spaxel set, which is used in the production of the ``3x3YES'' product, below; it is an attempt to correct for the shape of the continuum, picking a width large enough to account for pointing jitter but not so wide as to smooth out the actual continuum shape.  By using the flux of the ``3x3'' central 9 spaxels, and comparing to a true point source distribution, we correct for the movement of the flux centroid throughout the observation, referred to as pointing jitter.  The number is essentially based on the sampling and the significance of pointing jitter, for a particular observation; we settled on a value of 35 through trial-and-error testing.

\item We add a ``garbage collection'' line at the end of the loop.  This allows us to run at least 200 OBSIDs within a single command, given sufficient disk space.  It takes about 1 hour/OBSID to run at these settings on a (early 2014) 12-core processor with 96 GB of RAM dedicated to the process, excluding data download time; for a total of 200 OBSIDs, we required 200 CPU hours (e.g., 2 weeks) to reduce all the targets in our sample.

\end{itemize}

The default pipeline reduction shows significant mismatches between red and blue modules, and it frequently did not align with photometry or with SPIRE observations of the same position.  The early pipelines did not fully account for source spatial extent or pointing jitter, and the relative spectral response function did not work well at the edges of each order
(see e.g., \citealt{green13b}).  
All of these effects are accounted for in the new pipeline, which has dramatically improved the continuum calibration for PACS in particular.  Our new pipeline includes updated calibration data, improved flatfielding, and jitter correction, and other small improvements.

\begin{figure}[h]
\centering
\includegraphics[scale=0.7]{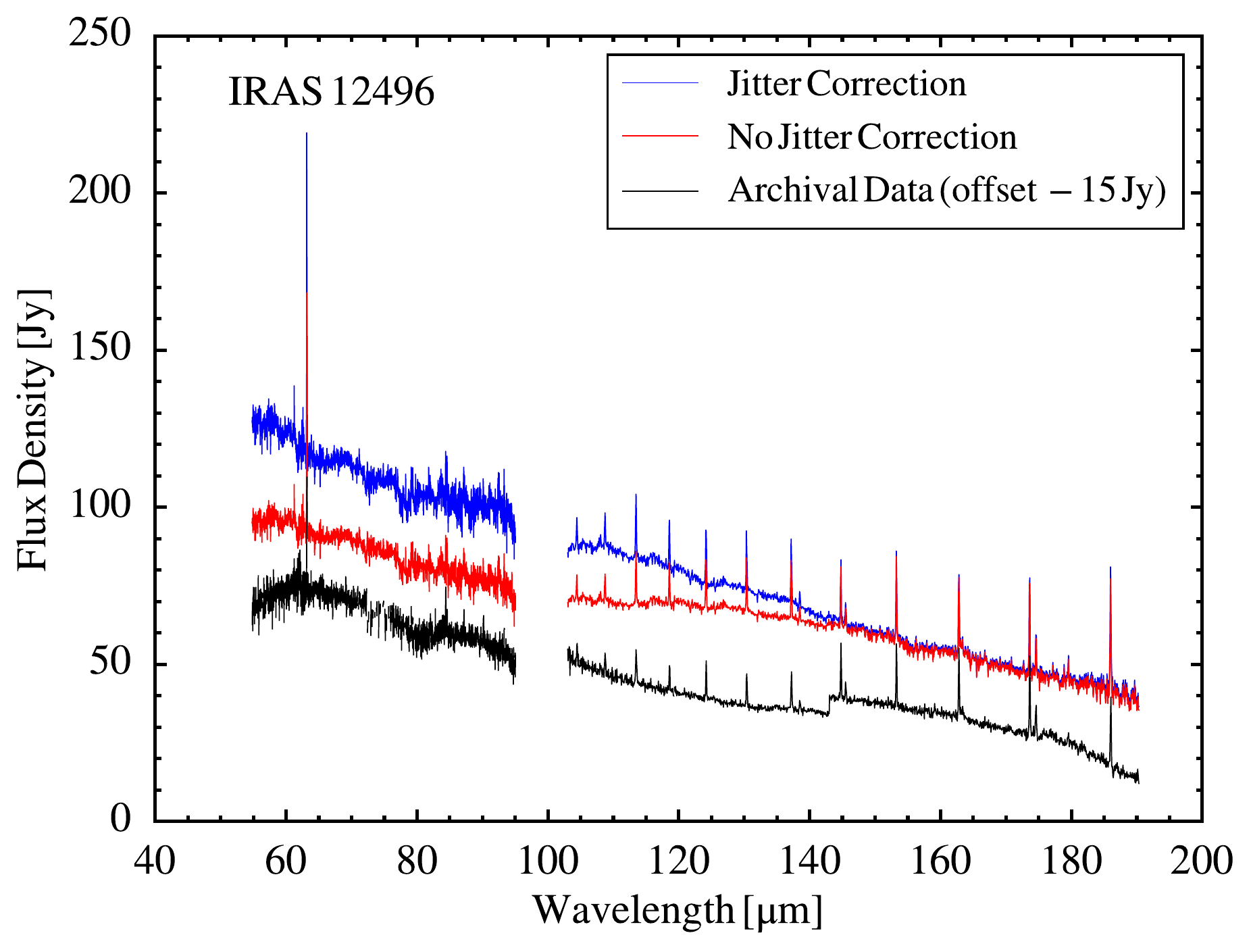}
\caption{The comparison of spectra of IRAS~12496 with different corrections applied.  
The archival data from HSA processed with the default pipeline are shown in black, offset by $-15$ Jy to separate it from the others. 
The offsets between modules and the incorrect shape at long wavelengths make
searches for broad solid-state features impossible. Our reduction without 
jitter correction is shown in red; it improves the long wavelength behavior and
module matching, but the no-jitter corrected spectrum shows a flattening around 130 \micron\ that could be
mistaken for a solid-state feature. The blue curve shows the result after
jitter correction; the broad feature around 130 \micron\ has been eliminated. 
}
\label{jitter}
\end{figure}

\begin{figure*}[ht!]
\centering
\includegraphics[width=0.7\textwidth]{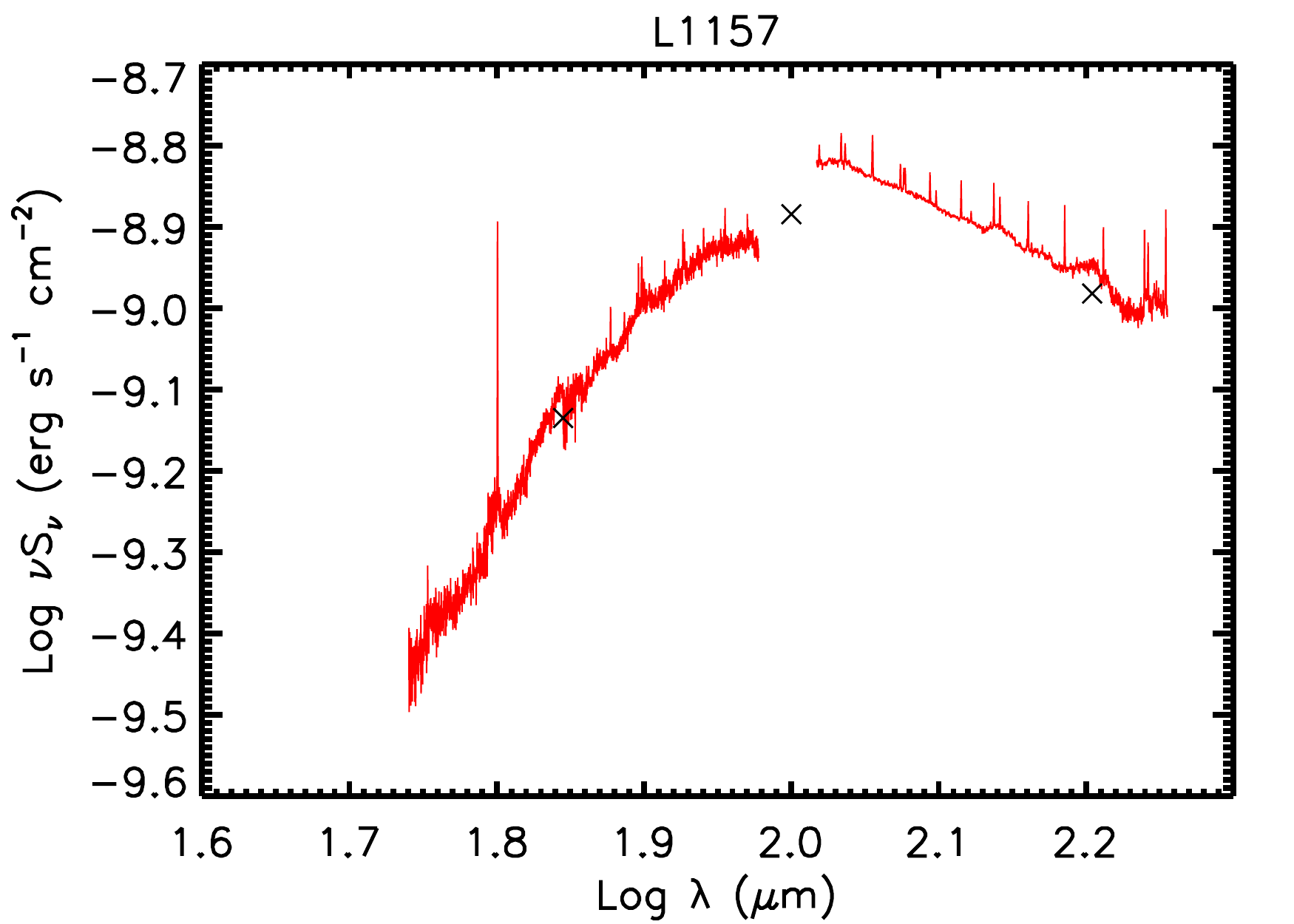}
\includegraphics[width=0.7\textwidth]{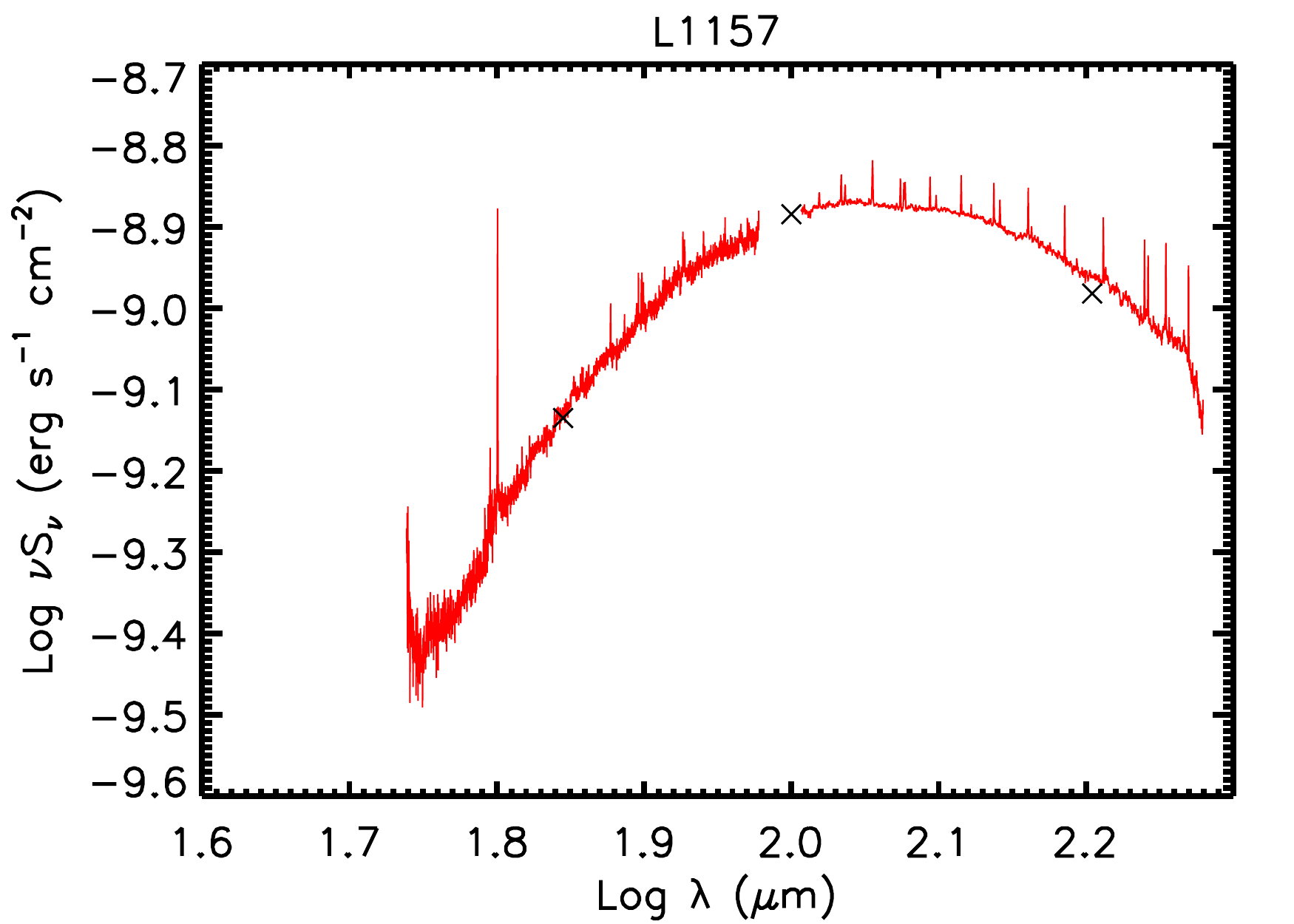}
\caption{Improvement in data quality using new techniques, illustrated for L1157.
The top plot shows the old (i.e. the 2012-13 era pipeline and calibration, without jitter correction and flatfielding correction) reduction without scaling. 
Photometric data (labeled by $\times$ symbols) obtained later shows a substantial discrepancy.
The bottom plot  shows the results of the new reduction which
produces smoother spectra and better agreement with photometry, without applied scaling.
The flatfield corrections have also removed most of the false ``solid-state 
features.''  Note that published spectra from 2012-13 typically have scaling applied in post-processing.}
\label{bhr71}
\end{figure*}

An example of the improvement in spectral/continuum shape, and inter-module flux calibration from the newer pipeline and the jitter correction is shown (for IRAS 12496) in Figure \ref{jitter}. The observations of this source were mis-pointed by about 0.5 spaxel, so it provides a good test. The jitter correction (blue) has removed several spurious broad features and improved overall flux calibration when compared with the non-corrected spectrum (red) or the default pipeline spectrum (black).  
 The absolute flux calibration has also improved considerably.  As an example, we consider the improvement to the continuum shape and absolute flux calibration shown for the protostar L1157, a source we will use for most of the following comparisons.  Figure \ref{bhr71} demonstrates the improvement.  Our best products from the new pipeline results match much better with photometry extracted from archival {\it Herschel}-PACS and SPIRE imaging of our sample objects in the Herschel Science Archive, from a variety of programs, including:   KPGT\_pandre\_1, KPGT\_okrause\_1, OT1\_jtobin\_1, OT1\_jgreen02\_1, and OT1\_mdunham\_1. These data were less subject to saturation problems than data from earlier missions or large beam sizes (e.g. {\it Spitzer}-MIPS, IRAS).  The photometry data are extracted with the  source size fitted by the spectroscopic (HIPE) pipeline, at the appropriate waveband (70, 100, 160, 250, 350, or 500 $\mu$m).  The data products from which we extracted our photometry have absolute flux calibration uncertainties of $<$ 7\% for PACS against stellar models \citep{balog14} and asteroid standards \citep{muller14} and 4\% for SPIRE when compared against models of Neptune \citep{bendo13}, with a 5\% uncertainty in cross-calibration between PACS and SPIRE \citep{muller14}.

\begin{figure}[h]
\centering
\includegraphics[width=0.65\textwidth]{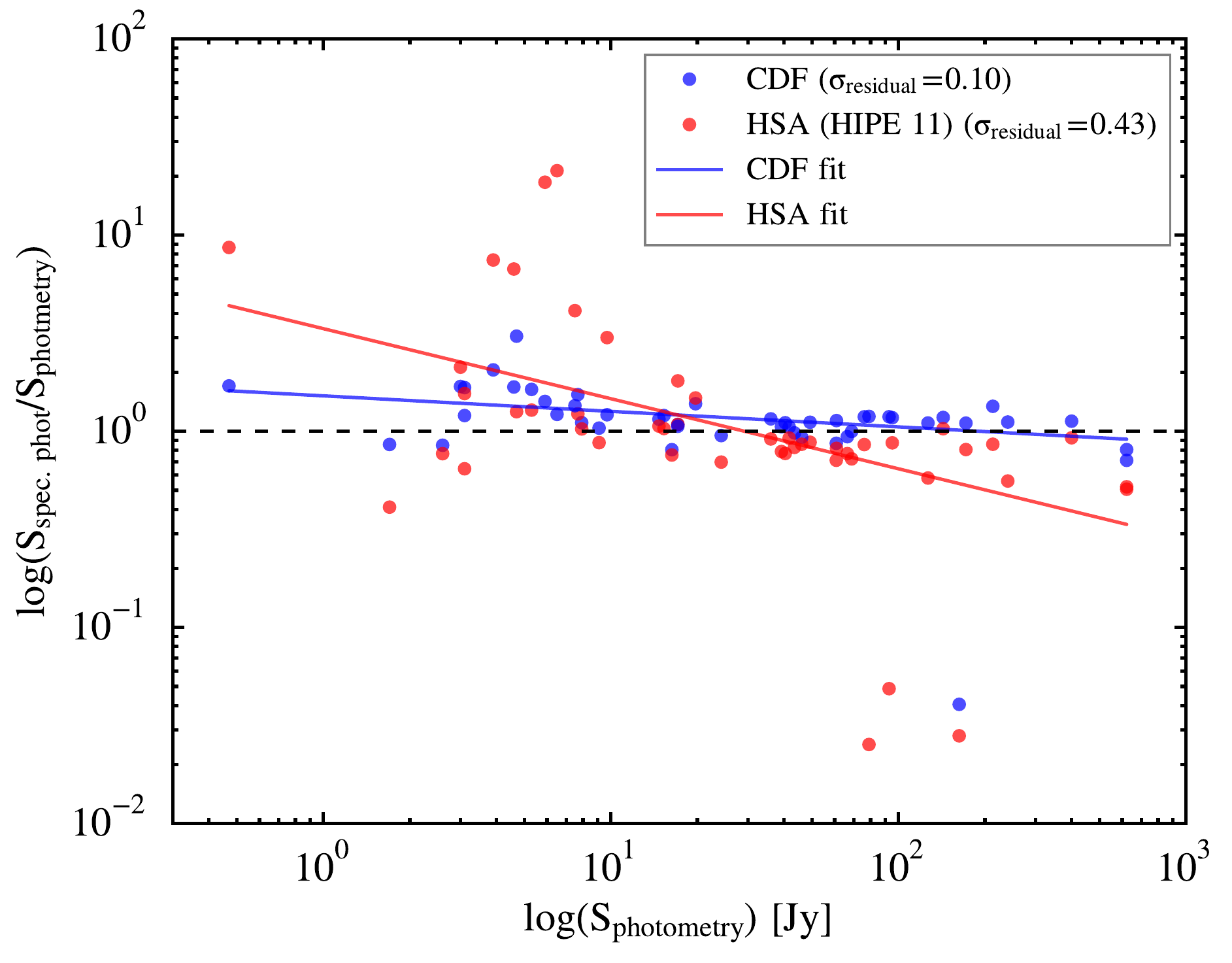}
\includegraphics[width=0.65\textwidth]{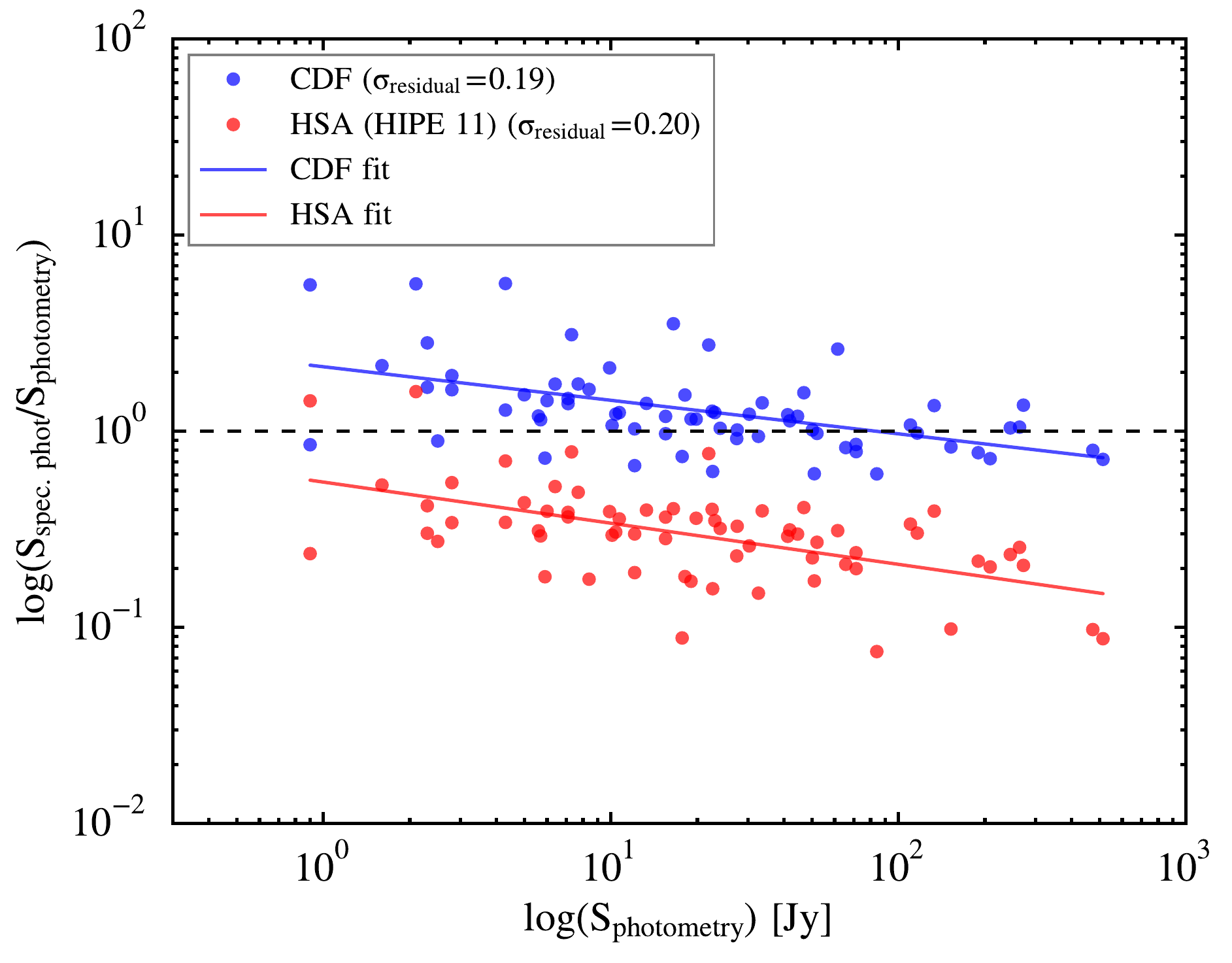}
\caption{Comparison of spectral flux density with photometric flux density, for both the COPS-DIGIT-FOOSH archive and {\it Herschel} Science Archive (HSA) products.  For PACS, shown in the top figure, data are collected from $\sim$1/3 of total sources in our archive -- all non-confused protostars for which we have gathered photometry data.   For SPIRE, shown in the bottom figure, data are collected from 21 sources among 31 COPS sources for which we have gathered photometry data.  Data from our archive are shown in blue, while data from the HSA are shown in red.  The data from the HSA were processed in the HIPE 11 pipeline and collected in mid-2014.  The $\sigma$ parameter presented in the box is a measure of the deviation from the line of equality, normalized to the mean flux density.  Spectra in our archive agree better with the photometric data, while the spectra from the HSA show larger deviations.  The blue line is a fit to the data from the CDF archive only; the red line is a fit to the HSA data only.  The significant outlier at the bottom right corner is excluded in this fitting.  For both PACS and SPIRE, the CDF results are consistent with the equality line, while the HSA archive products are not.}
\label{phot_com}
\end{figure}

Although L1157 is a typical example, we compared the absolute flux calibration of our final PACS products with the default pipeline products in Figure \ref{phot_com} (top), relative to photometric observations, for all non-confused, protostellar sources of which the photometry was available in the {\it Herschel} Science Archive -- about one third of our sources.  The spectra in this archive are convolved with corresponding photometry filters before comparing with photometric data.  The data from {\it Herschel} Science Archive were processed with HIPE~11 by {\it Herschel} Science Center and collected at mid-2014.  The flux is scaled from the spectra of central spaxel to the total flux within the central 3x3 spaxel to match with the flux of our corrected 1-D spectra.  The black line indicates a perfect agreement between photometry and spectroscopy; the blue points are values derived from our new products, and the red points are derived from default 2014 HSA products; the enhanced products show much smaller deviation from the photometric values.  The products from our archive have a better agreement between the photometry and spectrophotometry.  A straight line fitting with data (blue) from our archive shows only a small deviation toward faint sources and without a significant offset to the equality line. Compared to the default HSA products, the dispersion in the residual of line fitting of the CDF products is greatly reduced, from 0.42 to 0.1. The improved reduction has produced reliable spectrophotometry for most sources that are not in confused regions.
%
%

Overall, the jitter correction was successfully performed for $\sim$ 80\% of our sources.  The advanced pipeline with jitter correction matches the fluxes by matching the minor mispointings at different wavelengths.  However, not every spectrum was improved by the correction.  Among the embedded sources in our sample, we got very poor results from the jitter correction on 14 sources: IRAM~04191, L1014, L1455-IRS3, RCrA-IRS5A, RCrA-IRS7C, IRS~46, Serpens-SMM4, EC~82, HD~98922, HD~245906, HD~203024, HT~Lup, HD~142666, and HD~35187.  The common threads between these sources are twofold: all have either A) complicated PACS fields in which the central source is not the brightest object, or B) weak continuum ($<$ 3 Jy at 60 $\mu$m).  For these sources, we include only the ``non-jitter corrected'' versions in our archive.  This method assumes that the lines and continuum are distributed similarly.  Lines with very different distribution require special attention \citep[e.g.,][]{je15}.

In two embedded sources, we found significant (10\%) flux mismatches between the two OBSIDs merged to form the spectrum: L1551-IRS5 (in which the observations were separated by 1.5 yr); and GSS30-IRS1 (which shows complicated behavior at long wavelengths; \citealt{je15}).  In addition, slight mismatches are noted in L1448-MM (a partially blended chain of sources; \citealt{lee13}).

We find excellent flux matches between modules in all other sources, including all disk/point sources, even in cases with up to 0.5 spaxel mispointings (e.g. TMC1A, IRAS 12496, TMR1, and L1527). Note that we apply no ``by-hand'' manual scaling after the pipeline reduction.

\subsection{SPIRE Pipeline for Pointlike and Semi-Extended Sources}
\label{sec:spire_pipe}

SPIRE used an FTS spectrometer with an onboard calibration source; separate
modules were used for short (SSW) and long (SLW) wavelengths. 
Each module was reduced separately within HIPE version 13.0 using the standard pipeline for extended sources, including apodization.  Originally, SPIRE had a spectral resolution of 1.2 GHz.  Because the signals are recorded in the Fourier domain, the line would appear as a sinc function with a spectral resolution of
1.2 GHz.  After apodization, the spectral response becomes Gaussian-like, 
but the spectral resolution is degraded to 1.44 GHz. 

\begin{figure*}[htbp!]
\centering
\includegraphics[width=0.8\textwidth]{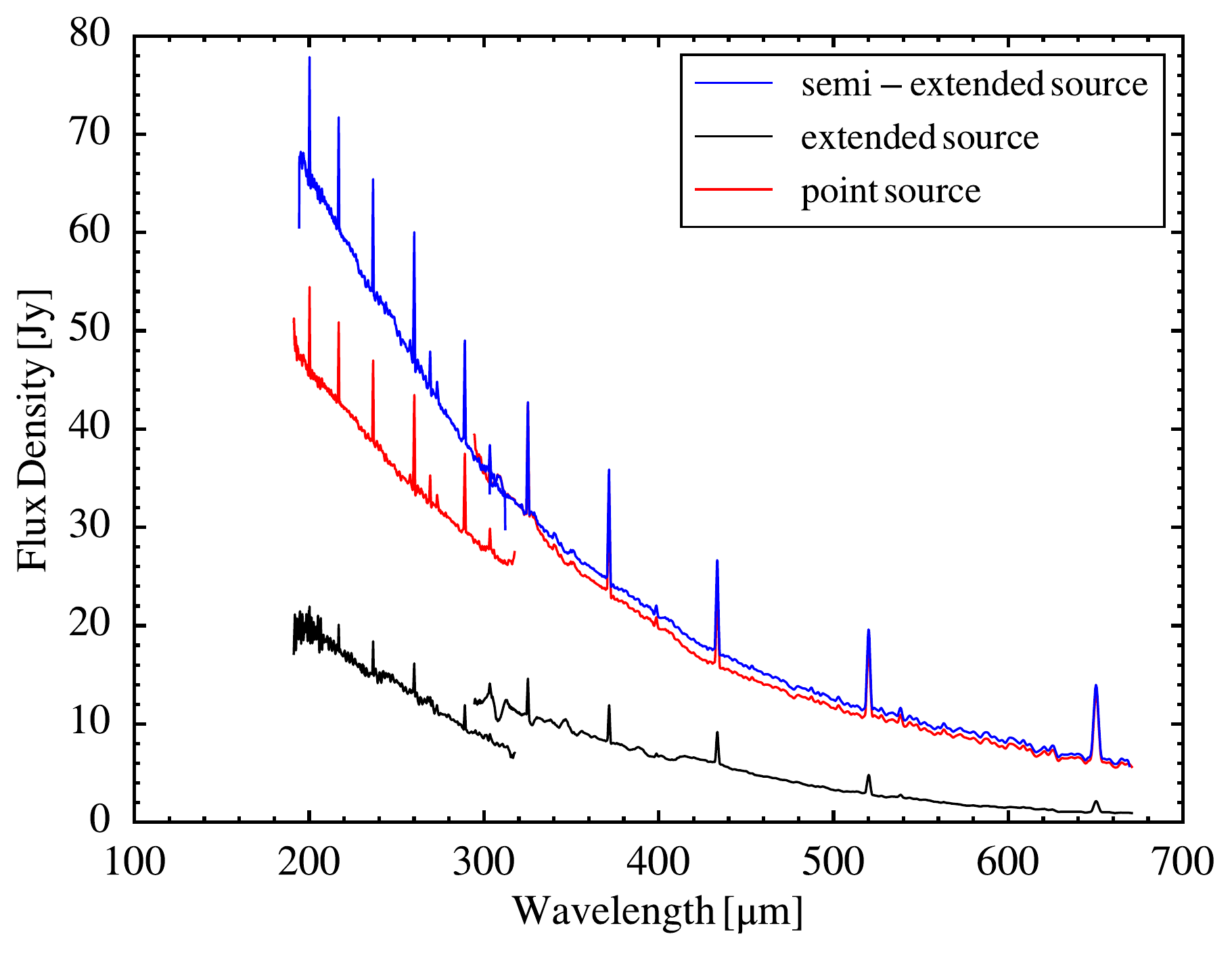}
\caption{Comparison of L1157 SPIRE spectroscopy calibrated with different correction methods: extraction as a point source (red), an extended source (black), and after the semi-extended source correction (blue).}
\label{fig:spire_correction_com}
\end{figure*}

Unlike the case of the PACS pipeline, no significant modifications were made to the basic pipeline script, except for allowing it to run in a loop and for the separate post-processing step to improve the calibration for 
semi-extended sources. The SPIRE data were extracted using the ``extended source'' calibration pipeline, as this produced a smoother continuum between modules, better signal-to-noise ratio (SNR), and fewer spectral artifacts than the ``point source'' pipeline (Figure~\ref{fig:spire_correction_com}).  We performed post-processing within HIPE version 13.0.1006, using SPIRE calibration dataset 12.  Starting from the baseline data products, we used the ``SemiExtendedSourceCorrector'' script, which minimizes the mismatch between SLW and SSW.  This semi-extended source correction calibrates sources that are partially resolved by the SPIRE beam.  The correction fits a source size based on the assumed source profile to minimize the ratio of SLW and SSW at the overlapped wavelength and then calculates the spectrum within a 40\arcsec\ Gaussian reference beam, which is independent of wavelength \citep{2013A&A...556A.116W}. This procedure greatly improves
the match of the two modules within SPIRE (as it was designed to) and
also smooths the continuum (compare Figure~\ref{fig:spire_correction_com}
and Figure~\ref{tmr1} (bottom) for L1157).
We applied the ``semi-extended source'' calibration to all but a few sources (HH~100, IRS~46, and V1735~Cyg) that were excluded owing to their poor spectral quality.

The SPIRE spectrophotometric and photometric fluxes, when extracted in the same reference beam, match to within 5\% in 21 of 32 protostars and 2 of 3 FUors.  The other 12 sources show varying degrees of mismatch with the photometry. The spectra of B1-a, GSS~30-IRS1, IRAS~03301, and L723-MM overestimate the photometry by 10-20\%; the reverse is true for L483 and IRAS~15398; these are within formal uncertainties for the spectral flux calibration, and are likely attributable to complicating extended emission. V1331 Cyg shows a larger mismatch at 250 $\mu$m, exceeding the photometric point by 50\%; the arclike extended structure associated with this compact source may be the cause.  The 250 $\mu$m photometry for WL~12 and L1455-IRS3 seems suspect.  Finally, the IRS~44/46 combined field cannot be fully disentangled; the ``IRS~46'' SED is contaminated by IRS 44 for $\lambda$ $>$ 70 $\mu$m; the ``IRS~44'' SED is missing substantial flux at all wavelengths due to the position of IRS~44 near the edge of the PACS field-of-view.

Figure~\ref{phot_com} (bottom) shows the comparison of 21 sources for which we gathered photometry data.  The spectrophotometry from the spectra from the Herschel Science Archive default product (collected in mid-2014) are roughly a factor of 4 lower than the photometry.  The spectrophotometry from our archive (benefitting from the semi-extended source correction processing), shown in blue, agrees significantly better with the actual photometry.  Separate fits to the HSA product spectrophotometry and our product spectrophotometry have similar slopes, which do not match the line of equality.  The increasing deviation at faint sources is therefore not caused by the semi-extended correction but derives from the uncorrected spectral data.  Note that the scatter in the fit residual is nearly identical -- 0.19 for HSA products and 0.18 for CDF products.

We also find improved agreement with previous ground-based submillimeter data.  We compared the filter-convolved spectrophotometry from this archive with SPIRE photometry for almost all SPIRE sources in this archive and SCUBA/SHARC-II total flux for the  sources from \citet{shirley00}, \citet{young06}, and \citet{wu07}.   The mean ratio of spectrophotometry from this archive over the flux measured by SCUBA/SHARC-II is 1.19, while the mean ratio measured from the spectrophotometry from the Herschel Science Archive is only 0.38.  The scatter in the residual subtracted with the fitted line is 0.11 for HSA products and 0.09 for CDF products. 


The semi-extended source correction has greatly improved the SPIRE
spectrophotometry, but it has introduced an offset between the
longest wavelength in the PACS data and the shortest wavelength in the
SPIRE data, which usually has more emission than the extrapolated PACS
data (Fig. \ref{fig:full_spec}). 
This problem arises from the fact that the semi-extended source
corrector assumes a source size that is constant with wavelength and
because no background subtraction was performed. In bright extended sources, the mismatch is systematic as noted above.  In fainter sources (those with 200 $\mu$m intensity $\lesssim$~25~Jy, the SPIRE flux exceeds the PACS flux by a factor of three on average. 
The protostars in our sample are typically resolved (marginally) by SPIRE, and their spatial extent increases  with wavelength.  As the corrector
assumes a source model that is independent of wavelength, it overpredicts the flux at short SPIRE wavelengths; thus the SPIRE flux for $\lambda \lesssim$ 300 $\mu$m is too high for extended sources.  Improving this
result requires  knowledge of source intensity distribution as a function
of wavelength.  We apply a 3-D radiative transfer model to explore this issue for one protostar in an upcoming paper (Yang et al,. in prep.), but,
lacking models for all sources, our archival products cannot account
for this effect. The SPIRE spectrum should be considered an upper limit to the true flux density at the shortest wavelengths.


\section{Automated Line-Fitting}
\label{sec:line_fitting}

We use an automated emission line fitting process (available at GitHub\footnote{\href{https://github.com/yaolun/Her_line_fit}{https://github.com/yaolun/Her\_line\_fit}}) that tries to fit lines from a pre-existing database of laboratory wavelengths, identifies the threshold for detection, then generates tables of line flux, width, centroid, and uncertainties for every detected line, along with an upper limit to the flux for every undetected line for all PACS and SPIRE spectra in the sample. This information is then combined with spatial information for each pixel to produce contour maps including all spatial positions where significant detections are found. The line-fitting algorithm uses {\tt mpfit}, a Levenberg-Marquardt non-linear least squares minimization \citep{markwardt09}, to determine emission line properties.  After producing a line detection database, we test the integrity of the line fits to better characterize the SNR, decouple any blended line, and produce easily searchable products for further science applications.  The automated results have been spot checked by eye but should be used with caution.  In particular, lines that are distributed very differently in space from the continuum emission, or sources in complex regions will need special consideration.  The detailed procedures are described in the following sections and an overview of the process is shown in Figure 
\ref{threepass}.

\subsection{Line List}
\label{sec:line_list}
A comprehensive and pre-selected line list is used to supply initial fit parameters for the automated routine. We require an expected line list to restrict our line fitting to specific regions of the spectrum, pre-select line-free continuum regions, and identify the lines after fitting. For this dataset, we verified that there were no significant unidentified lines. If the input spectrum contains a line that is not included in our list, one can easily check whether every significant feature has been extracted from the spectrum:  any line not included in the linelist will then appear in the residual spectrum (see \S \ref{sec:format}).  The line list includes lines of \co, \thirteenco, \water, OH, \hcoplus, \chplus, and atomic fine-structure transitions in the wavelength range of 55-670 \micron\ (Tables \ref{oh2o_list}, \ref{ph2o_list}, \ref{co_list}, \ref{oh_hco_list}, and \ref{atomic_list}). Table \ref{dg_list} lists potentially blended lines
that are fitted with double Gaussians (see \S \ref{sec:blend}).  
Two \chplus\ lines, \chul{4}{3} and \chul{2}{1}, were excluded because they are heavily blended with nearby potentially strong water lines. 

\begin{figure}[ht!]
	\centering
	\includegraphics[width=0.45\textwidth]{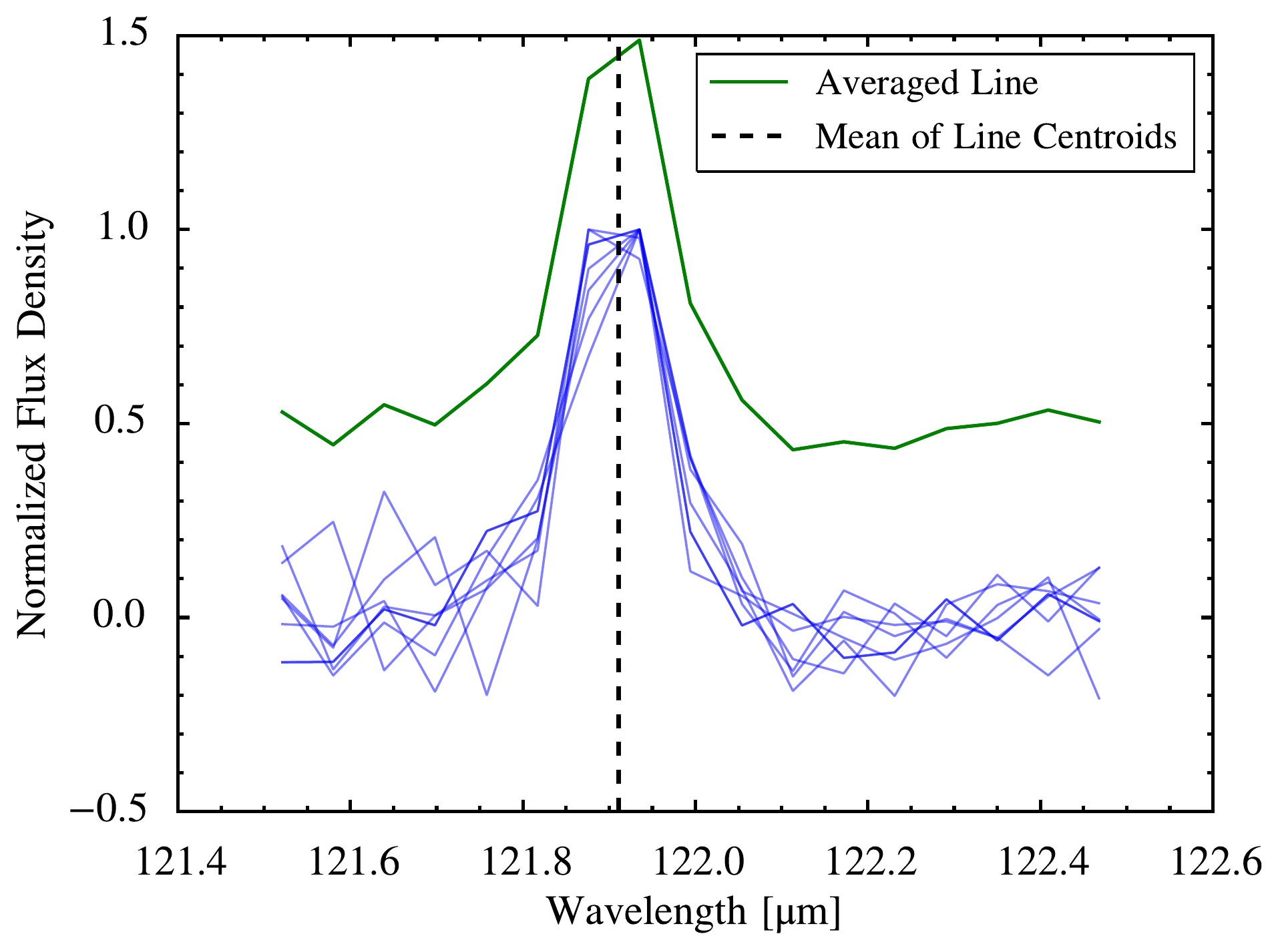}
	\includegraphics[width=0.45\textwidth]{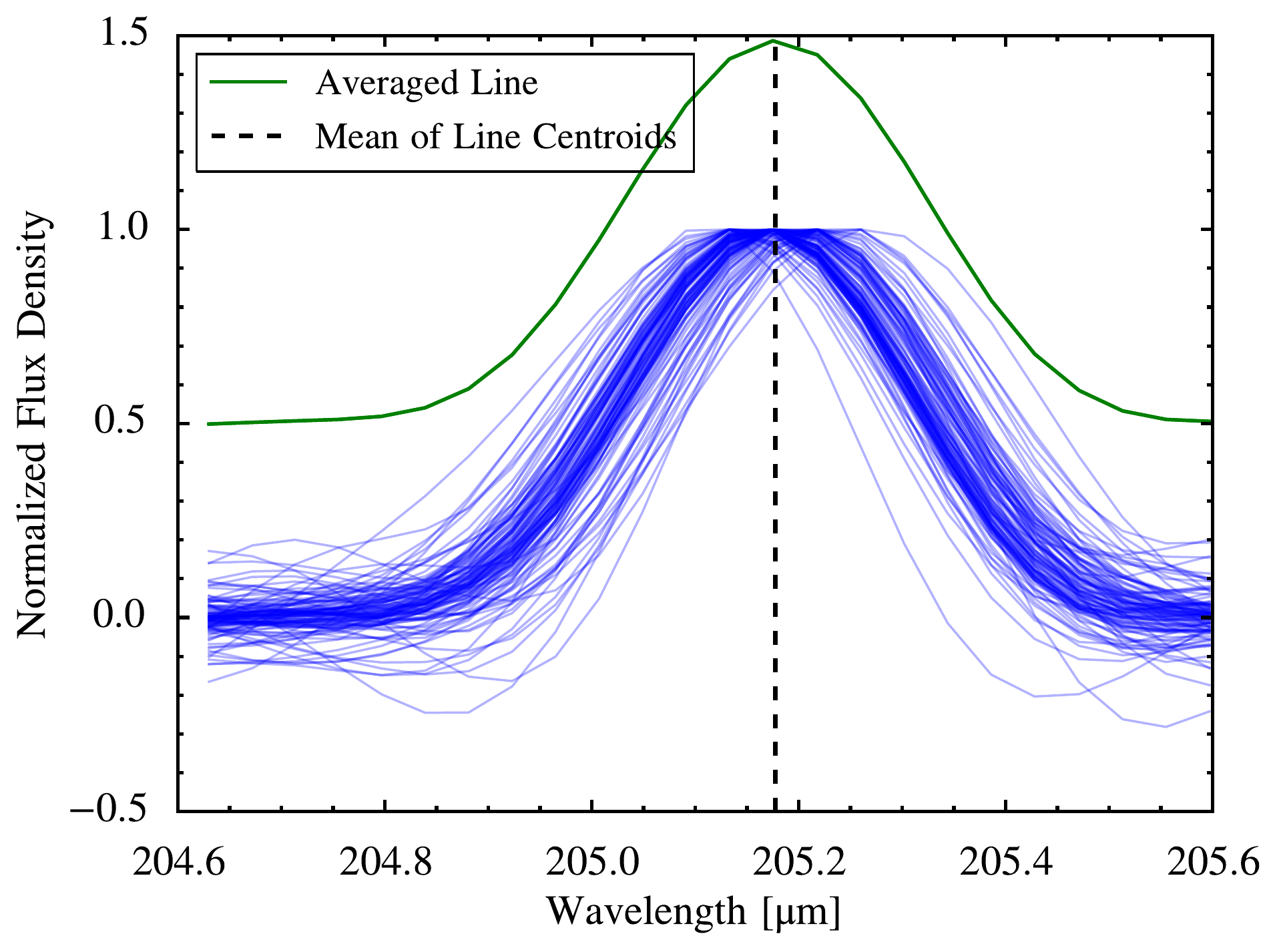}
	\caption{The line profiles of \NII~122~\micron\ and \NII~205~\micron\ lines identified in this archive with SNR greater than 10.  The individual line profiles are shown in blue, while the averaged line profile is shown in green.  The line centroids of the averaged line profiles are reported in Table~\ref{atomic_list}.}
	\label{fig:NII}
\end{figure}

The laboratory data for each line were collected from the Leiden Atomic and Molecular Database (LAMDA; \citealt{schoier05}) and the Cologne Database for Molecular Spectroscopy (CDMS; \citealt{2005JMol...742.215}). However, the \coul{48}{47} to \coul{41}{40} lines were not included in LAMDA until recently; thus we calculated the frequency and Einstein-A coefficient using a method from \citet{Authier1993590} and the model of the dipole matrix element in \citet{1994ApJS...95..535G}. An updated CO extrapolation added to LAMDA (\textcolor{blue}{Neufeld, in prep.}) agrees with our calculations.  We use slightly different line centroids of OH lines, adopted from \citet{wampfler13}.  The only difference is the line centroid of OH\ohul{1/2}{3/2}{+}$-$\ohul{1/2}{1/2}{-}; we use 163.12~$\mu$m instead of the value of 163.019~$\mu$m from LAMDA.  We use only subsets of the line list within the wavelength range of the observations for each specific source.  Finally, we note that \NII\ 122~\micron\ and 205~\micron\ do not appear in the LAMDA database, and there is a discrepancy between the centroid reported in the ISO-LWS line list (taken from the SMART data reduction package, \citealt{higdon04}) and the line list from Peter van Hoof (http://www.pa.uky.edu/$\sim$peter/atomic/).  We used the line centroids from the ISO line list as initial guess, but allowed the line centroids to vary up to $\pm2$ spectral resolution to determine better values.  The averaged fitted line centroids of two \NII\ lines from all detections with SNR $>$ 10 are 121.911~\micron\ and 205.170~\micron.  In Table~\ref{atomic_list}, we report the mean values and the standard deviation of the mean of two \NII\ lines.  Figure~\ref{fig:NII} shows the detected \NII~122~\micron\ and \NII~205~\micron\ lines found in this archive.  

\subsection{Baseline Selection and Fitting}

Before fitting each line, we select a nearby spectral region without line emission, from wavelengths near the theoretical line centroid (hereinafter referred as ``line-region'') and use this region for baseline fitting. The baseline fitting region is 10 times the spectral resolution at the wavelength of the line centroid. However, some of these wavelengths may include other emission lines.  Thus we use only line-free wavelengths (determined from our previously constructed line list) until we have accumulated 5 times the spectral resolution, both blue-ward and red-ward of the line centroid. Then, we fit the selected baseline region using a second order polynomial. Once we obtain a well-constrained baseline, we subtract the baseline from the identified wavelength region.  This continuum-subtracted spectrum is then used for line-fitting.

\subsection{Line Selection and Fitting}

We begin with the baseline-subtracted spectrum. The next step is to extract the line profile, using a Gaussian model. Several constraints are applied to the fitting process for PACS and SPIRE spectra. First, the line centroid is allowed to vary within $\pm$2 times the spectral resolution at the theoretical line centroid except for the [O~\textsc{i}] line at 63.18~{\micron}.  Because this line appears to have real velocity shifts in some sources, presumably due to wind motions, the line centroid  is allowed to vary within $\pm$3 times the spectral resolution instead.  Secondly, the linewidth is flexible in the fitting for SPIRE spectra, while the linewidth is fixed at the instrumental resolution for PACS spectra.  The spectral resolution of PACS was insufficient to resolve most of the lines except for the [O~\textsc{i}] line at 63.18~{\micron}.  The linewidth of \OI~63~\micron\ line is allowed to vary from $-30$\%\ to 50\%\ because this line may be broadened by the associated outflows.  After comparing the fitting results with and without the linewidth fixed at the PACS instrumental resolution, we found that both strong lines and weak lines can be identified in the fixed-width setting with moderate SNR, while the flexible width setting produces better SNR on strong lines but might fit a narrower than reasonable feature at some wavelengths.  The SNR of a strong emission line fitted with a flexible width shows only a 6\%\ increase compared to the SNR of the same line fitted with fixed-width setting.  Therefore, for all lines other than [O~\textsc{i}], we fix the FWHM of the line to the PACS spectral resolution.  Users might be able to achieve a higher SNR, but we adopt the fixed width setting to avoid fake lines in this archive. For SPIRE, similar problems are not found due to a higher oversampling rate; thus we keep all line widths flexible within $\pm$30\%\ of the SPIRE instrumental resolution. The instrumental resolution of PACS was obtained from personal communication with Dr.~Helmut~Feuchtgruber from MPE/Garching, and the instrumental resolution of SPIRE was obtained from the SPIRE data reduction guide released by {\it Herschel}.

\subsection{Blended Lines Fitting: Double Gaussian Fitting}
\label{sec:blend}
Some emission lines are sufficiently blended that our default single Gaussian profile does not characterize the line shape. For known blended lines (listed in Table \ref{dg_list}), we use a double Gaussian function to fit the blended profile.  We only fit a double Gaussian model to two blended lines, while cases of more than two blended lines are fitted with a single Gaussian model for each component and labeled with blending condition (see next paragraph).  The width of each component of blended lines are fixed to the instrumental spectral resolution.  For blended lines, we require that both lines are in emission; if an absorption line is blended with an emission line, it cannot be fitted.  When the fitting routine reached the boundary and tried to fit a negative value to the line height,  this value would be set to zero, resulting in a zero line flux in the report.  Note that line fitting with a single Gaussian profile does allow negative flux values; this restriction on negative line flux values was only necessary for blended lines.

In the cases that a double Gaussian profile is not sufficient to distinguish two lines, the fitting process fits each line with a  single Gaussian profile.  If a detected single line can feasibly be identified with either of two lines in our linelist, depending on the physical properties or chemical composition of the source, we use a simple criterion to identify the possible blended regions as well as the likely dominant line. For any pair of lines with centroids located within one resolution element of each other, we classify the fit as a possible blended line, label the two options with {\tt Red} and {\tt Blue} indicating whether we attribute the blended line to the longer or shorter wavelength candidate. If a given line is not blended, we label it as {\tt x}; if a line is blended with two other possible lines, the line is flagged with {\tt Red/Blue}. Our {\it recommended} line identification is simply the line with the highest Einstein-A coefficient, as the level population dependence on physical conditions can be complicated. These blended lines are typically water lines
with similar excitation requirements. These recommendations for line identification are recorded as {\tt 1} in the column {\tt Validity}. These suggested identifications should be used with caution.

\begin{figure*}[ht!]
\centering
\includegraphics[height=8cm]{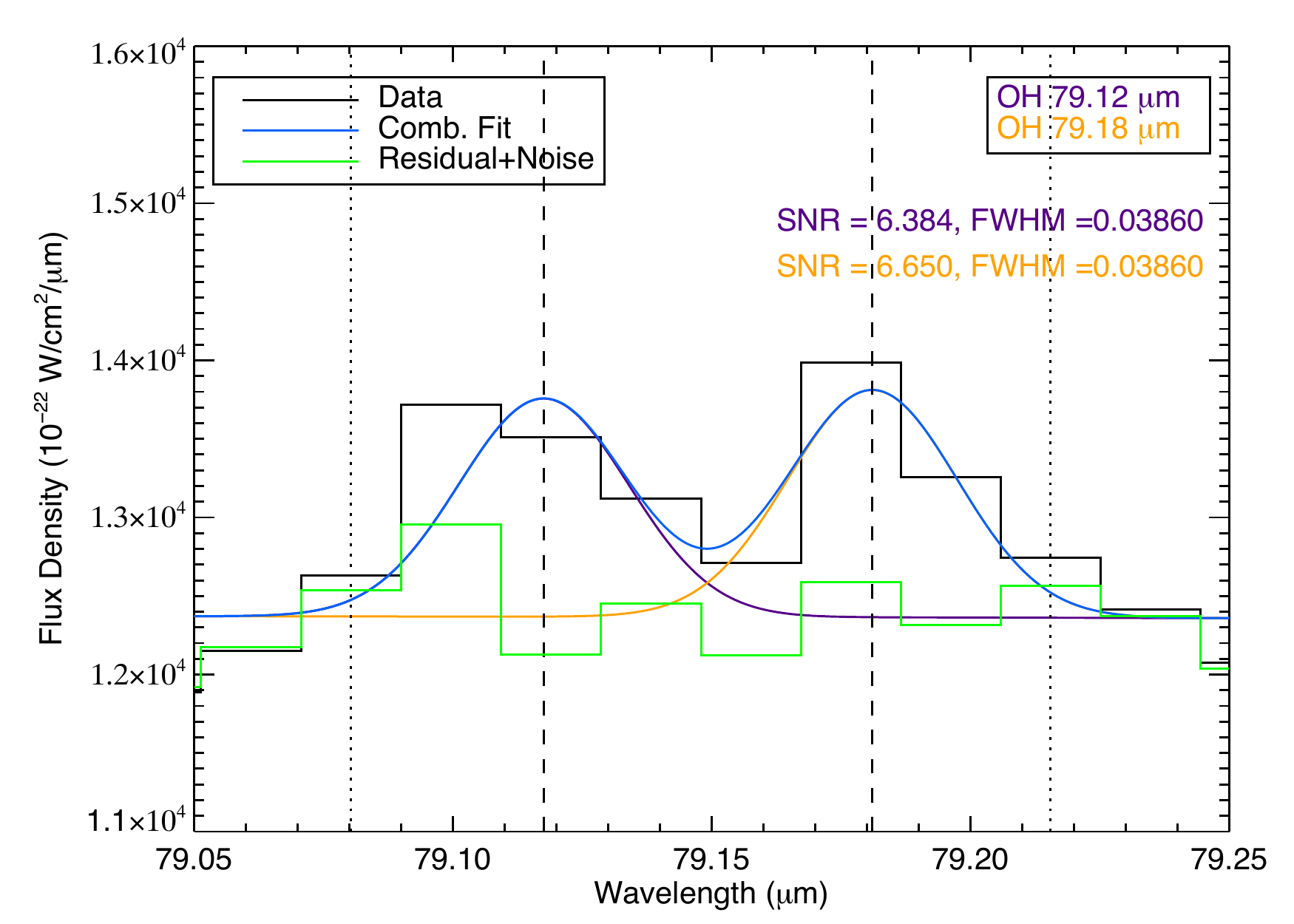}
\includegraphics[height=8cm]{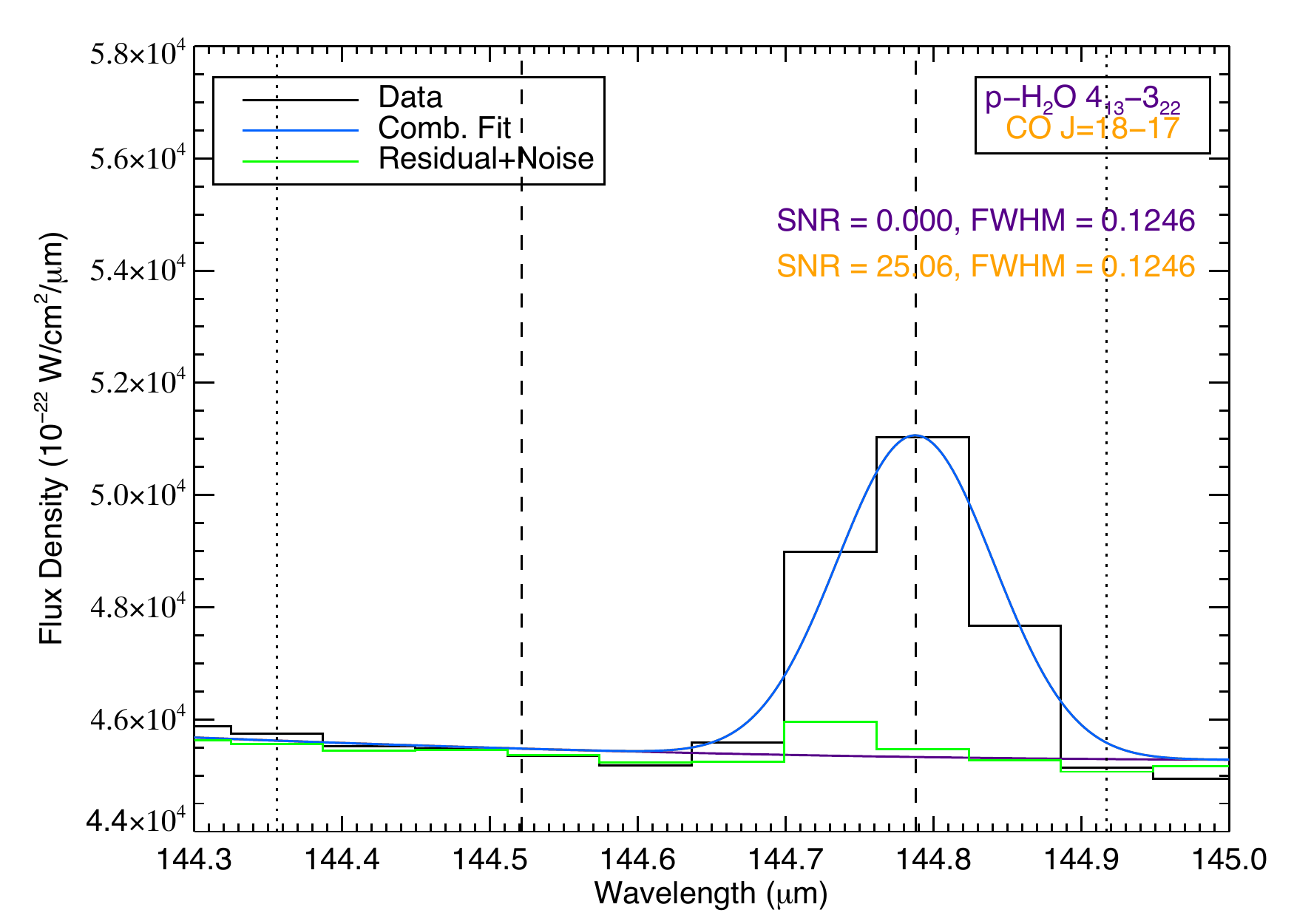}
\caption{Examples of double-gaussian fitting. {\bf Top}: For the blended OH\ohul{1/2}{1/2}{-}$-$\ohul{3/2}{3/2}{+} line and the OH\ohul{1/2}{1/2}{+}$-$\ohul{3/2}{3/2}{-} line, for L1157, using fixed line centroids and FWHM constraints. The lines plotted are data (black), combined fit (blue), OH\ohul{1/2}{1/2}{-}$-$\ohul{3/2}{3/2}{+} fit (purple), OH\ohul{1/2}{1/2}{+}$-$\ohul{3/2}{3/2}{-} fit (yellow), and residual (green) (See \S \ref{sec:noise}). The SNR of the two OH lines are 6.38 and 6.65, respectively. Both lines are considered as detections. {\bf Bottom}: An example of a double fit in which only a single component was detected. We show the blended $\mathrm{\rm p-H_{2}O~4_{13}-3_{22}}$ and \coul{18}{17} lines, for BHR71, using fixed line centroids and FWHM constraints. The color code is the same as the top figure. The SNR of the CO line is 25 and is significant; the \pwater\ line is not detected.}
\label{doublegauss}
\end{figure*}

We show two examples of double Gaussian fitting in Figure \ref{doublegauss}. In the top panel, we fit partially blended CO and \CI\ emission lines. The bottom panel shows \pwater\ and CO emission lines. In the first case, the fitting algorithm successfully resolved the blended lines, and both will be listed in the table of fitting results 
(see \S~\ref{sec:report}).
In the second case, only the \coul{18}{17} line is detected, so
the CO line will be reported as a detection and the \pwater\ line
as a non-detection.

In some spectral regions, there are three possible line candidates.  
We do not fit any triple Gaussian profiles due to the instability of such fits.  Those lines are labeled in the same fashion.  These regions will
require individual fits with some a priori knowledge of the likely relative
strengths of the three lines.

\subsection{Noise Estimation}
\label{sec:noise}
The problem of noise estimation is two-fold.  An accurate evaluation of the uncertainties of the fitted parameters needs a reliable uncertainty for each spectral data point. However, no uncertainties are provided for the SPIRE data, and those supplied for PACS are unrealistic. Therefore, the uncertainties of the spectral data points need to be estimated from the spectrum itself.  After the baseline fitting, the fitting process proceeds in two ways, one for full SED scans and one for short scans (\S \ref{obs}). We will describe the process first for
the full SED scans (the majority of the sources) and then describe the changes needed for the sources with only short scans.

\begin{figure}[htbp!]
	\centering
	\includegraphics[width=0.7\textwidth]{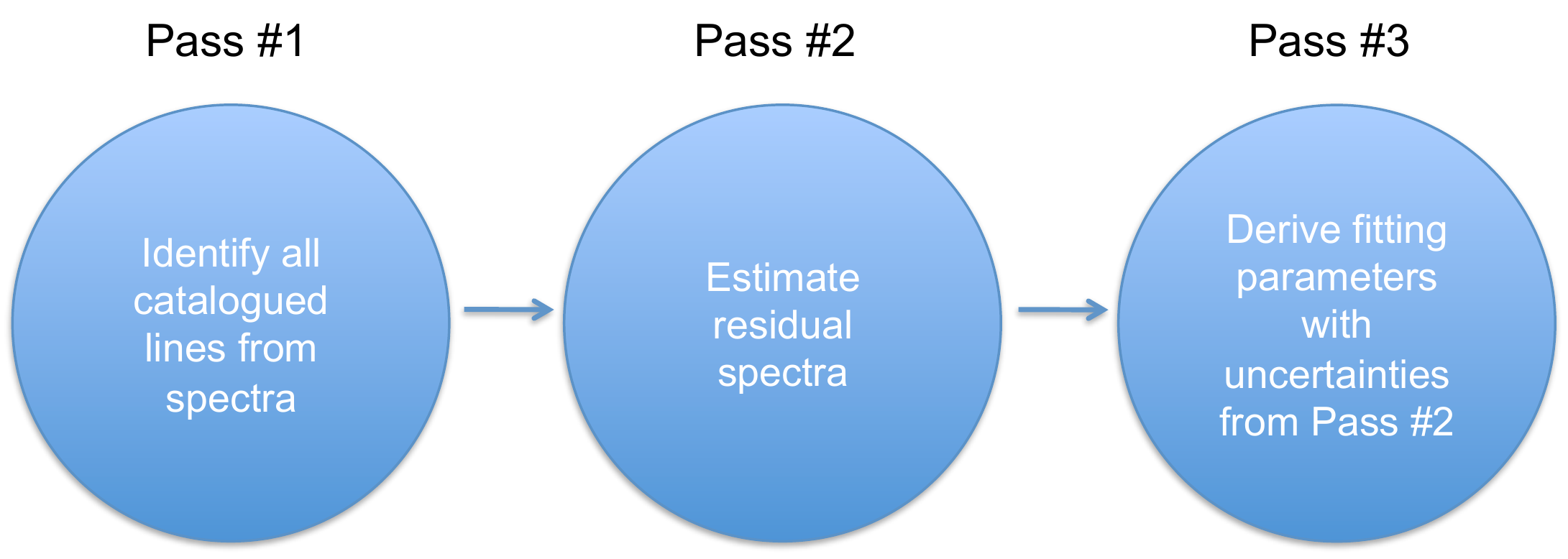}
	\caption{Summary of the three passes to fit spectral lines from an input 1-D spectrum.}
	\label{threepass}
\end{figure}

For sources with full SED scans, we feed the fitting routine with the spectra three times, two to precisely measure the uncertainties from the fluctuations of the baseline (hereafter baseline noise) and one to apply the baseline noise as the uncertainties of the spectral data points to get an accurate evaluation of the uncertainties of fitted line parameters (Figure \ref{threepass}). The baseline noise is fed into the final fitting process to get the final fitted parameters with uncertainties. Each time the lines are fitted, the SNR of the line flux ($F_{\rm l}$) is calculated by taking the line strength divided by the standard deviation of the modified baseline (the baseline with continuum removed) times the square root of the oversampling rate and the FWHM of the line and 1.064 for a Gaussian profile (Equation \ref{snr}).  
\begin{equation}
{{\rm SNR} = \frac{F_{\rm l}}{\rm 1.064~\sigma_{baseline}\sqrt{oversampling}{\times}FWHM}}
\label{snr}
\end{equation} 

During the first pass through the data (left panel of Figure \ref{threepass}), the
routine attempts to identify and remove all lines in the line list and calculates
a first estimate of the SNR using a restricted baseline range
 ($\pm$ 10 spectral resolution elements around the line centroid).  This modification is designed to include enough wavelength channels in the baseline while avoiding contamination from nearby strong lines.  It is a rough estimation, which will be refined if  the full noise estimation process can be
performed. 
 
In the first fitting process, every line is fitted without any prior knowledge of other lines.  This first estimate occasionally yields anomalous uncertainties for blended lines or lines with other lines nearby, as the residual after subtracting the fitted line is not only the noise but also other, not-yet-fitted, emission lines. Thus in the second pass of the data, we subtract all lines identified with SNR $> 2$ (SNR calculated from the noise biased by not-yet-fitted lines) from the whole spectrum. Then we use a top-hat smoothing function to approximate the continuum shape.  We then subtract the continuum to produce a residual spectrum without any contamination, which is the main goal of the second pass (middle panel of Figure \ref{threepass}). 
 
\begin{figure}[htbp!]
	\centering
	\includegraphics[width=0.7\textwidth]{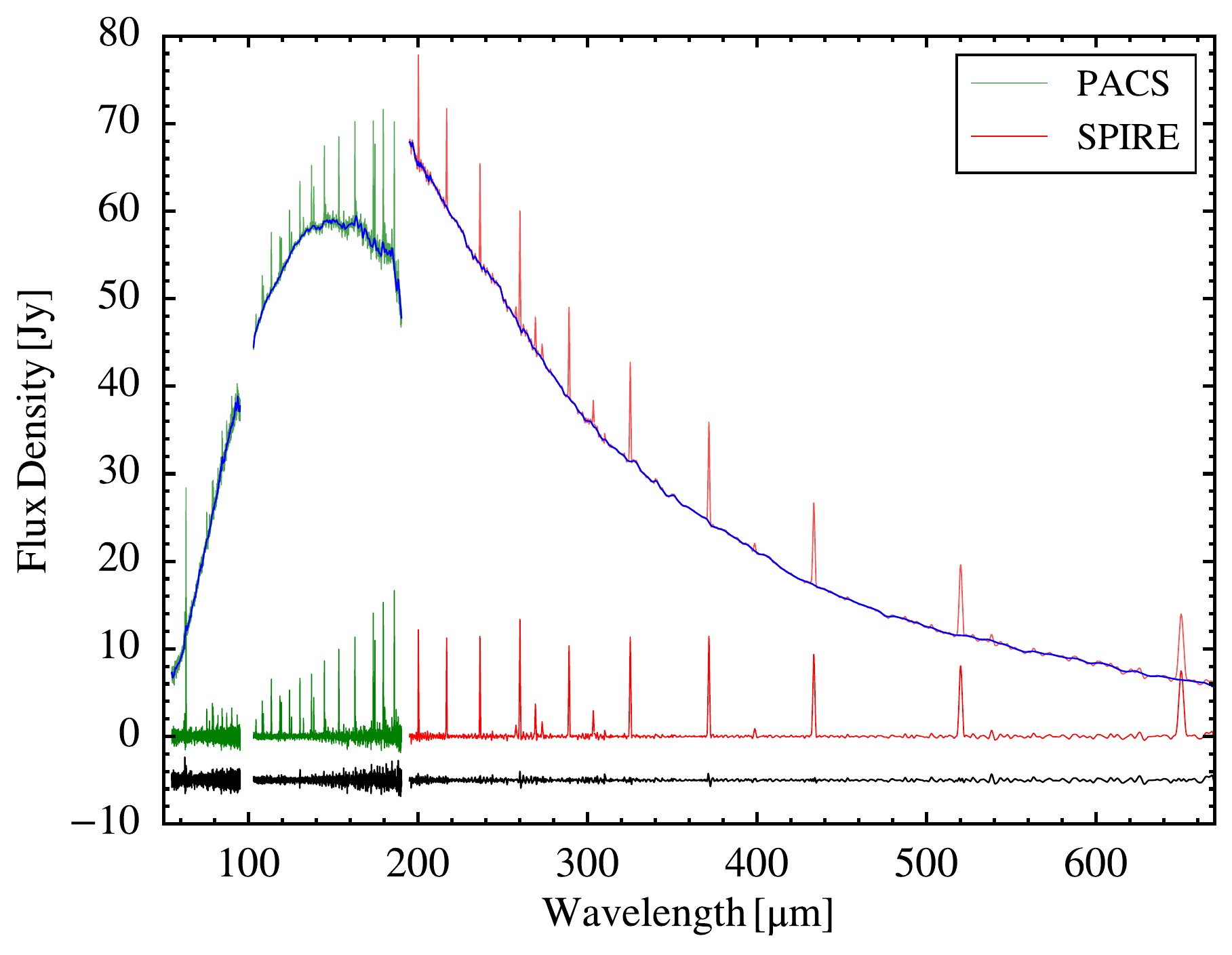}
	\caption{The spectrum of L1157 with flat spectrum (green/red for PACS/SPIRE, respectively), continuum (blue), and residual (black, offset by -5~Jy) overplotted. The continuum is the result of the smoothing function applied to the line-subtracted spectrum. The flat spectrum is produced by removing the continuum from the original spectrum. The residual spectrum is resulted from subtracting the continuum and fitted lines from the original spectrum. The comparison of the continuum and the flat spectrum at the edge of the PACS band shows that the fitting algorithm works well even in the region with severe instrumental effects.  The offset between PACS and SPIRE spectra is discussed in Section~\ref{sec:spire_pipe}.}
	\label{fig:full_spec}
\end{figure}

With the improved estimation of the uncertainties of the data points in hand, a third line fitting is executed (right panel of Figure \ref{threepass}), producing reliable uncertainties of fitted parameters. The extra two steps in the line fitting routine increase the SNR by 23\% (38\%\ for strong lines).
The smoothing function and continuum subtraction are also applied during the third fitting.  The SNRs are calculated from the residual spectra from the third fitting.  The fitting results produced from the third fitting, as well as the continuum, flat, and residual spectra, are those reported in the archive (examples in Figure \ref{fig:full_spec}). 
The continuum spectrum is the best product for modeling dust continuum emission. 
The flat spectrum provides an overview of the line emission, and the residual
spectrum can be examined for unidentified lines or anomalies in the line
fitting process.

For the short scan sources, the continuum coverage is insufficient to 
derive a complete continuum spectrum; consequently the estimate of the baseline
noise is less precise and nearby lines could be adding to the noise. 
 The second fitting process for those data uses the noise estimated from the first pass as the uncertainties in the fitting process to derive the fitted parameters with uncertainties. There is no third pass for those sources.
The fits with all three fitting processes are labeled as ``{\tt Global}'', while the ones with only two fitting processes are labeled as ``{\tt Local}'' in the ``{\tt full\_source\_list.txt}'' file (see \S \ref{sec:report}).

\begin{figure}
	\centering
	\includegraphics[width=0.63\textwidth]{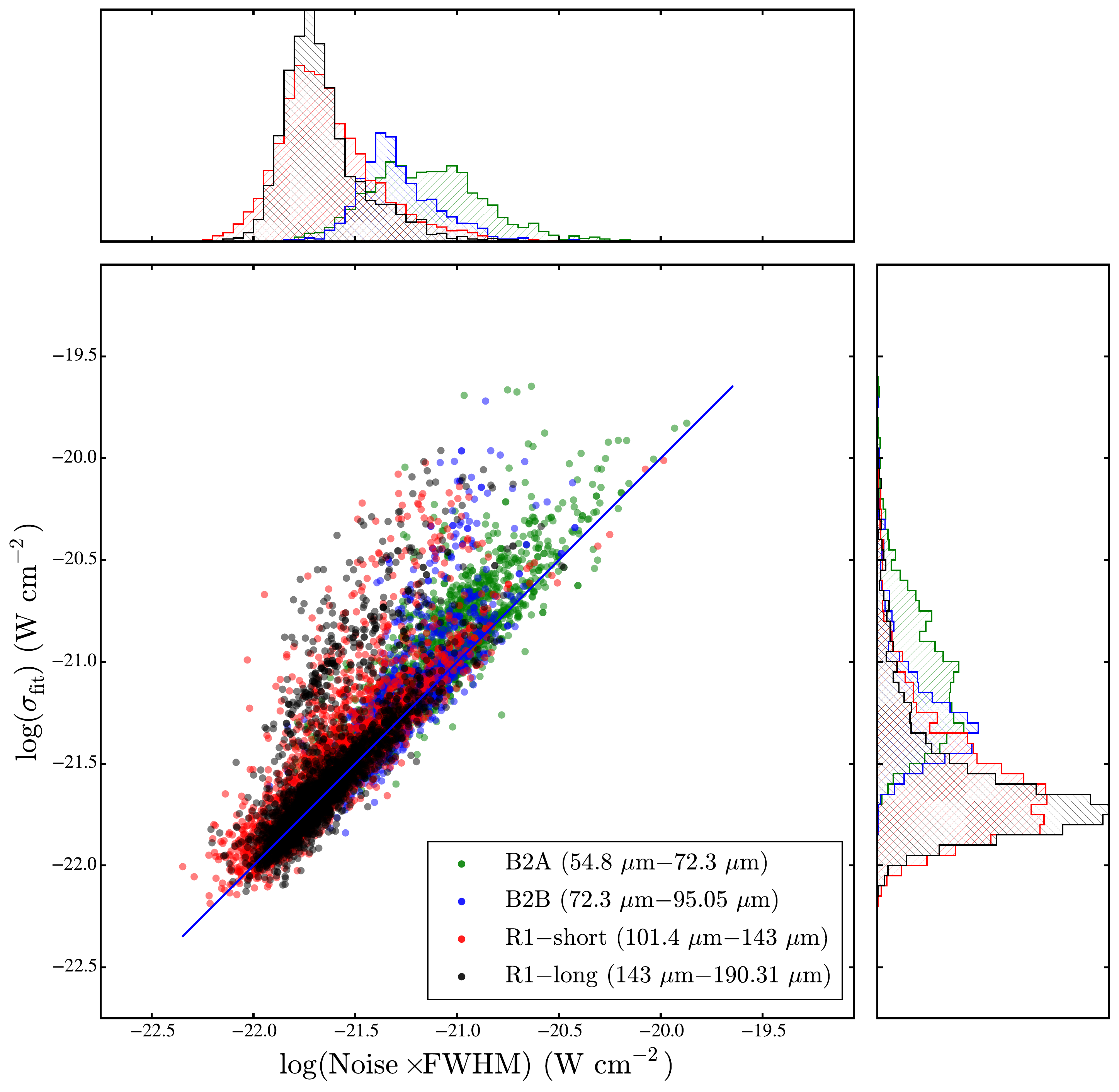}
	\includegraphics[width=0.63\textwidth]{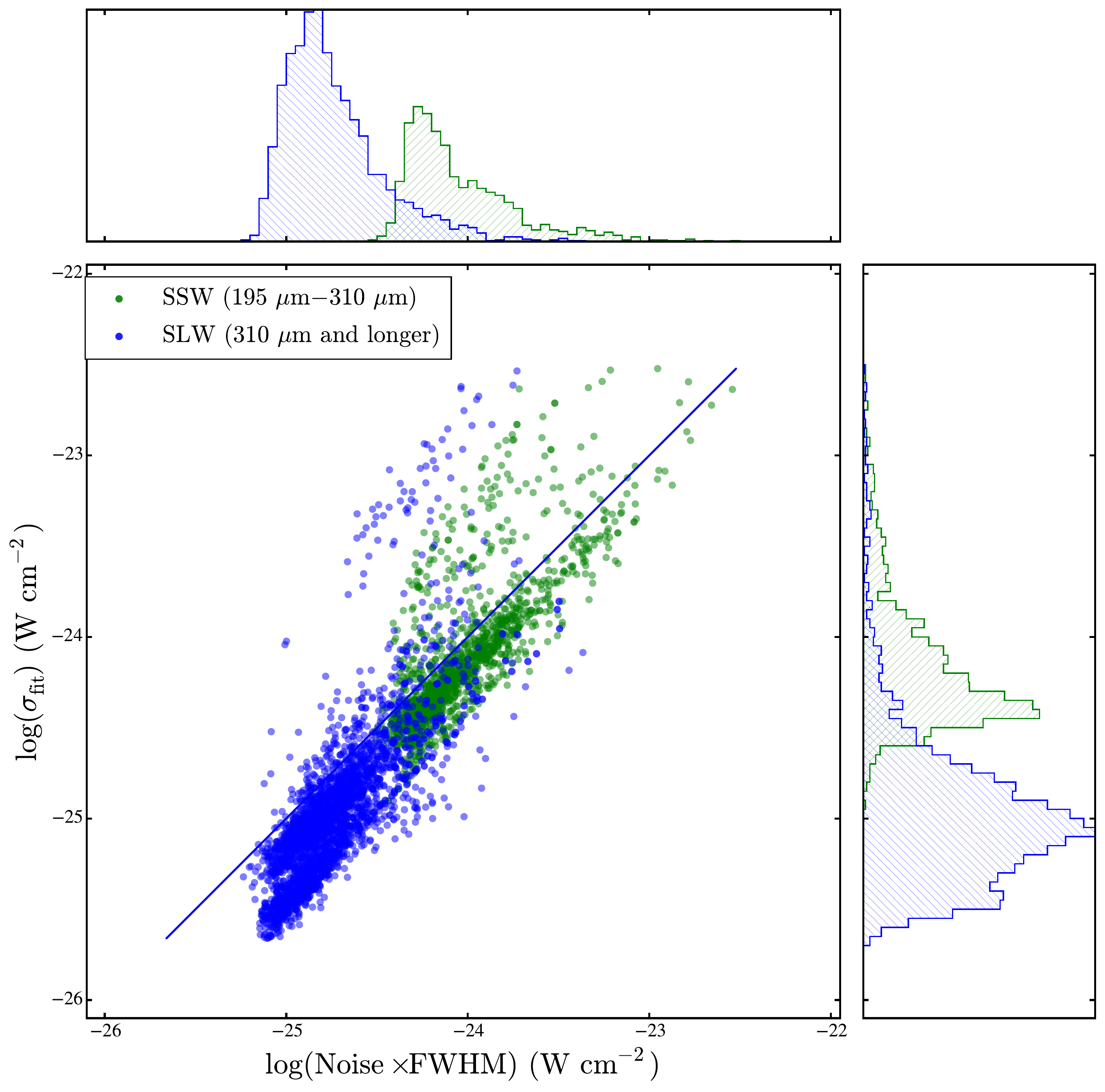}
	\caption{The correlations of the uncertainties measured from the baseline and the uncertainties of the fitted parameters (Top: PACS, Bottom: SPIRE).  Detections with different modules are presented in different colors (see legends).  The blue line indicates the equality of two quantities with zero offset.}
	\label{fig:error}
\end{figure}

By feeding in realistic uncertainties of the spectra, we can calculate reasonable uncertainties in the fitted parameters of the line.  In Figure \ref{fig:error}, we show the relation between the two different uncertainties, the uncertainty from the baseline and the uncertainty in the fitted line flux. Most uncertainties calculated by both methods correlate well with each other, for both PACS and SPIRE data, verifying that the uncertainties are correctly processed in the fitting routine. 

While the uncertainties determined by each method (baseline fitting and line fitting) are similar (marked by a ``line of equality''), there are some interesting features. In both figures, the points spread over about two orders of magnitude, partly reflecting the higher noise in some modules (see Figure \ref{bhr71} for example) and partly due to the changing FWHM in wavelength space, especially for SPIRE.  The distributions of the uncertainties measured in different modules of both PACS and SPIRE show a Gaussian-like distribution with a tail toward higher uncertainty.  For PACS, the uncertainty distribution of each module centers at different values with B2A highest, B2B in middle, and R1 lowest.  The two R1 modules have a similar distribution, although the data are processed separately.  For SPIRE, the SLW module has an uncertainty distribution peaking at a lower value compared to SSW.  Finally, there is a spread of fits that lie above the line of equality reflecting the fact the the assumed line profile is not always a good fit to the data.  
\subsection{Reporting of Line-fitting Results}
\label{sec:report}
The results from the fitting routine are written in the folder called {\tt advanced\_products}, part of the archive structure described in \S \ref{sec:format}. The results include information on the fitted line parameters, as well as the flat spectrum, pure continuum spectrum, and residual spectrum (Figure \ref{fig:full_spec}).  The offset between PACS and SPIRE spectra is discussed in Section~\ref{sec:spire_pipe}.  Guidelines for properly interpreting the report are discussed below.  

These text files are named {\tt object\_reduction\_trim\_lines.txt}, and they contain all of the information necessary to reconstruct the line fits. These ASCII table contains either 18 or 19 columns (for the 1-D spectrum or the spectrum of a single spaxel, respectively).

There are some caveats that should be considered before applying the results. 
Note that every line checked for a fit is listed, so many will be non-detections. Choosing a suitable minimum SNR would be necessary to select likely detections. Then further quality control checks should be applied. First, the ``{\tt Validity}'' column should be {\tt 1} for each unique plausible line detection. If an uncertainty column indicates {\tt -999}, the fit was unsuccessful and the fitted parameter and its uncertainty value should not be used; all lines flagged with {\tt -999} in any column should be discarded. The {\tt -999} flag is applied to the columns of {\tt Sig\_Cen\_WL(um)} and {\tt Sig\_FWHM(um)} if the quality of the fit is poor. This flag indicates that the line profile reached the local $\chi^2$ minimum by moving to the edge of the allowed parameter space, stopping only due to imposed fitting constraints. In this case, the fitting routine is probably not identifying the input emission line. We do not recommend using any line flagged as {\tt 0} in {\tt Validity} or {\tt -999} in other columns, except for the [O~\textsc{i}] 63 $\mu$m line, which can be wider than expected.  We {\it strongly encourage} users to check the fitting plots of [O~\textsc{i}] 63 $\mu$m lines before applying the results.

If an uncertainty column is flagged as {\tt -998}, the fitted parameters can be used, but the uncertainties will need to be extrapolated from other nearby line fits. The {\tt -998} flag appears only in the uncertainty columns {\tt Sig\_FWHM(um)}, {\tt Sig\_Cen\_WL(um)} and {\tt Sig\_Str(W/cm2)}, and only if the line width is fixed to a particular value (the instrumental resolution at that wavelength).  In this case uncertainties cannot be generated, but the fitted parameters are considered reliable.  We fix most line widths in PACS line fits (an exception is [O~\textsc{i}] 63 $\mu$m, which has been observed to be broader than the instrumental line width), but only fix the width for cases of double-Gaussian fitting in SPIRE line fits. In these cases of fixed line width, we cannot calculate uncertainties for the line width. If an uncertainty is needed, we recommend using values for a nearby line if available. A portion of the fitting results table is shown in Table \ref{table:report}; the full table is available in the online manuscript.

A summary of sources and how they were processed can be found at \linebreak``{\tt full\_source\_list\_refine.txt}''. It lists the names of the processed sources in order and the status of access to different reductions. Additional information can be found in the file ``{\tt full\_source\_list.txt}'', which includes the information about the noise re-estimation (see \S \ref{sec:noise}). 

In total, we performed line fitting on 67 and 33 objects in the PACS and SPIRE bands, respectively.  2-D PACS datacubes consist of four modules of 25 spaxels each, and SPIRE datacubes consist of two modules of 33 and 19 spaxels, for SSW and SLW respectively. Including all spaxels in the 2-D datacube, totaling 264590 PACS and 60330 SPIRE line fitting processes, we found 10474 and 4985 detections without any anomaly in PACS and SPIRE fits, compared to 254116 and 55345 non-detections, including anomalies (fits with SNR $>$ 3 but a flag for poor fitting quality applied). Any line fit in which an unphysical broad component was fitted, the line centroid diverged and reached our pre-defined boundary, or the fitted line was blended with another line with lower Einstein-A value, was determined to be an anomaly.  There were a total of 322 and 1334 anomalies  in the PACS and SPIRE fits. On average, we found 156 and 151 lines per object in PACS and SPIRE bands, including all spatial positions. The fitting results of all PACS and SPIRE sources are also stored in four ASCII files for 1-D and 2-D spectra (``{\tt CDF\_archive\_pacs[spire]\_1d[cube]\_lines.txt}'').

\subsection{Contour Plots}
\begin{figure*}[ht!]
\centering
\includegraphics[width=0.7\textwidth]{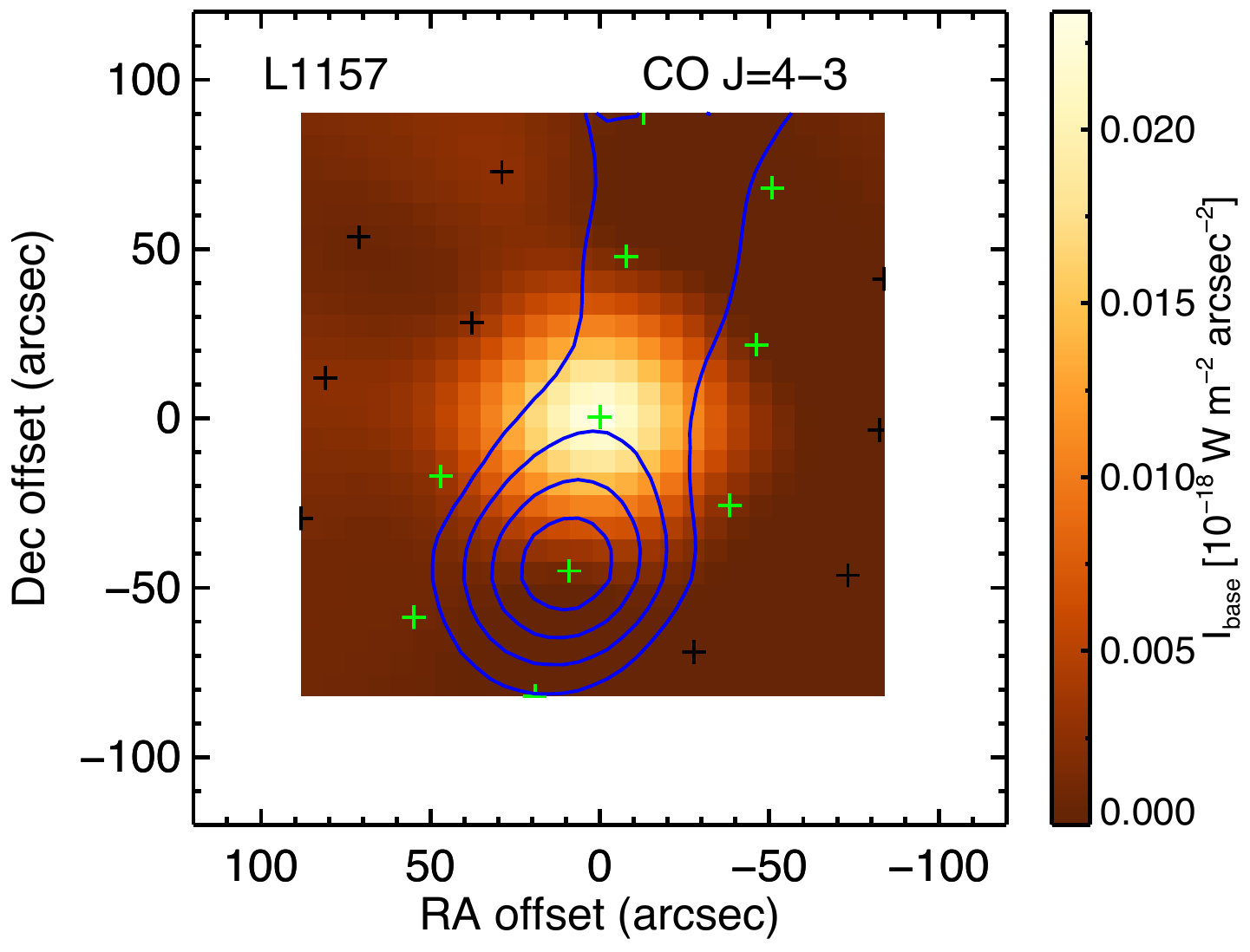}
\includegraphics[width=0.7\textwidth]{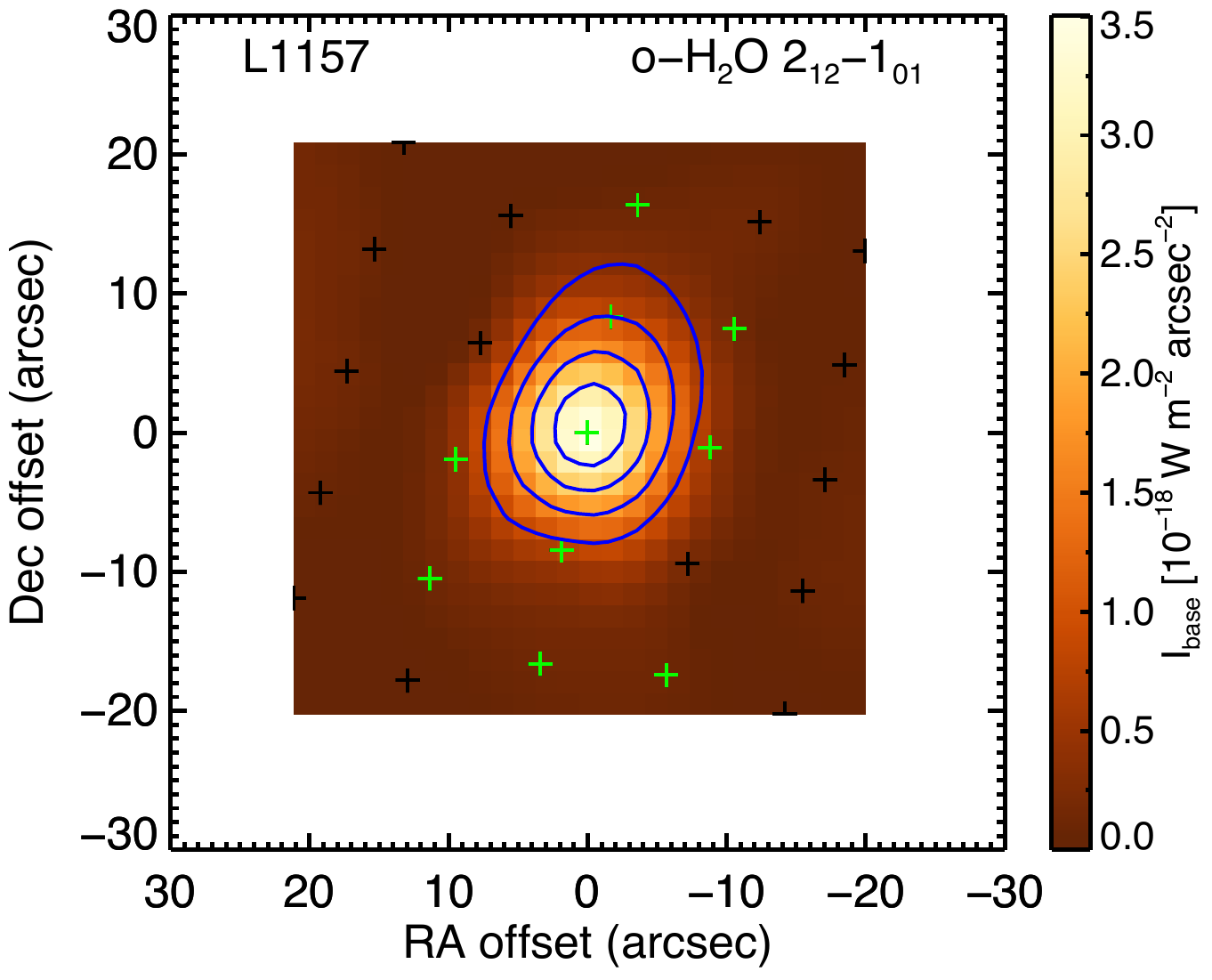}
\caption{An example of 2-D contour maps of L1157 \coul{4}{3} and $\owater~2_{12}-1_{01}$ line emission (blue contours) with local continuum (filled background) observed in SPIRE and PACS respectively. The green crosses indicate the spaxels that have detections.}
\label{fig:contour}
\end{figure*}

To further visualize the fitting results, we produce contour plots of each line (line contours) overplotted on the local continuum at the wavelength of the line centroid  (e.g., Figure \ref{fig:contour}). We plot contours only for maps with detections in {\it more} than 2 spaxels; however, plots with continuum emission are provided for maps with fewer detections.  The contours are the minimum curvature surface interpolated from lines with SNR greater than 3. The adjacent contours are in steps of 20\%\ of the peak, while the lowest 20\%\ contour is not shown. The crosses (white) in each figure indicate the physical position of each spaxel and the center cross indicates the pointing position. The crosses marked in green indicate spaxels with significant line detections; users can judge the reliability of detailed structure in an individual contour plot based on the number and location of spaxels with detections.  

\clearpage

\section{Archive Structure and File Format}
\label{sec:format}

The archive can be accessed  through the Herschel Science Archive under User Provided Data Products\footnote{\href{http://www.cosmos.esa.int/web/herschel/user-provided-data-products}{http://www.cosmos.esa.int/web/herschel/user-provided-data-products}} or as a downloadable archive file\footnote{\href{ftp://hsa.esac.esa.int/URD_rep/DIGIT/}{ftp://hsa.esac.esa.int/URD\_rep/DIGIT/}}. The file and directory trees contained in the archive are indicated in Figure \ref{file_struc}. 
We provide the high-level data from the optimized pipeline and the products of line fitting results. The following section describes the file structure and how to interpret the individual formats.

\subsection{File Structure}

\begin{figure}[htbp!]
\centering
\includegraphics[width=\textwidth]{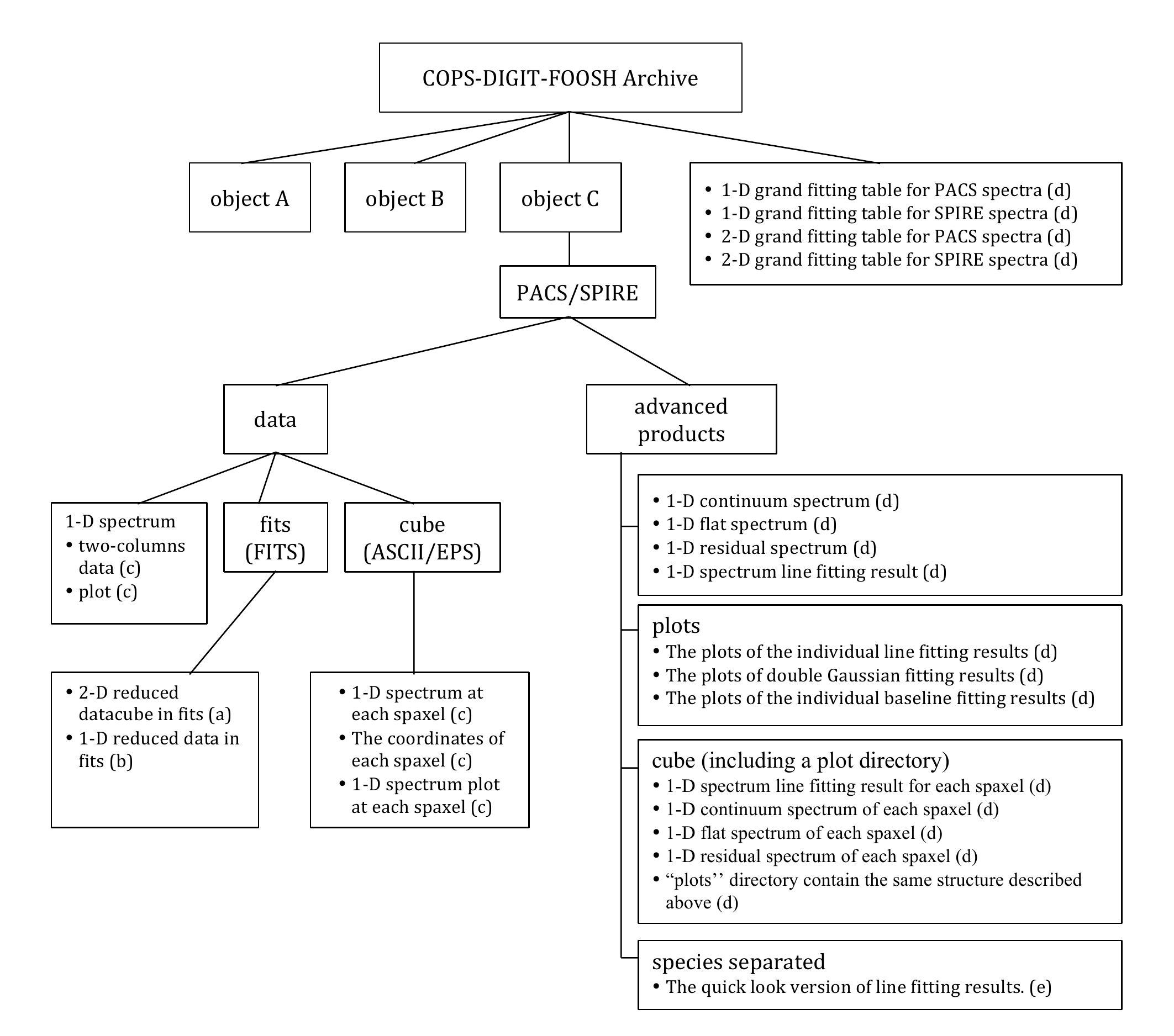}
\caption{The hierarchy structure of the archival file system. Files are organized primarily by the targets, and every target directory has the same substructure. The PACS and SPIRE high-level data and fitting results are stored in separate directories with the same structure. The letters refer to
the sub-headings describing the products in \S \ref{filedesc}.}
\label{file_struc}
\end{figure}

The schematic view of the file structure is shown in Figure \ref{file_struc}. If there is no PACS or SPIRE observation available for a target, then the corresponding directory is not present. The detailed description of each product is described below. The products are identified in Figure \ref{file_struc} with the same letter (a through e) used to identify the following sections.

\subsection{File Format}\label{filedesc}
\begin{enumerate}[label=(\alph*)]

\item 2-D reduced datacubes (FITS)
	\begin{itemize}
		\item PACS \\
		We generate three datacubes for PACS in FITS format: Source$-$Nod~A; Source$-$Nod~B; Average (Source$-$(Nod A$+$Nod B)~/~2).  These can be read as dataplanes for each spaxel, with appropriate RA/Dec, and flux vs. wavelength.  They are generated with wavelength grid oversample $=$ 2 and upsample $=$ 1 by default.  These cubes include all corrections (in particular jitter correction).

		{\noindent Filename: }
		\medskip

		{\noindent {\tt OBSID\_1342xxxxxx\_Targetname\_blue/red\_rebinnedcubesnoda\_os\#sf\#.fits} \\
				  {\tt OBSID\_1342xxxxxx\_Targetname\_blue/red\_rebinnedcubesnodb\_os\#sf\#.fits} \\
				  {\tt OBSID\_1342xxxxxx\_Targetname\_blue/red\_finalcubes\_os\#sf\#.fits}
		}
		\smallskip \\
		We also generate these same three cubes {\it without} the pointing offset correction, with the addition of ``{\tt \_nojitter.fits}'' to the filename.  We include the offset-corrected product only if the correction was successful.  We include the non-offset corrected version for {\it all} sources. {\it If a jitter-corrected version is supplied, it means that we recommend using it.}

		\item SPIRE \\
		We generate one datacube for SPIRE in FITS format.  These can be read as dataplanes for each spaxel, with appropriate RA/Dec, and flux vs. wavelength.  Note that both SLW and SSW modules are included in a single FITS file.  These cubes include all corrections.

		{\noindent Filename:}
		\medskip

		{\noindent {\tt 1342xxxxxx\_spectrum\_extended\_HR\_aNB\_15.fits}}
		\smallskip \\
		There is no offset-corrected product being generated for SPIRE data.
	\end{itemize}

\item 1-D reduced spectra (FITS)

\begin{itemize}
	\item PACS \\
	We generate FITS files for three cases listed below, each with and without jitter correction; all six resulting files have the point source (PSF) correction applied.  We also provide files in ASCII and EPS formats (see next section).

	{\noindent Filename: }
	\medskip

	{\noindent {\tt OBSID\_1342xxxxxx\_Targetname\_blue/red\_centralSpaxel\_\linebreak PointSourceCorrected\_Corrected3x3NO\_slice00\_os\#sf\#.fits}}
	\smallskip

	This is the simplest product, which is just the spectrum of the central spaxel with only the PSF correction applied.  The ``{\tt slice}'' is only non-zero for multiple rangescan or linescan observations, representing different wavelength settings within the original OBSID.

	{\noindent Filename: }
	\medskip

	{\noindent {\tt OBSID\_1342xxxxxx\_Targetname\_blue/red\_centralSpaxel\_\linebreak PointSourceCorrected\_Corrected3x3YES\_slice00\_os\#sf\#.fits}}
	\medskip

	This is the same as the above product but with the jitter/pointing correction for the loss of light between pixels included in addition to the PSF correction.   That is, we start with the central spaxel, and apply both the PSF and pointing/jitter correction. 

	We also generate the straight sum of the 3x3 spaxels, including PSF and pointing/jitter corrections:

	{\noindent Filename: }
	\medskip

	{\noindent {\tt OBSID\_1342xxxxxx\_Targetname\_blue/red\_central9Spaxels\_ \linebreak PointSourceCorrected\_slice00\_os\#sf\#.fits}}
	\medskip

	A summation of the central 9 spaxels (the ``{\tt 3x3}'' case) with PSF correction applied.

	So in summary, all three files have the point source (PSF) correction applied.  The ``{\tt 3x3NO}'' is just this raw product.  The ``{\tt 3x3YES}'' is the flux of the central spaxel after correcting for the PSF and the pointing/jitter.  The ``{\tt central9Spaxels}'' is the sum of the flux in the central 3x3 spaxels, still including the PSF/pointing jitter corrections.

	We also output a version of these files without the pointing/jitter corrections.  The filename is similar to the previous version, but with ``{\tt \_nojitter.fits}'' appended; ``{\tt nojitter}'' refers to no jitter corrections applied.

	\begin{figure*}[htbp!]
		\centering
		\includegraphics[width=0.7\textwidth]{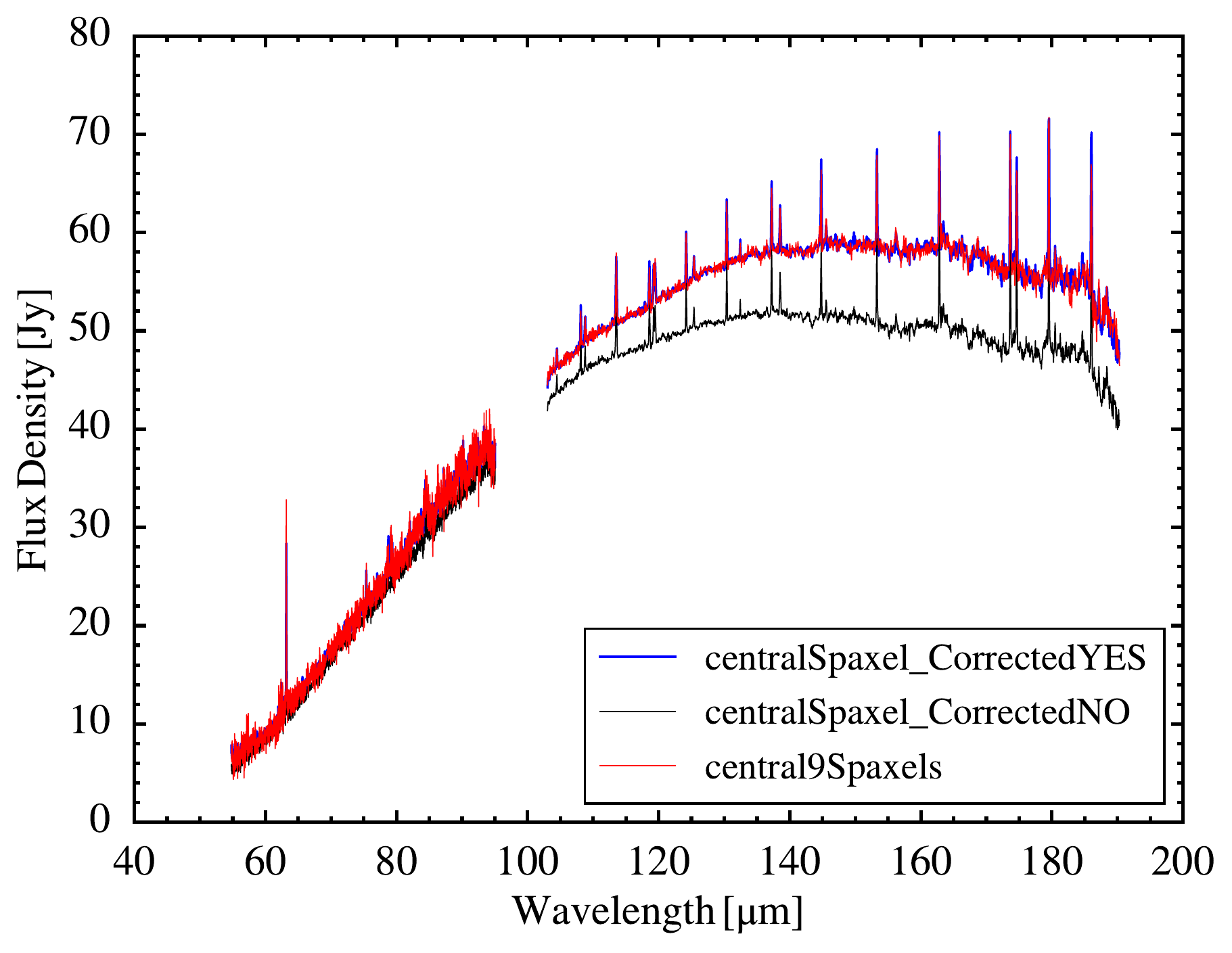}
		\caption{Plot of our three PACS pipeline products, for protostar L1157.  The blue spectrum shows the central spaxel product, with best SNR without accounting for source extent.  The red spectrum is the sum of the central 9 spaxels in the PACS array; this likely includes all of the source flux, but also added noise from low SNR spaxels.  The black spectrum is the best compromise product, flux calibrated to the red spectrum that has most of the source flux, but with SNR consistent with the blue spectrum.  Note that lines are stronger in black than red with only modest increase of noise, therefore higher SNR.}
		\label{3products}
	\end{figure*}

	We compare the three products for the case of L1157 in Figure \ref{3products}.  The blue spectrum shows the centralSpaxel product without the absolute flux calibration of the 3x3 aperture extraction; the red spectrum is the straight sum of the central 3x3 spaxels, and the black spectrum is the best product, combining the SNR of the central spaxel with the absolute flux calibration of the 3x3 aperture extraction.

	\item SPIRE \\
	We generate one FITS file containing the SPIRE 1-D spectrum for each target.  The spectra are processed by the pipeline setting described in \S \ref{sec:spire_pipe} including ``SemiExtendedSourceCorrector'' routine.

	{\noindent Filename:}
	\medskip

	{\noindent {\tt Targetname\_spire\_corrected.fits}}
	\smallskip

	In this FITS file, all of the dataplanes are ordered as in the 2-D datacube.  But only dataplanes 5 and 8 containing the central spaxel in SLW and SSW are the ones with corrected values.  The lines are fitted from these spectra as well.
\end{itemize}

\item 1-D reduced spectra (two-columns ASCII/EPS)

For both the 1-D reduced spectra and 2-D datacubes, we extract the information (wavelength and flux) from the FITS files and report them in two-column ASCII format with plots of the corresponding spectrum for each spaxel or source.  The same plotting method is applied to both 1-D spectra and spaxels within the datacubes.  In PACS, we combine all of the modules (B2A, B2B and R1, see \S 2) together in ASCII files and plots.  In addition, because the pointing coordinate of each spaxel varied slightly with wavelength,  the files ending with ``{\tt coord}'' contain the coordinates (RA/Dec) of each spaxel at each wavelength point.  The coordinates reported in the 1-D spectra fitting results table are derived from the mean values of the coordinates across all wavelengths. The standard deviations of the coordinates across all wavelengths range from from 0.0073\arcsec\ to 0.354\arcsec\ for RA, and 0.0233\arcsec\ to 0.4386\arcsec\ for Dec.  The mean values of the standard deviation are 0.129\arcsec\ and 0.0675\arcsec\ in RA and Dec respectively.  But different modules are reported separately for SPIRE due to their different spaxel configurations.  \textbf{We recommend using the coordinates from the files we provided instead of reading in from the header of the FITS files.}

\noindent Filename: 
\medskip

\noindent (1-D) \\
\noindent {\tt Targetname\_reduction\_trim.txt} \\
\noindent {\tt Targetname\_reduction\_trim.eps} \\
\noindent (2-D) \\
\noindent {\tt Targetname\_pacs\_pixel\#\#\_os\#sf\#.txt} \\
\noindent {\tt Targetname\_pacs\_pixel\#\#\_os\#sf\#\_coord.txt} \\
\noindent {\tt Targetname\_pacs\_pixel\#\#\_os\#sf\#.eps} \\
\noindent {\tt Targetname\_pixelname.txt} \\
\noindent {\tt Targetname\_pixelname.eps} \\

\medskip

These ASCII files (and the corresponding EPS plots) are trimmed consistently at specific wavelengths.  Here we describe trimming details for PACS and SPIRE separately. 
\begin{itemize}
\item PACS \\
For the shortest wavelength module (B2A), we remove all wavelengths short of 54.80 $\mu$m, and long of 72.3 $\mu$m.  For the next shortest (B2B), we trim all wavelengths short of 72.3 $\mu$m and long of 95.05 $\mu$m.  For the third (R1; 100-145 $\mu$m), we remove all wavelengths short of 101.4 $\mu$m and long of 143.0 $\mu$m.  For the fourth (R1; 145-210 $\mu$m), we trim all wavelengths short of 190.31 $\mu$m and long of 143.0 $\mu$m. 
\item SPIRE \\
For the shorter wavelength module (SSW), we remove all wavelengths short of 195 $\mu$m and long of 310 $\mu$m.  For the longer wavelength module (SLW), we remove all wavelengths short of 310 $\mu$m. 
\end{itemize}

We find this method gives the best overall continuum fit between modules, with the highest SNR and consistent continuum in the overlap regions, stitched into a single 1-D spectrum.  Figure~\ref{trimmed_detail} shows the original spectra from all modules, and the shaded regions indicate the part preserved after the trimming.  The regions where the continuum behaves abnormally and/or the noise increases significantly are excluded via the trimming process.  Note that the original,
untrimmed spectra are still present in our archive.  The FITS files previously described do not use these trim points and present the full spectra from all modules.  An example of the final trimmed spectra of the protostar L1157 (PACS and SPIRE) is shown in Figure \ref{tmr1}, with each module highlighted by color.

\medskip

\begin{figure}[ht!]
	\centering
	\includegraphics[width=0.7\textwidth]{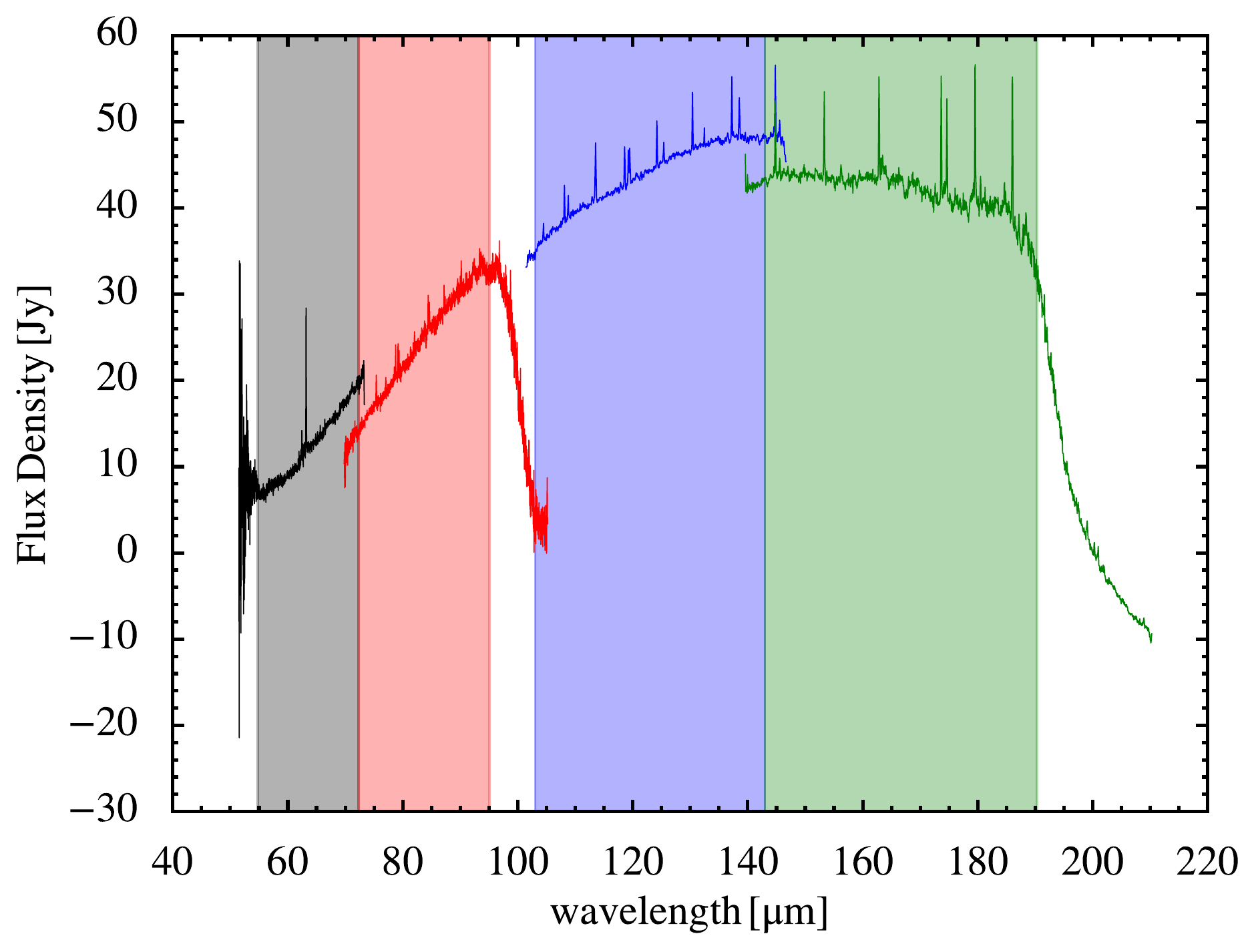} \\
	\includegraphics[width=0.7\textwidth]{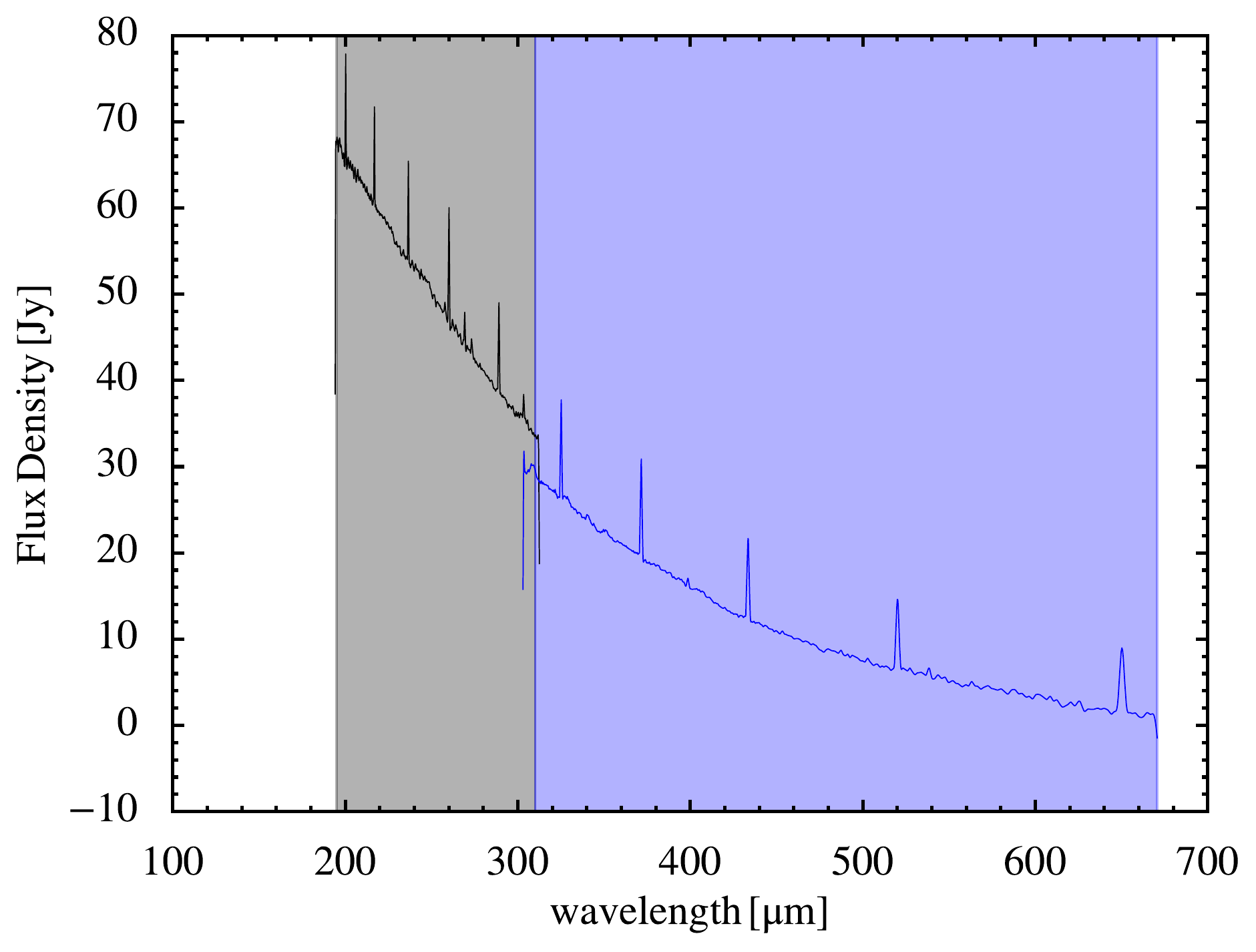}
	\caption{Original spectra of L1157, each module shown in a different color.  The PACS spectra are shown at the top, while the SPIRE spectra are shown at the bottom.  Note that the PACS spectra are shifted by -5~Jy between each module and the SPIRE spectra are shifted by 5~Jy between two modules for better visualization of the overlap regions.  The shaded area indicates the region where the spectrum with the same color is preserved after trimming.}
	\label{trimmed_detail}
\end{figure}

\begin{figure}[ht!]
\centering
\includegraphics[width=0.7\textwidth]{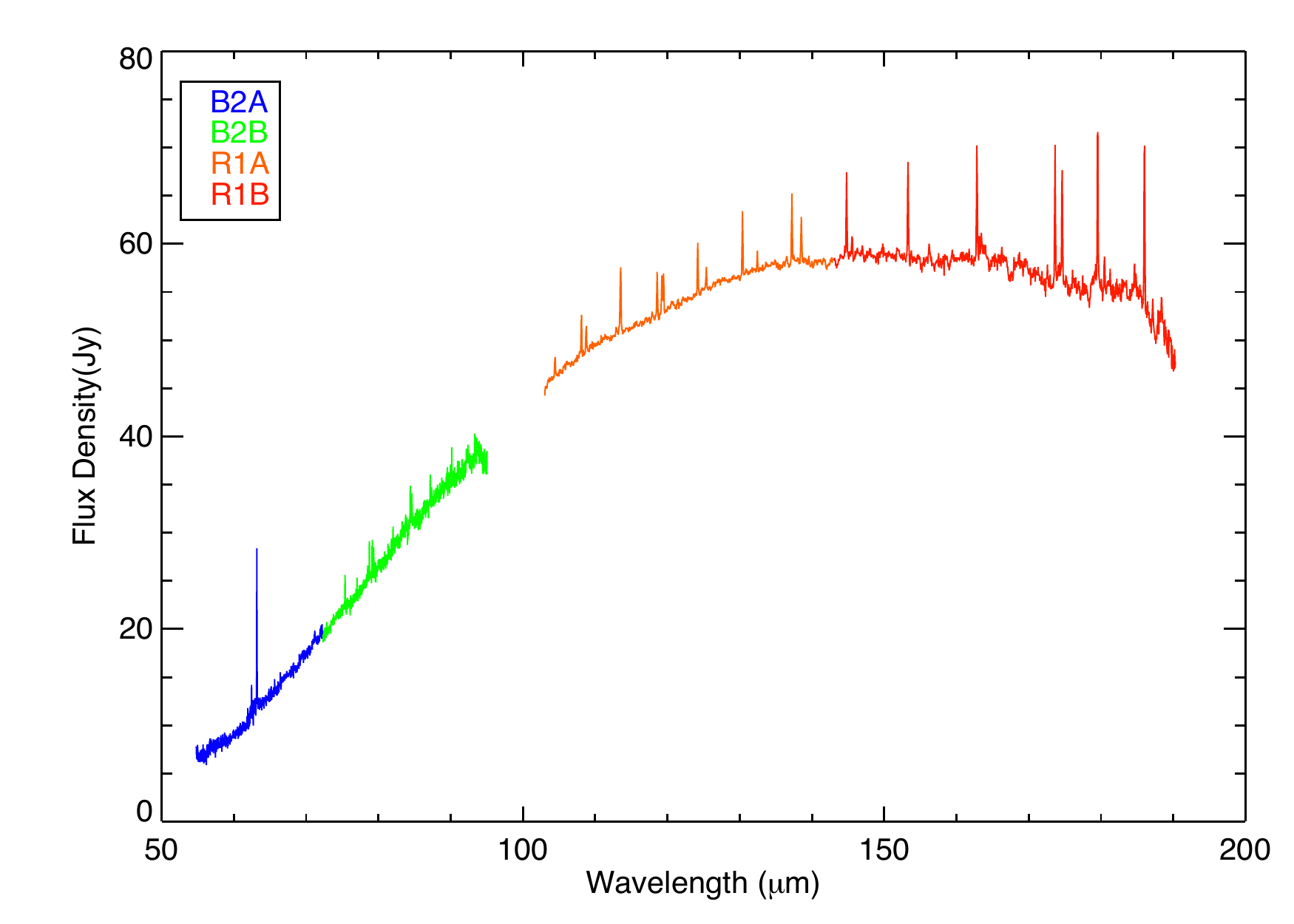} \\
\includegraphics[width=0.7\textwidth]{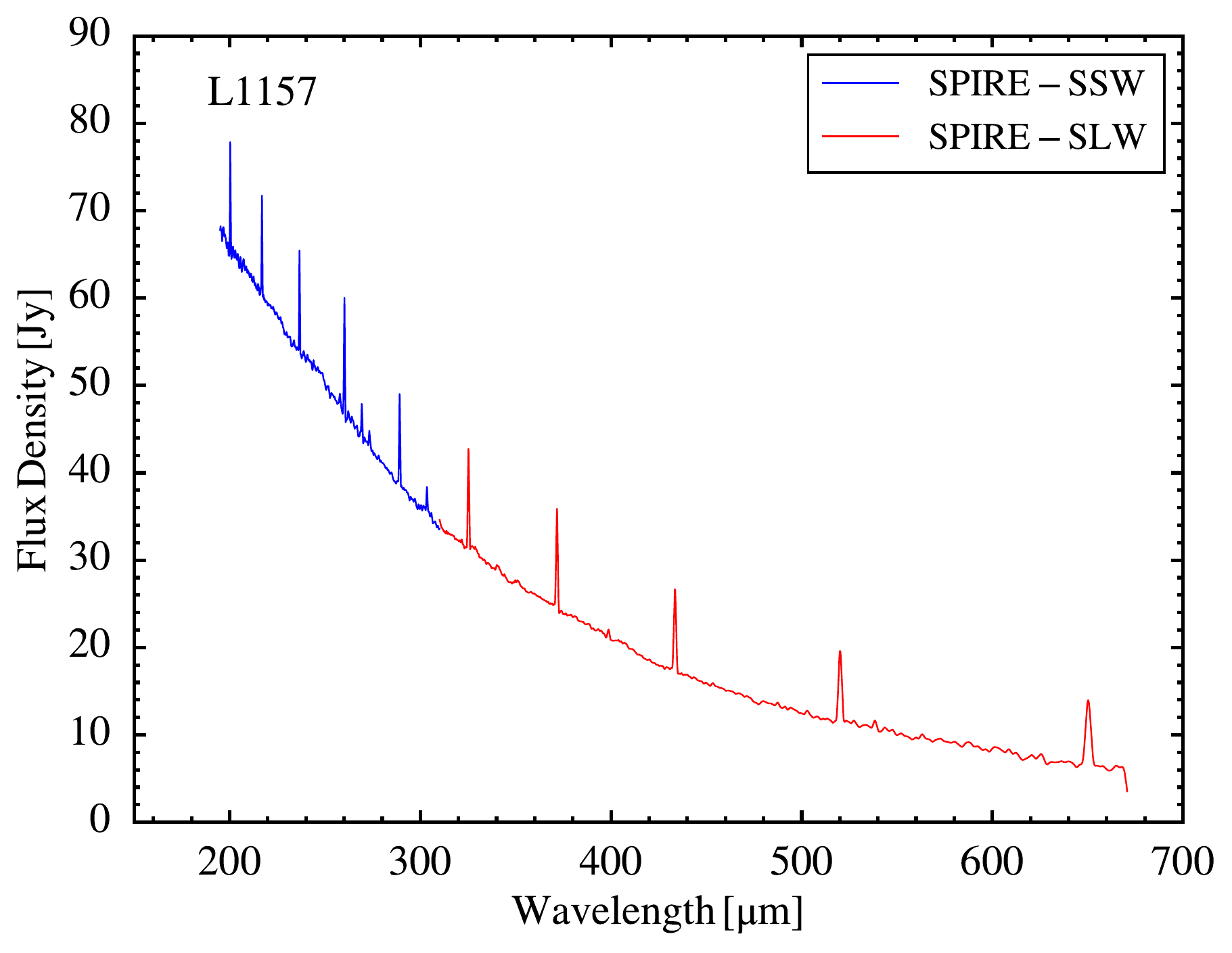}
\caption{{\bf Top}: An example 1-D PACS spectrum of L1157, with the four modules highlighted by color.  {\bf Bottom}: An example 1-D SPIRE spectrum of L1157, with the two modules highlighted by color.}
\label{tmr1}
\end{figure}

\clearpage

\item Line fitting results \\
The line fitting results are reported in text and figures format.  The ASCII files provide the tables of the fitting results so that users can read them in with any language they prefer.  The figures visualize the fitting results with the original data, fitted line(s), and residual.  The users can visually check the fitting results with the figures easily.

\begin{minipage}{\textwidth}
\noindent Filename:
\smallskip 

(PACS)\\
\noindent {\tt Targetname\_reduction\_trim\_lines.txt} \\
\noindent {\tt Targetname\_reduction\_trim\_continuum.txt} \\
\noindent {\tt Targetname\_reduction\_trim\_flat\_spectrum.txt} \\
\noindent {\tt Targetname\_reduction\_trim\_residual\_spectrum.txt} \\
\noindent {\tt Targetname\_pacs\_pixel\#\#\_os\#sf\#\_lines.txt} \\
\noindent {\tt Targetname\_pacs\_pixel\#\#\_os\#sf\#\_continuum.txt} \\
\noindent {\tt Targetname\_pacs\_pixel\#\#\_os\#sf\#\_flat\_spectrum.txt} \\
\noindent {\tt Targetname\_pacs\_pixel\#\#\_os\#sf\#\_residual\_spectrum.txt} \\
\end{minipage}

\begin{minipage}{\textwidth}
(SPIRE)\\
\noindent {\tt Targetname\_spire\_corrected\_lines.txt} \\
\noindent {\tt Targetname\_spire\_corrected\_continuum.txt} \\
\noindent {\tt Targetname\_spire\_corrected\_flat\_spectrum.txt} \\
\noindent {\tt Targetname\_spire\_corrected\_residual\_spectrum.txt} \\ 
\noindent {\tt Targetname\_pixelname\_lines.txt} \\
\noindent {\tt Targetname\_pixelname\_continuum.txt} \\
\noindent {\tt Targetname\_pixelname\_flat\_spectrum.txt} \\
\noindent {\tt Targetname\_pixelname\_redsidual\_spectrum.txt} \\
\end{minipage}
The ASCII format line fitting results are presented in four output forms (lines, continuum, flat\_spectrum, and noise\_spectrum), as listed above.  The first contains the full report of the line fitting results using the method described in \S \ref{sec:line_fitting}, containing the fitted line parameters (Table \ref{table:report}) of each line in the full input list -- including non-detections. Users can simply read this report and apply the fitting results.  The second file contains the continuum spectrum of the source produced by subtracting the fitted lines from the original spectrum. The third file is the counterpart of the continuum file, a continuum-subtracted spectrum in which only the spectral lines and flat baseline remain (Figure \ref{fig:full_spec}).  The fourth file contains  the residual spectrum after the subtraction of fitted lines and smoothed continuum.  The file names of PACS spectra are listed in the first block with the first four as the products of 1-D spectra and last four as the products of 2-D spectra, while file names of SPIRE spectra are listed in the second block with the same fashion. 
\medskip

\begin{minipage}{\textwidth}
(Fitting table including all sources) \\
\noindent {\tt CDF\_archive\_pacs\_1d\_lines.txt} \\
\noindent {\tt CDF\_archive\_pacs\_cube\_lines.txt} \\
\noindent {\tt CDF\_archive\_spire\_1d\_lines.txt} \\
\noindent {\tt CDF\_archive\_spire\_cube\_lines.txt} \\
\end{minipage}
The fitting results of 1-D PACS, 1-D SPIRE, 2-D PACS, and 2-D SPIRE of all sources are stored in separate ASCII files with similar tables to those described above and in \S \ref{sec:line_fitting} with an additional column with the object name, and a summary table that includes ALL fitting results (PACS, SPIRE, 1-D, and 2-D) is provided with a name of {\tt CDF\_archive\_lines.txt}, in which the pixel column is labeled as {\tt ``c''} for 1-D spectra.
\medskip

\begin{minipage}{\textwidth}
(Individual line fitting plot in PACS) \\
\noindent {\tt spectrum\_line\_subtracted\_Targetname\_reduction\_trim.eps} \\
\noindent {\tt Targetname\_reduction\_trim\_linename.eps} \\
\noindent {\tt Targetname\_reduction\_trim\_line1+line2.eps} (in {\tt ``double\_gauss'' folder}) \\
\\
(Individual line fitting plot in SPIRE) \\
\noindent {\tt Targetname\_spectrum\_line\_subtracted\_spire\_corrected.eps} \\
\noindent {\tt Targetname\_spire\_corrected\_linename.eps} \\
\noindent {\tt Targetname\_spire\_corrected\_line1+line2.eps} (in {\tt ``double\_gauss'' folder})\\
\end{minipage}
The figures are all presented for typical spectra with various levels of data, as listed above.  The first three of the six items are figures for PACS, while the last three of six are for SPIRE.  The files with ``{\tt spectrum\_line\_subtracted}'' provide overviews for line fitting results of each object and each instrument.  It includes the original data (black), continuum (blue), and residual spectrum (green).  The files with ``{\tt linename}'' provide the visualization of the fitting result for each line (Figure \ref{fig:line_base} top).  It includes original data (black), fitted line (blue), and residual/noise (green).  The vertical lines represent the allowed region for line (dash line) and the region taken for the baseline fitting (dotted line).  And the SNR and FWHM are printed within the figures.  The fitting results of the double Gaussian, stored in {\tt ``double\_gauss'' folder}, are plotted in figures with two lines in their file names (``{\tt line1+line2}'') with an additional fitted line in orange (Figure \ref{doublegauss}).  We also report the fitting results of the baseline in the subdirectory named ``base'' (Figure \ref{fig:line_base} bottom).  The plot includes the original data (black), the fitted baseline (purple), residual (red), and the points taken in the baseline fitting (asterisks).  There is another directory named ``cannot\_fit'' which includes the very rare case when the fitting failed to coverage.

\begin{figure}
	\centering
	\includegraphics[width=0.7\textwidth]{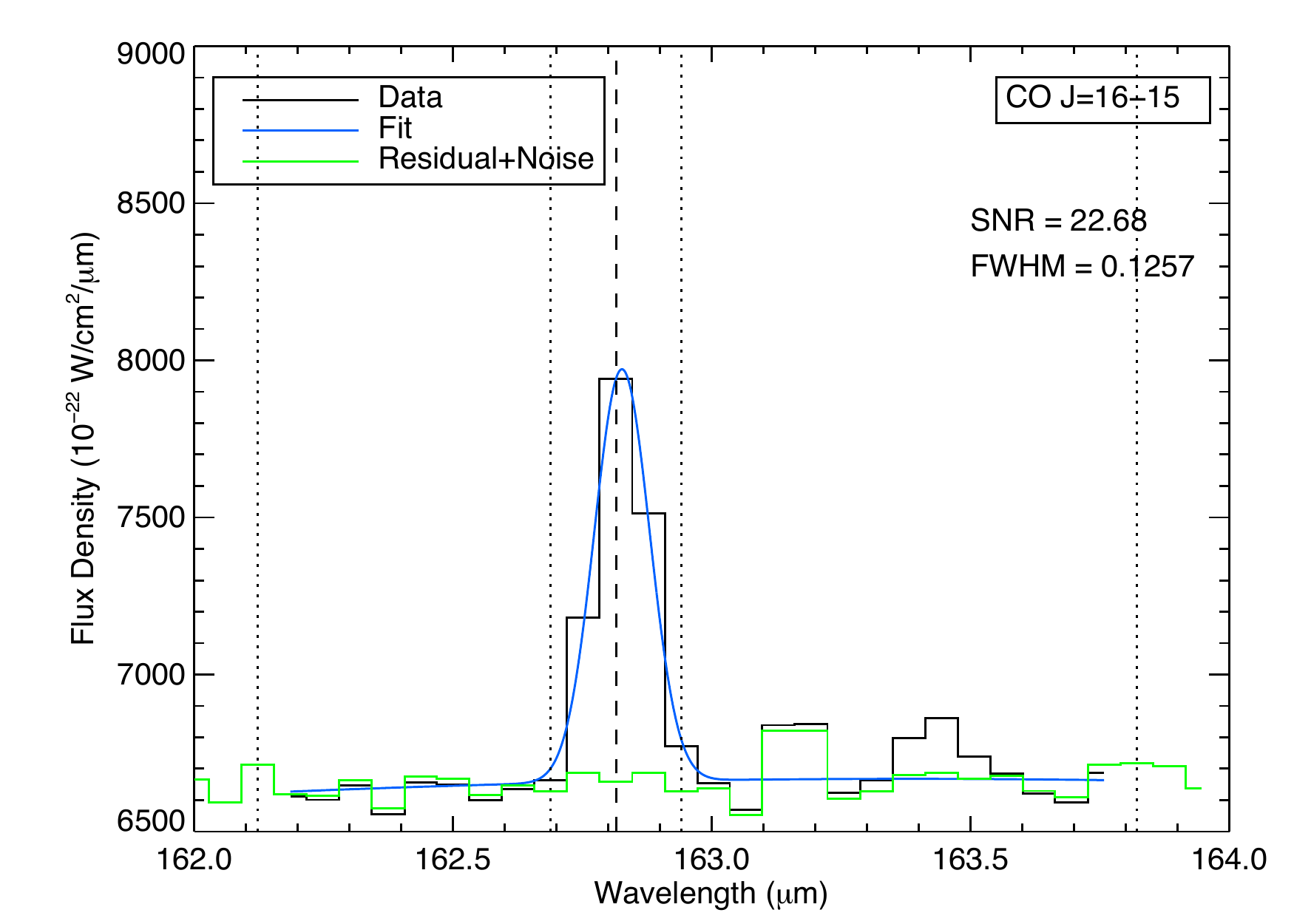}
	\includegraphics[width=0.7\textwidth]{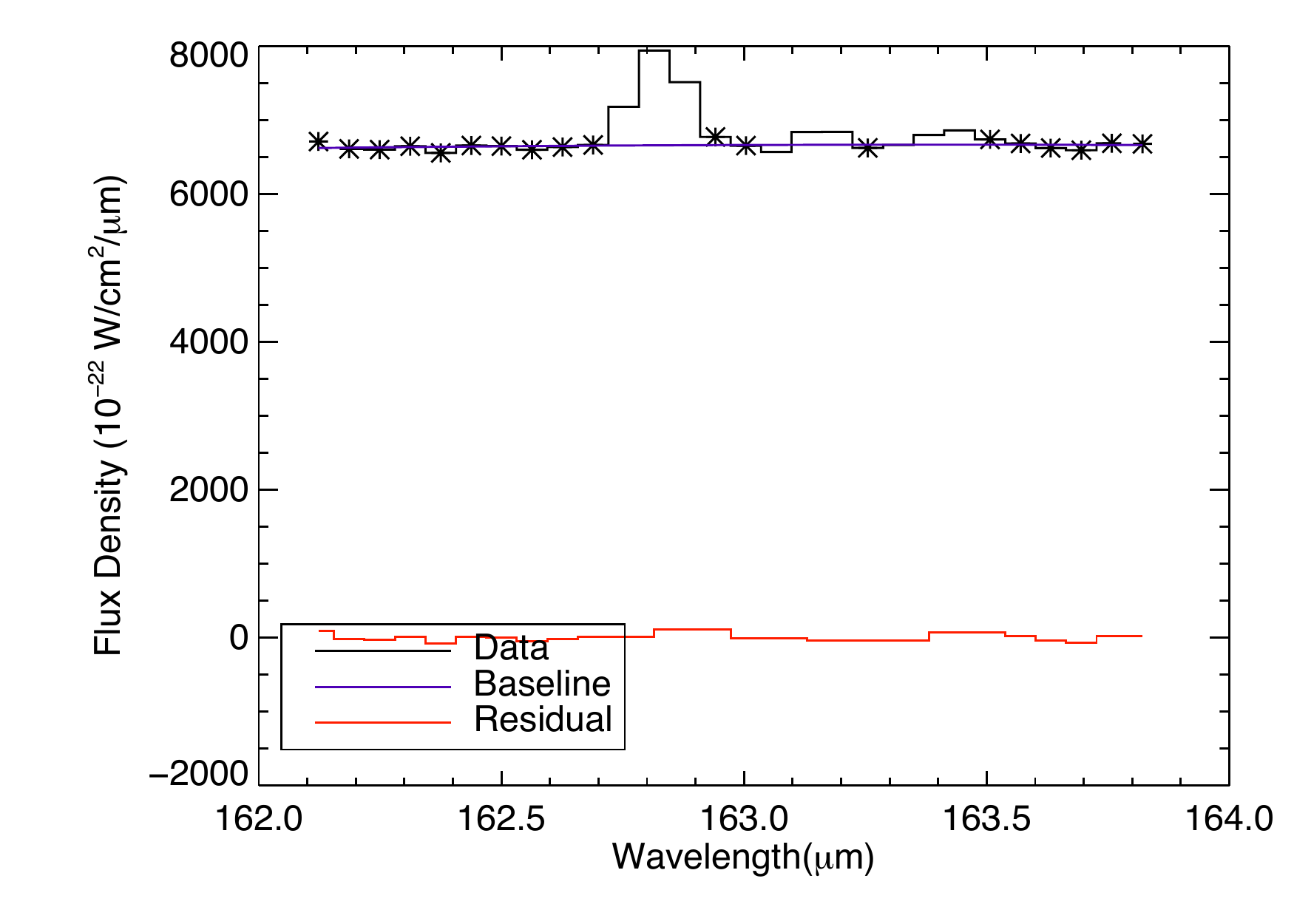}
	\caption{{\bf Top}: An example of single line fitting result of \coul{16}{15} in L1157. {\bf Bottom}: An example of the baseline fitting result at the region of \coul{16}{15} in L1157.  The asterisks indicate the points used for the baseline fitting.}
	\label{fig:line_base}
\end{figure}

\item Quick look output for the fitting results

\noindent Filename:
\medskip \\
\noindent {\tt Targetname\_reduction\_trim\_lines\_species.txt} \\
\noindent {\tt Targetname\_spire\_corrected\_species.txt}
\medskip

This output provides the user a convenient quick look at the source spectral properties and evaluation of the fitting quality. These files contain the sorted line information of either the 1-D spectrum or the spectrum of each spaxel for given molecular/atomic species. The line profiles of seven different species are separated including \co, \thirteenco, \hcoplus, OH, \owater, \pwater, and atomic fine-structure transitions. The data for each molecule are stored in the corresponding files.

\end{enumerate}

\subsection{Comparison to Previously Published Results}

The scientific implications of the reprocessed data archive will be reported in other papers.  We note here the impact of the new archive on rotational temperatures and densities reported in \citet{green13b} and other works.  We compared a sample of 18 sources with measured (PACS only) rotational temperatures and densities, divided into ``warm'' ($\sim$ 300 K) and ``hot'' ($\sim$ 1000 K) components as defined in previous papers, measured using the 2012-13 pipeline and again with the current pipeline.    The rotational temperatures changed insignificantly: the average warm component temperature increased by 1.4\% (ranging between -18\% and 25\%), while the average hot component temperature decreased by 1.5\% (almost entirely due to one source, GSS30-IRS1, decreasing by 7\%).  In contrast, the derived number densities were altered substantially.  The average warm component increased in count by 17.6\% (ranging from -27.5\% to 62.5\%, with outlier GSS30-IRS1 increasing by 198\%).    The hot component increased by 22\%, but this was entirely because of GSS30-IRS1, which increased 107\%; with that
source omitted, the mean of the rest changed by less than 2\%.

Thus the measured temperature values have changed only minimally, and the hot component was only slightly different except in cases where we properly captured the extended nature of the source in the new data products.  In GSS30-IRS1 in particular, we previously used the central spaxel only, and did not include extended emission; in this new analysis, we include the larger region.  However, we do not attempt to disentangle GSS30, as was done using earlier data products; \citet{je15} deconvolved GSS30 into three separate IR sources. However, the increase in the warm component molecule number count is widespread in our sample, and is due to the calibration improvements in the long wavelengths of the PACS pipeline (100-190 $\mu$m).

In general, the adjustment in line flux does not impact previous results qualitatively.  \citet{green13b} noted in Figure 23 that the low density solution in \citet{neufeld12} was a good fit to the CO rotational diagram for B335; this conclusion is unchanged.

\citet{lindberg14} and \citet{dionatos13} considered complicated regions (RCrA and Serpens, respectively) and created prescriptions to deconvolve the sources therein.  Our pipeline does not consider the deconvolution problem and we recommend using the earlier works.  For example, \citet{dionatos13} find that the lines peak at the position of outflow shocks rather than the protostellar sources.  This can be identified from the contour plots in our dataset, and would require custom extraction techniques.  We provide a well-calibrated spectrum for Serpens-SMM3 including the jitter correction; however the complex of shocks causes a different distribution of lines.  Serpens-SMM4 was not successfully corrected for pointing jitter as noted in Table \ref{obslog}.  Similarly, RCrA-IRS7B is calibrated via our pipeline, but RCrA-IRS7C is not successfully corrected. In each case, the earlier work uses line fluxes from spaxels associated with outflow shocks rather than the continuum (protostellar) sources.  Both find that the rotational temperatures of the spaxels dominated by shocks are similar to those toward the protostar, but the number of molecules is different.  In each case, we expect that using our new dataset would provide reduced absolute calibration and line flux uncertainty, but the deconvolution would still be required as post-processing.

The case of L1448-MM \citep{lee13} is similar, but the central source dominates the continuum and line fluxes.  Thus the deconvolution is required to separate L1448-C, N, and S, but the overall results for the dominant submillimeter source L1448-MM are not significantly changed.

Finally, the FOOSH dataset included both an earlier SPIRE and PACS reduction, and 4 of those objects appear in this dataset.  The warm (PACS) component was only detected in V1057 Cyg and the temperature has increased by a factor consistent with the DIGIT PACS sources, 17\%, with a similar decrease in number, consistent with the spread in changes seen in the DIGIT sample. However, some of the SPIRE data are significantly changed.  FU Ori still includes no significant CO detections in the SPIRE bands. V1057 Cyg exhibited a well-characterized cool component which was essentially unchanged ($\sim$ 3\%).  V1515 Cyg and V1331 Cyg had only 3 detections in the earlier work; V1331 Cyg now has 6 CO lines detected, and this has altered the rotational properties considerably.  The cool CO rotational temperatures increased in both sources, from $\sim$ 20-30 K to 75-100 K, and the number of molecules  dropped by a factor of 2.   We conclude that the improvement in absolute flux calibration can dramatically change the number of molecules detected in the SPIRE bands, and the correction for extended sources can alter the temperature.  

\section{Conclusions}
\label{sec:conclusion}

We have provided a dataset of {\it Herschel}-PACS and SPIRE spectroscopy to generate a broadly accessible archive with easily downloaded data products (grids of maps, continuum, lines, and full SEDs) of publishable quality. This dataset will complement and enhance the {\it Herschel} Science Archive, to serve the community with a Legacy Archive for young stellar objects.  Expected legacy investigations will reveal fundamental constraints on accretion processes, environments of protostars, and the conditions in emerging protoplanetary disks.

\acknowledgements

JDG would like to acknowledge numerous helpful discussions with the PACS and SPIRE ICC teams, via webinars, helpdesk queries, and contacts.   Support for this work, part of the  {\it Herschel} Open Time Key Project Program, was provided by NASA through  awards issued by the Jet
Propulsion Laboratory, California Institute of Technology.  J.-E.L. was supported by the 2013 Sabbatical Leave Program of Kyung Hee University (KHU-20131724) and also supported by Basic Science Research Program through the National Research Foundation of Korea(NRF) funded by the Ministry of Science, ICT and future Planning (2015R1A2A2A01004769).  JDG and Y-LY acknowledge support from NASA Herschel Science Center Cycle 2 grants.  RL acknowledges support from the Karl G. Henize Endowed Scholarship, the John W. Cox
Endowment for the Advanced Studies in Astronomy, and the College of Natural Sciences Summer
Undergraduate Research Fellowship at the University of Texas at Austin.  AK acknowledges support from the Foundation for Polish Science (FNP) and the Polish National Science Center grant 2013/11/N/ST9/00400.

\bibliographystyle{apj}

\section{Appendix: Sourcelist and Data Product Status}
\label{append}

The comprehensive source list, OBSID number, RA, Dec, and reduction type are listed in Table \ref{obslog}.

\begin{landscape}
\begin{longtable}[ht]{llp{2.3cm}p{2.3cm}llcp{1cm}p{1cm}}

\caption[Archive Sources]{Archive Sources} \\
\centering

Source & Other Name & PACS \quad OBSIDs & SPIRE \quad OBSID & RA & Dec & Prog. & Jitter Corr.& SPIRE 1-D \\ \toprule
\endfirsthead

Source & Other Name & PACS \quad \quad OBSIDs & SPIRE \quad OBSID & RA & Dec & Prog. & Jitter Corr.& SPIRE 1-D \\ \toprule
\endhead

AB Aur & ~ & 1342217842, 1342217843 & ~ & 04h55m45.8s & +30d33m04.3s & D & x & ~ \\
AS 205 & ~ & 1342215737, 1342215738 & & 16h11m31.4s & --18d38m26s & D & x & ~ \\
B1-a & ~ & 1342216182, 1342216183 & 1342249475 & 03h33m16.7s & +31d07m55.2s & D,C & x & x \\
B1-c & ~ & 1342216213, 1342216214 & 1342249476 & 03h33m17.9s & +31d09m31.9s & D,C & x & x \\
B335 & ~ & 1342208889, 1342208888 & 1342253652 & 19h37m00.9s & +07d34m09.7s & D,C & x & x \\
BHR 71 & ~ & 1342212230, 1342212231 & 1342248249 & 12h01m36.3s & --65d08m53.0s &  D,C & x & x  \\
Ced110-IRS4 & ~ & ~ & 1342248246 & 11h06m47.0s & --77d22m32.4s & C & & x \\
DG Tau & ~ & 1342225730, 1342225731 & ~ & 04h27m04.7s & +26d06m16s & D & x & ~ \\
DK Cha & IRAS 12496--7650 & 1342188039, 1342188040 & 1342254037 & 12h53m17.2s & --77d07m10.7s & D,C & x & x \\
EC 82 & [EC92] 82 & 1342192975, 1342219435 & ~ & 18h29m56.9s & +01d14m47s & D & ~ & ~ \\
Elias 29 & ~ & 1342228519, 1342228520 & ~ & 16h27m09.4s & --24d37m18.6s & D & x & x \\
FU Ori & ~ & 1342250907, 1342250908 & 1342230412 & 05h45m22.4s & +09d04m12s & F & x & x \\
GSS30-IRS1 & ~ & 1342215678, 1342215679 & 1342251286 & 16h26m21.4s & --24d23m04.3s & D,C & x & x \\
HD 100453 & ~ & 1342211695, 1342211696 & ~ & 11h33m05.6s & --54d19m29s & D & x & ~ \\
HD 100546 & ~ & 1342188037, 1342188038 & ~ & 11h33m25.4s & --70d11m41s & D & x & ~ \\
HD 104237 & ~ & 1342207819, 1342207820 & ~ & 12h00m05.1s & --78d11m35s & D & x & ~ \\
HD 135344 & ~ & 1342213921, 1342213922 & ~ & 15h15m49.0s & --37d08m56s & D & x & ~ \\
HD 139614 & ~ & 1342215683, 1342215684 & ~ & 15h40m46.4s & --42d29m54s & D & x & ~ \\
HD 141569 & ~ & 1342213913  & & 15h49m57.8s & --03d55m16s & D & x & ~ \\
HD 142527 & ~ & 1342216174, 1342216175 & ~ & 15h56m41.9s & --42d19m23s & D & x & ~ \\
HD 142666 & ~ & 1342213916 & ~ & 15h56m40.0s & --22d01m40s & D & ~ & ~ \\
HD 144432 & ~ & 1342213919 & ~ & 16h06m58.0s & --27d43m10s & D & x & ~ \\
HD 144668 & ~ & 1342215641, 1342215642 & & 16h08m34.3s & --39h06m18s & D & x & ~ \\
HD 150193 & ~ & 1342227068 & & 16h40m17.9s & --23d53m45s & D & x & ~ \\
HD 163296 & ~ & 1342217819, 1342217820 & ~ & 17h56m21.3s & --21d57m22s & D & x & ~ \\
HD 169142 & ~ & 1342206987, 1342206988 & ~ & 18h24m29.8s & --29d46m49s & D & x & ~ \\
HD 179218 & ~ & 1342208884, 1342208885 & ~ & 19h11m11.3s & +15d47m16s & D & x & ~ \\
HD 203024 & ~ & 1342206975 & & 21h16m03.0s & +68d54m52s & D & ~ & ~ \\
HD 245906 & ~ & 1342228528 & ~ & 05h39m30.5s & +26d19m55s & D & ~ & ~ \\
HD 35187 & ~ & 1342217846 & ~ & 05h24m01.2s & +24d57m38s & D & x & ~ \\
HD 36112 & ~ & 1342228247, 1342228248 & ~ & 05h30m27.5s & +25d19m57s & D & x & ~ \\
HD 38120 & ~ & 1342226212, 1342226213 & ~ & 05h43m11.9s & --04d59m50s & D & x & ~ \\
HD 50138 & ~ & 1342206991, 1342206992 & ~ & 06h51m33.4s & --06d57m59s & D & x & ~ \\
HD 97048 & ~ & 1342199412, 1342199413 & ~ & 11h08m03.3s & --77d39m18s & D & x & ~ \\
HD 98922 & ~ & 1342210385 & ~ & 11h22m31.7s & --53d22m12s & D & ~ & ~ \\
HH 100 & ~ & ~ & 1342252897 & 19h01m49.1s & --36d58m16.0s & C & & \\
HH 46 & ~ & ~ & 1342245084 & 08h25m43.9s & --51d00m36.0s & C & & x \\
HT Lup & ~ & 1342213920 & ~ & 15h45m12.9s & --34d17m31s & D & ~ & ~ \\
IRAM 04191+1522 & ~ & 1342216654, 1342216655 & & 04h21m56.9s & +15d29m45.9s & D & ~ & ~ \\
IRAS 03245+3002 & L1455-IRS1 & 1342214677, 1342214676 & 1342249053 &  03h27m39.1s & +30d13m03.1s &  D,C & x & x \\
IRAS 03301+3111 & Perseus Bolo76 & 1342215668, 1342216181 & 1342249477 & 03h33m12.8s & +31d21m24.2s & D,C & x & x \\
IRAS 15398--3359 & B228 & ~ & 1342250515 & 15h43m01.3s &  --34d09m15.0s & C & & x \\
IRS 46 / IRS44 & Oph-IRS44/46 & 1342228474, 1342228475 & 1342251289 & 16h27m29.4s & --24d39m16.1s & D,C & ~ &  \\
IRS 48 & Oph-IRS48 & 1342227069, 1342227070 & ~ & 16h27m37.2s & --24d30m35s & D & x & ~ \\
IRS 63 & Oph-IRS63 & 1342228473, 1342228472 &  & 16h31m35.6s & --24d01m29.3s & D & x &  \\
L1014 & ~ & 1342208911, 1342208912 & 1342245857 & 21h24m07.5s & +49d59m09.0s & D,C & ~ & x \\
L1157 & ~ & 1342208909, 1342208908 & 1342247625 & 20h39m06.3s & +68d02m16.0s &  D,C & x & x \\
L1448-MM & ~ & 1342213683, 1342214675 & ~ & 03h25m38.9s & +30d44m05.4s &  D & x & x \\
L1455-IRS3 & IRAS 03249+2957 & 1342204122, 1342204123 & 1342249474 & 03h28m00.4s & +30d08m01.3s & D,C & ~ & x \\
L1489 & IRAS 04016+2610 & 1342216216, 1342216215 &  & 04h04m42.9s & +26d18m56.3s & D & x &  \\
L1527 & IRAS 04368+2557 & 1342192981, 1342192982 &  & 04h39m53.9s & +26d03m09.8s & D & x & ~ \\
L1551-IRS5 & ~ & 1342192805, 1342229711 & 1342249470 & 04h31m34.1s & +18d08m04.9s & D,C & x & x \\
L483 & IRAS 18140--0440 & ~ & 1342253649 & 18h17m29.9s & --04d39m39.5s & C & ~ & x \\
L723-MM & ~ & ~ & 1342245094 & 19h17m53.7s & +19d12m20.0s &  C & & x \\
RCrA-IRS5A & ~ & 1342207806, 1342207805 & 1342253646 & 19h01m48.1s & --36d57m22.7s & D,C & ~ & x \\
RCrA-IRS7B & ~ & 1342207807, 1342207808  & 1342242620 & 19h01m56.4s & --36d57m28.3s & D,C & x & x \\
RCrA-IRS7C & ~ & 1342206990, 1342206989 & 1342242621 & 19h01m55.3s & --36d57m17.0s & D,C & ~ & x \\
RNO 90 & ~ & 1342228206 & ~ & 16h34m09.2s & --15d48m17s & D & x & ~ \\
RNO 91 & ~ & ~ & 1342251285 &  16h34m29.3s & --15d47m01.4s & C & & x \\
RU Lup & ~ & 1342215682 & ~ & 15h56m42.3s & --37d49m16s & D & x & ~ \\
RY Lup & ~ & 1342216171 & ~ & 15h59m28.4s & --40d21m51s & D & x & ~ \\
S Cra & ~ & 1342207809, 1342207810 & ~ & 19h01m08.6s & --36d57m20s & D & x & ~ \\
Serpens-SMM3 & & 1342193216, 1342193214 & ~ & 18h29m59.3s & +01d14m01.7s & D & x & ~ \\
Serpens-SMM4 & & 1342193217, 1342193215 & ~ & 18h29m56.7s & +01d13m17.2s & D & ~ & ~ \\
SR 21 & ~ & 1342227209, 1342227210 & ~ & 16h27m10.3s & --24d19m13s & D & x & ~ \\
TMC 1A & IRAS 04362+2535 & 1342192987, 1342192988 & 1342250510 & 04h39m35.0s & +25d41m45.5s & D,C & x & x \\
TMC 1 & IRAS 04381+2540 & 1342225803, 1342225804 & 1342250512 & 04h41m12.7s & +25d46m35.9s & D,C & x & x \\
TMR 1 & IRAS 04361+2547 &  1342192985, 1342192986 & 1342250509 & 04h39m13.9s & +25d53m20.6s & D,C & x & x \\
V1057 Cyg & ~ & 1342235853, 1342235852 & 1342221695 & 20h58m53.7s & +44d15m29s & F & x & x  \\
V1331 Cyg & ~ & 1342233446, 1342233445 & 1342221694 & 21h01m09.2s & +50d21m45s & F & x & x \\
V1515 Cyg & ~ & 1342235691, 1342235690 & 1342221685 & 20h23m48.0s & +42d12m26s &  F & x & x  \\
V1735 Cyg & Elias 1-12 & 1342235849, 1342235848 & 1342219560 & 21h47m20.7s & +47d32m04s & F & x & ~ \\
VLA1623-243 & ~ & 1342213918, 1342213917 & 1342251287 & 16h26m26.4s & --24d24m30.0s & D,C & x & x \\
WL 12 & ~ & 1342228187, 1342228188 & 1342251290 & 16h26m44.2s & --24d34m48.4s & D,C & x & x \\
\bottomrule
\caption{Contents of the archive organized by source. The `Prog.' column refers to the first program in which the source was observed, where D $=$ DIGIT, F $=$ FOOSH, and C $=$ COPS. The `Jitter Corr.' column indicates the accessibility of the jitter corrected data. Sources with an `x' in this column use the jitter-corrected products. The `SPIRE 1-D' column indicates the accessibility of SPIRE extended calibrated 1-D spectra.}
\label{obslog}
\end{longtable}
\end{landscape}

\begin{table*}[ht]
\small
\ra{1.15}
\caption{Line list for single Gaussian fitting: \owater}
\centering
\begin{tabular}{ll|ll}

\toprule
Wavelength~(\micron) & Line~name & Wavelength~(\micron) & Line~name \\
\hline 
55.13238 & $\owater~8_{27}-7_{16}$ & 55.84108 & $\owater~10_{29}-10_{110}$ \\
56.81777 & $\owater~9_{09}-8_{18}$ & 57.39511 & $\owater~7_{52}-8_{27}$ \\
58.70051 & $\owater~4_{32}-3_{21}$ & 61.31782 & $\owater~5_{41}-6_{16}$ \\
62.92979 & $\owater~9_{18}-9_{09}$ & 63.32514 & $\owater~8_{18}-7_{07}$ \\
63.91602 & $\owater~6_{61}-6_{52}$ & 63.95796 & $\owater~7_{61}-7_{52}$ \\
65.16779 & $\owater~6_{25}-5_{14}$ & 66.09434 & $\owater~7_{16}-6_{25}$ \\
66.43937 & $\owater~3_{30}-2_{21}$ & 67.27067 & $\owater~3_{30}-3_{03}$ \\
70.70435 & $\owater~8_{27}-8_{18}$ & 71.94881 & $\owater~7_{07}-6_{16}$ \\
74.94690 & $\owater~7_{25}-6_{34}$ & 75.38257 & $\owater~3_{21}-2_{12}$ \\
75.49740 & $\owater~8_{54}-8_{45}$ & 75.83188 & $\owater~6_{52}-6_{43}$ \\
75.91180 & $\owater~5_{50}-5_{41}$ & 77.76345 & $\owater~7_{52}-7_{43}$ \\
78.74431 & $\owater~4_{23}-3_{12}$ & 81.40747 & $\owater~9_{27}-9_{18}$ \\
82.03351 & $\owater~6_{16}-5_{05}$ & 82.97879 & $\owater~8_{36}-8_{27}$ \\
84.76907 & $\owater~7_{16}-7_{07}$ & 85.77088 & $\owater~8_{45}-8_{36}$ \\
92.81312 & $\owater~6_{43}-6_{34}$ & 94.64643 & $\owater~6_{25}-6_{16}$ \\
94.70758 & $\owater~4_{41}-4_{32}$ & 104.09629 & $\owater~6_{34}-6_{25}$ \\
108.07588 & $\owater~2_{21}-1_{10}$ & 112.51342 & $\owater~7_{43}-7_{34}$ \\
112.80576 & $\owater~4_{41}-5_{14}$ & 113.54021 & $\owater~4_{14}-3_{03}$ \\
114.45656 & $\owater~9_{27}-10_{110}$ & 116.35288 & $\owater~8_{36}-9_{09}$ \\
116.78196 & $\owater~7_{34}-6_{43}$ & 121.72477 & $\owater~4_{32}-4_{23}$ \\
123.46352 & $\owater~9_{36}-9_{27}$ & 127.88735 & $\owater~7_{25}-7_{16}$ \\
129.34220 & $\owater~9_{45}-9_{36}$ & 132.41173 & $\owater~4_{23}-4_{14}$ \\
133.55242 & $\owater~8_{36}-7_{43}$ & 134.93863 & $\owater~5_{14}-5_{05}$ \\
136.49943 & $\owater~3_{30}-3_{21}$ & 156.26908 & $\owater~5_{23}-4_{32}$ \\
159.05453 & $\owater~8_{45}-7_{52}$ & 159.40427 & $\owater~6_{34}-7_{07}$ \\
160.51410 & $\owater~5_{32}-5_{23}$ & 166.81885 & $\owater~7_{34}-7_{25}$ \\
174.63028 & $\owater~3_{03}-2_{12}$ & 174.92441 & $\owater~4_{32}-5_{05}$ \\
179.53118 & $\owater~2_{12}-1_{01}$ & 180.49281 & $\owater~2_{21}-2_{12}$ \\
187.81488 & $\owater~8_{54}-9_{27}$ & 212.53093 & $\owater~5_{23}-5_{14}$ \\
226.76647 & $\owater~6_{25}-5_{32}$ & 229.21129 & $\owater~8_{45}-9_{18}$ \\
231.25379 & $\owater~8_{27}-7_{34}$ & 234.53645 & $\owater~7_{43}-6_{52}$ \\
256.59928 & $\owater~8_{54}-7_{61}$ & 257.80114 & $\owater~3_{21}-3_{12}$ \\
258.82221 & $\owater~6_{34}-5_{41}$ & 259.98875 & $\owater~3_{12}-2_{21}$ \\
261.46379 & $\owater~7_{25}-8_{18}$ & 273.19988 & $\owater~3_{12}-3_{03}$ \\
483.00214 & $\owater~5_{32}-4_{41}$ & 538.30236 & $\owater~1_{10}-1_{01}$ \\
\bottomrule
\end{tabular}
\label{oh2o_list}
\end{table*}

\begin{table*}[ht]
\small
\ra{1.15}
\caption{Line list for single Gaussian fitting: \pwater}
\centering
\begin{tabular}{ll|ll}

\toprule
Wavelength~(\micron) & Line~name & Wavelength~(\micron) & Line~name \\
\hline 
55.85970 & $\pwater~6_{51}-7_{26}$ & 55.98480 & $\pwater~7_{71}-7_{62}$ \\
56.02824 & $\pwater~10_{19}-10_{010}$ & 56.32640 & $\pwater~4_{31}-3_{22}$ \\
56.77240 & $\pwater~9_{19}-8_{08}$ & 57.63798 & $\pwater~4_{22}-3_{13}$ \\
57.71080 & $\pwater~8_{17}-7_{26}$ & 58.37824 & $\pwater~6_{42}-7_{17}$ \\
59.98863 & $\pwater~7_{26}-6_{15}$ & 60.16365 & $\pwater~8_{26}-7_{35}$ \\
60.23082 & $\pwater~7_{62}-8_{35}$ & 61.81016 & $\pwater~4_{31}-4_{04}$ \\
61.91772 & $\pwater~4_{40}-5_{15}$ & 62.43311 & $\pwater~9_{28}-9_{19}$ \\
63.45961 & $\pwater~8_{08}-7_{17}$ & 63.88177 & $\pwater~7_{62}-7_{53}$ \\
67.09082 & $\pwater~3_{31}-2_{20}$ & 71.06907 & $\pwater~5_{24}-4_{13}$ \\
71.54146 & $\pwater~7_{17}-6_{06}$ & 71.78955 & $\pwater~5_{51}-6_{24}$ \\
72.03407 & $\pwater~8_{17}-8_{08}$ & 73.61471 & $\pwater~9_{37}-9_{28}$ \\
75.78324 & $\pwater~5_{51}-5_{42}$ & 75.81532 & $\pwater~7_{53}-7_{44}$ \\
76.42386 & $\pwater~6_{51}-6_{42}$ & 78.93042 & $\pwater~6_{15}-5_{24}$ \\
80.22431 & $\pwater~9_{46}-9_{37}$ & 80.55884 & $\pwater~8_{53}-8_{44}$ \\
81.21765 & $\pwater~7_{26}-7_{17}$ & 81.69220 & $\pwater~8_{35}-7_{44}$ \\
83.28606 & $\pwater~6_{06}-5_{15}$ & 89.99061 & $\pwater~3_{22}-2_{11}$ \\
90.05205 & $\pwater~7_{44}-7_{35}$ & 94.21193 & $\pwater~5_{42}-5_{33}$ \\
103.91892 & $\pwater~6_{42}-6_{33}$ & 103.94278 & $\pwater~6_{15}-6_{06}$ \\
111.63078 & $\pwater~5_{24}-5_{15}$ & 113.95081 & $\pwater~5_{33}-5_{24}$ \\
117.68692 & $\pwater~9_{46}-8_{53}$ & 118.40826 & $\pwater~9_{37}-8_{44}$ \\
122.52519 & $\pwater~8_{44}-8_{35}$ & 125.35683 & $\pwater~4_{04}-3_{13}$ \\
126.71711 & $\pwater~3_{31}-3_{22}$ & 130.32199 & $\pwater~7_{53}-8_{26}$ \\
137.68652 & $\pwater~7_{35}-8_{08}$ & 138.53127 & $\pwater~3_{13}-2_{02}$ \\
138.64412 & $\pwater~8_{44}-7_{53}$ & 144.52144 & $\pwater~4_{13}-3_{22}$ \\
146.92643 & $\pwater~4_{31}-4_{22}$ & 148.71156 & $\pwater~8_{35}-8_{26}$ \\
148.79410 & $\pwater~5_{42}-6_{15}$ & 156.19792 & $\pwater~3_{22}-3_{13}$ \\
158.31551 & $\pwater~3_{31}-4_{04}$ & 159.48926 & $\pwater~8_{26}-9_{19}$ \\
167.03912 & $\pwater~6_{24}-6_{15}$ & 169.74305 & $\pwater~7_{35}-6_{42}$ \\
170.14342 & $\pwater~6_{33}-6_{24}$ & 174.61126 & $\pwater~5_{33}-6_{06}$ \\
187.11543 & $\pwater~4_{13}-4_{04}$ & 208.08146 & $\pwater~7_{26}-6_{33}$ \\
208.91861 & $\pwater~9_{46}-10_{19}$ & 222.95328 & $\pwater~7_{44}-8_{17}$ \\
243.98004 & $\pwater~2_{20}-2_{11}$ & 248.25302 & $\pwater~4_{22}-4_{13}$ \\
251.75737 & $\pwater~8_{53}-7_{62}$ & 255.68729 & $\pwater~7_{44}-6_{51}$ \\
269.27908 & $\pwater~1_{11}-0_{00}$ & 303.46381 & $\pwater~2_{02}-1_{11}$ \\
308.97175 & $\pwater~5_{24}-4_{31}$ & 327.23126 & $\pwater~4_{22}-3_{31}$ \\
330.82984 & $\pwater~9_{28}-8_{35}$ & 398.65260 & $\pwater~2_{11}-2_{02}$ \\
613.72660 & $\pwater~6_{24}-7_{17}$ & 631.57098 & $\pwater~5_{33}-4_{40}$ \\
636.66801 & $\pwater~6_{42}-5_{51}$ & ~ & ~ \\
\bottomrule
\end{tabular}
\label{ph2o_list}
\end{table*}

\begin{table*}[ht]
\ra{1.15}
\caption{Line list for single Gaussian fitting: \co\ and \thirteenco}
\centering
\begin{tabular}{ll|ll}

\toprule
\multicolumn{2}{c}{\co} & \multicolumn{2}{c}{\thirteenco} \\
Wavelength~(\micron) & Line~name & Wavelength~(\micron) & Line~name \\
\hline 
54.98622  & \coul{48}{47} & 209.48144 & \tcoul{13}{12} \\
56.12190  & \coul{47}{46} & 226.90368 & \tcoul{12}{11} \\
57.30773  & \coul{46}{45} & 247.49662 & \tcoul{11}{10} \\
58.54705  & \coul{45}{44} & 272.21148 & \tcoul{10}{9} \\
59.84349  & \coul{44}{43} & 302.42221 & \tcoul{9}{8} \\
61.20106  & \coul{43}{42} & 340.18978 & \tcoul{8}{7} \\
62.62410  & \coul{42}{41} & 388.75282 & \tcoul{7}{6} \\
64.11741  & \coul{41}{40} & 453.50906 & \tcoul{6}{5} \\
65.68791  & \coul{40}{39} & 544.17444 & \tcoul{5}{4} \\
67.33814  & \coul{39}{38} & & \\
69.07614  & \coul{38}{37} & & \\
70.90902  & \coul{37}{36} & & \\
72.84469  & \coul{36}{35} & & \\
74.89194  & \coul{35}{34} & & \\
77.06064  & \coul{34}{33} & & \\
79.36181  & \coul{33}{32} & & \\
81.80787  & \coul{32}{31} & & \\
84.41284  & \coul{31}{30} & & \\
87.19261  & \coul{30}{29} & & \\ \cline{3-4}
90.16527  & \coul{29}{28} & 173.63580 & \coul{15}{14} \\
93.35147  & \coul{28}{27} & 186.00397 & \coul{14}{13} \\
104.44758 & \coul{25}{24} & 200.27751 & \coul{13}{12} \\
108.76555 & \coul{24}{23} & 216.93275 & \coul{12}{11} \\
113.46045 & \coul{23}{22} & 236.61924 & \coul{11}{10} \\
118.58370 & \coul{22}{21} & 260.24634 & \coul{10}{9} \\
124.19648 & \coul{21}{20} & 289.12761 & \coul{9}{8} \\
130.37221 & \coul{20}{19} & 325.23335 & \coul{8}{7} \\
137.19978 & \coul{19}{18} & 371.65974 & \coul{7}{6} \\
144.78783 & \coul{18}{17} & 433.56713 & \coul{6}{5} \\
153.27056 & \coul{17}{16} & 520.24412 & \coul{5}{4} \\ 
162.81572 & \coul{16}{15} & 650.26787 & \coul{4}{3} \\
\bottomrule
\end{tabular}
\label{co_list}
\end{table*}

\begin{table*}[ht]
\ra{1.15}
\caption{Line list for single Gaussian fitting: OH ,\hcoplus, and \chplus}
\centering
\begin{tabular}{ll|ll}

\toprule
\multicolumn{2}{c}{OH} & \multicolumn{2}{c}{\hcoplus\ \&\ \chplus} \\
Wavelength~(\micron) & Line~name & Wavelength~(\micron) & Line~name \\
\hline
55.89231 & OH\ohul{1/2}{9/2}{+}$-$\ohul{1/2}{7/2}{-}   & 210.28817 & \hcoul{16}{15} \\
55.95141 & OH\ohul{1/2}{9/2}{-}$-$\ohul{1/2}{7/2}{+}   & 224.28136 & \hcoul{15}{14} \\
65.13337 & OH\ohul{3/2}{9/2}{-}$-$\ohul{3/2}{7/2}{+}   & 240.27541 & \hcoul{14}{13} \\
65.28048 & OH\ohul{3/2}{9/2}{+}$-$\ohul{3/2}{7/2}{-}   & 258.73207 & \hcoul{13}{12} \\
71.17262 & OH\ohul{1/2}{7/2}{-}$-$\ohul{1/2}{5/2}{+}   & 280.26712 & \hcoul{12}{11} \\
71.21723 & OH\ohul{1/2}{7/2}{+}$-$\ohul{1/2}{5/2}{-}   & 305.71952 & \hcoul{11}{10} \\
79.11754 & OH\ohul{1/2}{1/2}{-}$-$\ohul{3/2}{3/2}{+}   & 336.26531 & \hcoul{10}{9} \\
79.18106 & OH\ohul{1/2}{1/2}{+}$-$\ohul{3/2}{3/2}{-}   & 373.60195 & \hcoul{9}{8} \\
84.42237 & OH\ohul{3/2}{7/2}{+}$-$\ohul{3/2}{5/2}{-}   & 420.27521 & \hcoul{8}{7} \\
84.59877 & OH\ohul{3/2}{7/2}{-}$-$\ohul{3/2}{5/2}{+}   & 480.28810 & \hcoul{7}{6} \\
115.15057 & OH\ohul{1/2}{5/2}{+}$-$\ohul{3/2}{7/2}{-}  & 560.30913 & \hcoul{6}{5} \\
115.38441 & OH\ohul{1/2}{5/2}{-}$-$\ohul{3/2}{7/2}{+}  & 60.24659  & \chul{6}{5} \\
119.23740 & OH\ohul{3/2}{5/2}{-}$-$\ohul{3/2}{3/2}{+}  & 72.13950  & \chul{5}{4} \\
119.44450 & OH\ohul{3/2}{5/2}{+}$-$\ohul{3/2}{3/2}{-}  & 119.85466 & \chul{3}{2} \\
134.84152 & OH\ohul{1/2}{7/2}{-}$-$\ohul{3/2}{9/2}{+}  & 358.99894 & \chul{1}{0} \\
135.95981 & OH\ohul{1/2}{7/2}{+}$-$\ohul{3/2}{9/2}{-}  & ~ & ~ \\
154.78349 & OH\ohul{1/2}{9/2}{+}$-$\ohul{3/2}{11/2}{-} & ~ & ~ \\
157.80984 & OH\ohul{1/2}{9/2}{-}$-$\ohul{3/2}{11/2}{+} & ~ & ~ \\
163.12467 & OH\ohul{1/2}{3/2}{+}$-$\ohul{1/2}{1/2}{-}  & ~ & ~ \\
163.40013 & OH\ohul{1/2}{3/2}{-}$-$\ohul{1/2}{1/2}{+}  & ~ & ~ \\
\bottomrule
\multicolumn{4}{p{0.9\textwidth}}{Note\textemdash The notation of OH transition follows the form of ${\rm ^{2}\Pi_{J_{e},J_{total}}^{parity}}$. ${\rm J_{e}}$ is the electron angular momentum quantum number, while ${\rm J_{total}}$ is the total angular momentum quantum number including electron and nuclear rotation.  Please see more detail of the 163.12467 $\mu$m OH line in \S \ref{sec:line_list}.} \\
\end{tabular}
\label{oh_hco_list}
\end{table*}

\begin{table*}[ht]
\ra{1.15}
\caption{Line list for single Gaussian fitting: atomic fine-structure lines}
\centering
\begin{tabular}{ll|ll}

\toprule
Wavelength~(\micron) & Line~name & Wavelength~(\micron) & Line~name \\
\hline
63.18367  & $\rm [O~\textsc{i}]~\tensor*[^{3}]{P}{_{1}}-\tensor*[^{3}]{P}{_{2}}$ & 121.911$\pm 0.006^{a}$ & $\rm [N~\textsc{ii}]~^{3}P_{2}-\tensor*[^{3}]{P}{_{1}}$ \\
145.48056 & $\rm [O~\textsc{i}]~\tensor*[^{3}]{P}{_{0}}-\tensor*[^{3}]{P}{_{1}}$ & 157.69228 & $\rm [C~\textsc{ii}]~\tensor*[^{2}]{P}{_{3/2}}-\tensor*[^{2}]{P}{_{1/2}}$ \\
205.170$\pm 0.004^{a}$ & $\rm [N~\textsc{ii}]~\tensor*[^{3}]{P}{_{1}}-\tensor*[^{3}]{P}{_{0}}$ & 230.34913 & $\rm [C~\textsc{i}]~^{3}P_{2}-\tensor*[^{3}]{P}{_{0}}$ \\
370.42438 & $\rm [C~\textsc{i}]~\tensor*[^{3}]{P}{_{2}}-\tensor*[^{3}]{P}{_{1}}$ & 609.15069 & $\rm [C~\textsc{i}]~\tensor*[^{3}]{P}{_{1}}-\tensor*[^{3}]{P}{_{0}}$ \\
\bottomrule
\multicolumn{4}{L{\textwidth}}{$^{a}$ The mean values and the standard deviation of the mean of line centeroids are measured from the corresponding fitted lines with SNR $>$ 10.  The initial guesses were 121.9~\micron, and 205.178~\micron.}
\end{tabular}
\label{atomic_list}
\end{table*}

\begin{table*}[ht]
\ra{1.15}
\caption{Blended lines fit with double Gaussian profile}
\centering
\begin{tabular}{rlll}

\toprule
\multicolumn{2}{c}{Wavelength~(\micron)} & \multicolumn{2}{c}{Line~names} \\
\hline
55.85970 & 55.89231 & $\pwater~6_{51}-7_{26}$ & OH\ohul{1/2}{9/2}{+}$-$\ohul{1/2}{7/2}{-} \\
55.95141 & 55.98480 & OH\ohul{1/2}{9/2}{-}$-$\ohul{1/2}{7/2}{+} & $\pwater~7_{71}-7_{62}$ \\
56.77240 & 56.81777 & $\pwater~9_{19}-8_{08}$ & $\owater~9_{09}-8_{18}$ \\
65.13337 & 65.16779 & OH\ohul{3/2}{9/2}{-}$-$\ohul{3/2}{7/2}{+} & $\owater~6_{25}-5_{14}$ \\
71.17262 & 71.21723 & OH\ohul{1/2}{7/2}{-}$-$\ohul{1/2}{5/2}{+} & OH\ohul{1/2}{7/2}{+}$-$\ohul{1/2}{5/2}{-} \\
75.81532 & 75.83188 & $\pwater~7_{53}-7_{44}$ & $\owater~6_{52}-6_{43}$ \\
79.11754 & 79.18106 & OH\ohul{1/2}{1/2}{-}$-$\ohul{3/2}{3/2}{+} & OH\ohul{1/2}{1/2}{+}$-$\ohul{3/2}{3/2}{-} \\
84.41284 & 84.42237 & \coul{31}{30} & OH\ohul{3/2}{7/2}{+}$-$\ohul{3/2}{5/2}{-} \\
113.46045 & 113.54021 & \coul{23}{22} & $\owater~4_{14}-3_{03}$ \\
118.40826 & 118.58370 & $\pwater~9_{37}-8_{44}$ & \coul{22}{21} \\
130.32199 & 130.37221 & \coul{20}{19} & $\pwater~7_{53}-8_{26}$ \\
134.84152 & 134.93863 & OH\ohul{1/2}{7/2}{-}$-$\ohul{3/2}{9/2}{+} & $\owater~5_{14}-5_{05}$ \\
156.19792 & 156.26908 & $\pwater~3_{22}-3_{13}$ & $\owater~5_{23}-4_{32}$ \\
144.52144 & 144.78783 & $\pwater~4_{13}-3_{22}$ & \coul{18}{17} \\
166.81885 & 167.03912 & $\owater~7_{34}-7_{25}$ & $\pwater~6_{24}-6_{15}$ \\
174.61126 & 174.63028 & $\pwater~5_{33}-6_{06}$ & $\owater~3_{03}-2_{12}$ \\
259.98875 & 260.24634 & $\owater~3_{12}-2_{21}$ & \coul{10}{9} \\
272.21148 & 273.19988 & \tcoul{10}{9} & $\owater~3_{12}-3_{03}$ \\
302.42221 & 303.46381 & \tcoul{9}{8} & $\pwater~2_{02}-1_{11}$ \\
370.42438 & 371.65974 & $\rm [C~\textsc{i}]~\tensor*[^{3}]{P}{_{2}}-\tensor*[^{3}]{P}{_{1}}$ & \coul{7}{6} \\
609.15069 & 613.72660 & $\rm [C~\textsc{i}]~\tensor*[^{3}]{P}{_{1}}-\tensor*[^{3}]{P}{_{0}}$ & $\pwater~6_{24}-7_{17}$ \\
\bottomrule
\end{tabular}
\label{dg_list}
\end{table*}


\begin{landscape}
	\begin{table*}[htcb!]
		\small
		\caption{A portion of the 1-D spectrum fitting results for BHR71.  The tables in the ASCII files have the same columns and style except that the rows are chopped into three parts here for better display.  Also this table  has selected lines from different parts of the original results to demonstrate different flags, etc. As mentioned in \S \ref{sec:report}, any column with {\tt -999} indicates a fitting result that is not well-constrained.  Therefore, the {\tt Validity} flag is set to be 0.  The {\tt Pixel\_No.} column lists ``c'' for the 1-D spectrum measurements, and the specific pixel number/name for cube measurements. This table (all line measurements for all sources) is published in its entirety in the electronic edition.}
		\tt
		\begin{tabular}{r r r r r r r}
			\toprule
			              Line &      LabWL(um) &        ObsWL(um) &     Sig\_Cen(um) &       Str(W/cm2) &  Sig\_str(W/cm2) \\
		               CO48-47 &    54.98621750 &      54.98154779 &    0.03082777749 & -2.277802385e-21 &  2.967605715e-21 \\
              p-H2O4\_31-3\_22 &    56.32640076 &      56.30309023 &   0.007689705798 &  8.970946845e-21 &  2.958435088e-21 \\
                       CO46-45 &    57.30772781 &      57.30463621 &     0.2850716473 & -2.699435981e-22 &  3.260885618e-21 \\
  			  o-H2O6\_61-6\_52 &    63.91602325 &      63.93649292 &     -999.0000000 &  2.197260987e-21 &  2.763702731e-21 \\
  			  o-H2O7\_61-7\_52 &    63.95696259 &      63.93729134 &    0.02943608129 &  2.198141514e-21 &  2.764042837e-21 \\
  		      p-H2O9\_37-8\_44 &    118.4082565 &      118.4959793 &     -999.0000000 &  3.497948853e-20 &  1.314679686e-20 \\
	                   CO22-21 &    118.5837021 &      118.5876394 &   0.003691032704 &  7.394416427e-20 &  3.865617628e-21 \\
              p-H2O8\_35-8\_26 &    148.7115631 &      148.7528381 &     -999.0000000 & -3.672127599e-21 &  1.629373707e-21 \\
                         CO4-3 &    650.2678833 &      650.2828864 &    0.01499258376 &  4.128831785e-20 &  6.530348726e-22 \\
                     CI3P2-3P1 &    370.4243774 &      370.4243774 &     -998.0000000 &      0.000000000 &     -998.0000000 \\

                \hline \hline
   				      FWHM(um) &  Sig\_FWHM(um) &   Base(W/cm2/um) &  Noise(W/cm2/um) &              SNR &          E\_u(K) \\
   				 0.03897858372 &   -998.0000000 &  5.860891006e-18 &  8.720767999e-20 &     0.6297869004 &      6457.230000 \\
   				 0.03906316840 &   -998.0000000 &  6.295500447e-18 &  8.226700068e-20 &      2.623635637 &      552.3000000 \\
   				 0.03912018226 &   -998.0000000 &  6.354233016e-18 &  1.049475567e-19 &    0.06179569179 &      5939.210000 \\
 				 0.03936703190 &   -998.0000000 &  7.381455348e-18 &  6.509757235e-20 &     0.8058283595 &      1503.600000 \\
 				 0.03936783131 &   -998.0000000 &  7.381706869e-18 &  7.165146509e-20 &     0.7323985066 &      1749.900000 \\
 				  0.1169993003 &   -998.0000000 &  6.777696602e-18 &  9.334557437e-20 &      3.010196356 &      1749.900000 \\
 				  0.1170676737 &   -998.0000000 &  6.772795551e-18 &  3.430364591e-20 &      17.30553393 &      1397.380000 \\
   				  0.1251977122 &   -998.0000000 &  4.240183287e-18 &  1.615138300e-20 &      1.706750436 &      1511.000000 \\
   				   2.951818452 &  0.03529221031 &  1.568093311e-20 &  3.629903522e-22 &      36.21602920 &      55.32000000 \\
   				  0.9943682530 &   -998.0000000 &  2.116988821e-19 &  1.348351580e-21 &      0.000000000 &      62.46200000 \\

   				  \hline \hline
   			            A(s-1) &          g &          RA(deg) &         Dec(deg) & Pixel\_No. &          Blend &  Validity \\
      		    0.006556000000 &         97 &      180.3982553 &     -65.14761521 &          c &              x &         1 \\
      		       1.452000000 &          9 &      180.3982553 &     -65.14761521 &          c &              x &         1 \\
      		    0.006091000000 &         93 &      180.3982553 &     -65.14761521 &          c &              x &         1 \\
			 	  0.3533000000 &         39 &      180.3982553 &     -65.14761521 &          c &            Red &         0 \\
			 	  0.5828000000 &         45 &      180.3982553 &     -65.14761521 &          c &       Red/Blue &         0 \\
 			     0.01724000000 &         19 &      180.3982553 &     -65.14761521 &          c &            Red &         0 \\
 			    0.001006000000 &         45 &      180.3982553 &     -65.14761521 &          c &       Red/Blue &         1 \\
      		      0.1235000000 &         17 &      180.3982553 &     -65.14761521 &          c &            Red &         0 \\
      		   6.126000000e-06 &          9 &      180.4010925 &     -65.14799500 &          c &              x &         1 \\
      		   2.650000000e-07 &          5 &      180.4010925 &     -65.14799500 &          c & DoubleGaussian &         1 \\
       		\bottomrule
       		\label{table:report}
		\end{tabular}
	\end{table*}
\end{landscape}

\end{document}